\pdfoutput=1
\documentclass[11pt]{article}
\usepackage[margin=1in]{geometry}

\usepackage{mathtools}
\usepackage{amssymb,amsthm}
\usepackage{thmtools}
\usepackage[varbb]{newpxmath}
\usepackage{physics} 
\usepackage{nicefrac}

\usepackage{graphicx}
\usepackage{enumerate}
\usepackage{enumitem}
\usepackage{bbm} 
\usepackage{appendix}
\usepackage{caption}
\RequirePackage{subfigure}

\usepackage{algorithm}
\usepackage{multicol}
\usepackage[framemethod=TikZ]{mdframed}
\mdfsetup{%
backgroundcolor=gray!10, , %royalBlue!20, 
roundcorner=4pt,
linewidth=1pt}
\usepackage[noend]{algpseudocode}

\usepackage[dvipsnames]{xcolor}
\usepackage{color}
\definecolor{niceRed}{RGB}{190,38,38}
\definecolor{niceYellow}{HTML}{f5b400}
\definecolor{blueGrotto}{HTML}{059DC0}
\definecolor{royalBlue}{HTML}{057DCD}
\definecolor{navyBlue}{HTML}{0B579C}
\definecolor{limeGreen}{HTML}{81B622}
\definecolor{nicePurple}{HTML}{9c27b0}
\definecolor{lightRoyalBlue}{HTML}{def2ff}  
\definecolor{gold}{HTML}{ffa300}

\usepackage{hyperref} 
\usepackage[capitalize,nameinlink]{cleveref}

\hypersetup{
  colorlinks = true, 
  urlcolor = {blueGrotto},
  linkcolor = {royalBlue},
  citecolor = {navyBlue}
}
\usepackage[
  backref=true,
  backend=biber,
  natbib=true,
  style=alphabetic,
  sorting=alphabeticlabel,
  sortcites=true,
  minbibnames=3,
  maxbibnames=999,
  mincitenames=4,
  maxcitenames=4,
  minalphanames=4,
  maxalphanames=4
]{biblatex}

\usepackage{enumitem}

\DeclareSortingTemplate{alphabeticlabel}{
  \sort[final]{%
    \field{labelalpha} % Sort by the alphabetic label (\eg{}, DKL, DLL)
  }
  \sort{%
    \field{year}   % If labels are the same, sort by year
  }
  \sort{%
    \field{title}  % If both label and year are the same, sort by title
  }
}

\AtBeginRefsection{\GenRefcontextData{sorting=ynt}}
\AtEveryCite{\localrefcontext[sorting=ynt]}
\usepackage{algpseudocode}

\usepackage[suppress]{color-edits}

\addauthor[Shuchen]{sl}{blue}
\addauthor[Alkis]{ak}{niceRed}
\addauthor[Manolis]{mz}{limeGreen}
\addauthor[Pravesh]{pk}{nicePurple}

\DeclareMathOperator{\lap}{\mathrm{Lap}}
\DeclareMathOperator{\sym}{\mathrm{Sym}}
\newcommand{\fnorm}[1]{\left\| #1 \right\|_{\mathrm{F}}}

\usepackage{tikz}
\usepackage{bm}

\theoremstyle{plain} %%%%%% Theorem style 
\newtheorem{theorem}{Theorem}[section]
\newtheorem{corollary}[theorem]{Corollary}

\newtheorem{lemma}[theorem]{Lemma}
\newtheorem{fact}[theorem]{Fact}

\newtheorem{definition}{Definition}

\newtheorem*{definition*}{Definition}

\theoremstyle{definition} %%%%%% Definition style              

\theoremstyle{remark} %%%%%% Remark Style 

\newtheorem{remark}{Remark}
\AfterEndEnvironment{definition}{\noindent\ignorespaces}
\AfterEndEnvironment{infdefinition}{\noindent\ignorespaces}
\AfterEndEnvironment{example}{\noindent\ignorespaces}
\AfterEndEnvironment{assumption}{\noindent\ignorespaces}
\AfterEndEnvironment{lemma}{\noindent\ignorespaces}
\AfterEndEnvironment{theorem}{\noindent\ignorespaces}
\AfterEndEnvironment{proposition}{\noindent\ignorespaces}
\AfterEndEnvironment{fact}{\noindent\ignorespaces}
\AfterEndEnvironment{question}{\noindent\ignorespaces}
\AfterEndEnvironment{corollary}{\noindent\ignorespaces}
\AfterEndEnvironment{model}{\noindent\ignorespaces}
\AfterEndEnvironment{remark}{\noindent\ignorespaces}
\AfterEndEnvironment{proof}{\noindent\ignorespaces}
\AfterEndEnvironment{fact}{\noindent\ignorespaces}
\AfterEndEnvironment{minftheorem}{\noindent\ignorespaces}
\AfterEndEnvironment{inftheorem}{\noindent\ignorespaces}
\AfterEndEnvironment{maintheorem}{\noindent\ignorespaces}
\AfterEndEnvironment{restatable}{\noindent\ignorespaces}
\AfterEndEnvironment{observation}{\noindent\ignorespaces}

\crefname{section}{Section}{Sections}
\crefname{theorem}{Theorem}{Theorems}
\crefname{theorem*}{Theorem}{Theorems}
\crefname{inftheorem}{Informal Theorem}{Informal Theorems}
\crefname{assumption}{Assumption}{Assumptions}
\crefname{lemma}{Lemma}{Lemmas}
\crefname{definition}{Definition}{Definitions}
\crefname{infdefinition}{Informal Definition}{Informal Definitions}
\crefname{conjecture}{Conjecture}{Conjectures}
\crefname{corollary}{Corollary}{Corollaries}
\crefname{construction}{Construction}{Constructions}
\crefname{conjecture}{Conjecture}{Conjectures}
\crefname{claim}{Claim}{Claims}
\crefname{observation}{Observation}{Observations}
\crefname{proposition}{Proposition}{Propositions}
\crefname{fact}{Fact}{Facts}
\crefname{question}{Question}{Questions}
\crefname{problem}{Problem}{Problems}
\crefname{remark}{Remark}{Remarks}
\crefname{example}{Example}{Examples}
\crefname{equation}{Equation}{Equations}
\crefname{appendix}{Appendix}{Appendices}
\crefname{algorithm}{Algorithm}{Algorithms}
\crefname{model}{Model}{Models}
\crefname{figure}{Figure}{Figures}
\crefname{condition}{Condition}{Conditions}

\AfterEndEnvironment{definition}{\noindent\ignorespaces}
\AfterEndEnvironment{infdefinition}{\noindent\ignorespaces}
\AfterEndEnvironment{assumption}{\noindent\ignorespaces}
\AfterEndEnvironment{problem}{\noindent\ignorespaces}
\AfterEndEnvironment{lemma}{\noindent\ignorespaces}
\AfterEndEnvironment{theorem}{\noindent\ignorespaces}
\AfterEndEnvironment{proposition}{\noindent\ignorespaces}
\AfterEndEnvironment{fact}{\noindent\ignorespaces}
\AfterEndEnvironment{question}{\noindent\ignorespaces}
\AfterEndEnvironment{corollary}{\noindent\ignorespaces}
\AfterEndEnvironment{model}{\noindent\ignorespaces}
\AfterEndEnvironment{remark}{\noindent\ignorespaces}
\AfterEndEnvironment{proof}{\noindent\ignorespaces}
\AfterEndEnvironment{fact}{\noindent\ignorespaces}
\AfterEndEnvironment{minftheorem}{\noindent\ignorespaces}
\AfterEndEnvironment{inftheorem}{\noindent\ignorespaces}
\AfterEndEnvironment{maintheorem}{\noindent\ignorespaces}
\AfterEndEnvironment{restatable}{\noindent\ignorespaces}
\AfterEndEnvironment{infassumption}
{\noindent\ignorespaces}
\AfterEndEnvironment{infcorollary}{\noindent\ignorespaces}

\crefname{section}{Section}{Sections}
\crefname{theorem}{Theorem}{Theorems}
\crefname{lemma}{Lemma}{Lemmas}
\crefname{problem}{Problem}{Problems}
\crefname{program}{Program}{Progams}
\crefname{definition}{Definition}{Definitions}
\crefname{conjecture}{Conjecture}{Conjectures}
\crefname{corollary}{Corollary}{Corollaries}
\crefname{construction}{Construction}{Constructions}
\crefname{conjecture}{Conjecture}{Conjectures}
\crefname{claim}{Claim}{Claims}
\crefname{observation}{Observation}{Observations}
\crefname{proposition}{Proposition}{Propositions}
\crefname{fact}{Fact}{Facts}
\crefname{question}{Question}{Questions}
\crefname{problem}{Problem}{Problems}
\crefname{remark}{Remark}{Remarks}
\crefname{example}{Example}{Examples}
\crefname{equation}{Equation}{Equations}
\crefname{appendix}{Section}{Sections}
\crefname{algorithm}{Algorithm}{Algorithms}
\crefname{model}{Model}{Models}
\crefname{figure}{Figure}{Figures}
\crefname{infassumption}{Informal Assumption}{Informal Assumptions}
\crefname{inftheorem}{Informal Theorem}{Informal Theorems}
\crefname{infdefinition}{Informal Definition}{Informal Definitions}
\crefname{minftheorem}{Main Informal Theorem}{Main Informal Theorems}
\crefname{maintheorem}{Main Theorem}{Main Theorems}
\crefname{assumption}{Assumption}{Assumptions}
\crefname{step}{Step}{Steps}
\crefname{result}{Result}{Results}
\crefname{event}{Event}{Events}
\crefname{none}{}{}

\definecolor{myC}{rgb}{0, 255, 255}
\definecolor{myY}{rgb}{204, 204, 0}
\definecolor{myM}{rgb}{255, 0, 255}
\definecolor{secinhead}{RGB}{249,196,95}
\definecolor{lgray}{gray}{0.8}

\usepackage{appendix}
\crefname{appsec}{Appendix}{Appendices}

\newcommand{\N}{\mathbb{N}}
\newcommand{\R}{\mathbb{R}}

\newcommand{\Z}{\mathbb{Z}}

\newcommand{\E}{\operatornamewithlimits{\mathbb{E}}}

\newcommand{\poly}{\mathrm{poly}}
\newcommand{\polylog}{\mathrm{polylog}}
\newcommand{\eps}{\varepsilon}
\renewcommand{\epsilon}{\varepsilon}

\makeatletter
\newcommand*{\tran}{{\mathpalette\@tran{}}}
\newcommand*{\@tran}[2]{\raisebox{\depth}{$\m@th#1\intercal$}}
\makeatother

\newcommand{\wh}[1]{\widehat{#1}}

\renewcommand{\bar}{\overline}

\renewcommand{\tilde}{\widetilde}
\newcommand{\wt}[1]{\widetilde{#1}}

\def\<{\langle}
\def\>{\rangle}

\newcommand{\C}{\mathbb{C}}

\DeclareMathAlphabet{\mathpzc}{OT1}{pzc}{m}{it}

\newcommand{\customcal}[1]{\euscr{#1}}

\newcommand{\cN}{\customcal{N}}

\DeclareMathAlphabet{\mathdutchcal}{U}{dutchcal}{m}{n}
\SetMathAlphabet{\mathdutchcal}{bold}{U}{dutchcal}{b}{n}
\DeclareMathAlphabet{\mathdutchbcal}{U}{dutchcal}{b}{n}

\DeclareMathAlphabet\urwscr{U}{urwchancal}{b}{n}%
\DeclareMathAlphabet\rsfscr{U}{rsfso}{m}{n}
\DeclareMathAlphabet\euscr{U}{eus}{m}{n}
\DeclareFontEncoding{LS2}{}{}
\DeclareFontSubstitution{LS2}{stix}{m}{n}
\DeclareMathAlphabet\stixcal{LS2}{stixcal}{m} {n}

\title{Learning Mixture Models via\\ {Efficient} High-dimensional Sparse Fourier Transforms  
}

\date{}

\author{
Alkis Kalavasis\\ Yale University
\and 
Pravesh K. Kothari \thanks{Supported by NSF CAREER Award no. 2047933, NSF Medium Grant no. 2211971, and an Alfred P. Sloan Fellowship.} 
\\Princeton University
\and Shuchen Li
\\Yale University
\and Manolis Zampetakis
\\Yale University
}

\begin{document}

\maketitle

\bigskip 
\bigskip 
\thispagestyle{empty}

\begin{abstract}
{
In this work, we give a $\poly(d,k)$ time and sample algorithm for efficiently learning the parameters (i.e., the means and the mixture weights) of a mixture of $k$ spherical distributions in $d$ dimensions. Unlike all previous methods, our techniques apply to heavy-tailed distributions 
and include examples that do not even have finite covariances. Our method succeeds whenever the component distributions have a characteristic function with sufficiently \emph{heavy} tails. \akedit{This includes for example the Laplace distribution but crucially excludes Gaussians.}

All previous methods for learning mixture models relied implicitly or explicitly on the low-degree \emph{method of moments}. Even for the special case of Laplace distributions, we prove that any such algorithm must necessarily use a super-polynomial number of samples. Our method thus adds to the short list of techniques that circumvent the limitations of the method of moments.

Somewhat surprisingly, our algorithms succeed in learning the parameters in $\poly(d,k)$ time and samples without needing any minimum separation between the component means. This is in stark contrast to the case of spherical Gaussian mixtures where a minimum $\ell_2$-separation is provably necessary even information-theoretically~\cite{regev_learning_2017}. Our methods compose well with existing techniques and allow obtaining ``best of both worlds" guarantees for mixtures of distributions where every component either has a heavy-tailed characteristic function or has a sub-Gaussian tail with a light-tailed characteristic function.

Our algorithm is based on a new approach to learning mixture models via efficient high-dimensional noisy sparse Fourier transforms. We believe that this method will find more applications to statistical estimation. As an example, we give an algorithm for {consistent} robust estimation of the mean of a distribution $D$ in the presence of a constant fraction of outliers introduced by a \emph{noise-oblivious} adversary. This model is practically motivated by the literature on multiple hypothesis testing, it was formally proposed in a recent Master's thesis {by one of the authors} \cite{li2023thesis}, and has already inspired follow-up works. 
}

\end{abstract}

\clearpage
\tableofcontents
\thispagestyle{empty}

\clearpage
\setcounter{page}{1}
\section{Introduction}

\label{sec:intro:mixture}

Learning mixture models has been a benchmark problem in statistical estimation. The algorithmic goal is to take an input independent sample from a high-dimensional mixture and find the mean and covariance of the underlying component distributions. The history of the problem dates back to the landmark work of Pearson from 1894 on learning \emph{Gaussian} mixture models \cite{pearson1894contributions} in one dimension. Learning high-dimensional Gaussian mixtures was a central question in statistical learning, starting with the pioneering work of Dasgupta \cite{dasgupta1999learning}. And starting with the same work, a significant effort has been focused on finding techniques that avoid ``overfitting" to the assumption of Gaussianity on the cluster distributions. 

Mixture models have also served as a testing ground and often the first striking application for some of the most versatile tools developed for algorithms in statistical estimation. Classical examples include {low-rank projections} and spectral methods~\cite{vempala2002spectral,kannan2005spectral,achlioptas2005spectral} (that apply more generally to all log-concave distributions), random projections and the method of moments~\cite{kalai2010efficiently,moitra2010settling,belkin2010polynomial} (that are restricted to distributions with known moment relations), and tensor decompositions~\cite{hsu2013learning} (that need milder but still Gaussian-like low-order moments). 

In recent years, with a renewed focus on robust statistics~\cite{diakonikolas2019robust,LaiRaoVempala}, learning mixture models served as the central challenge~\cite{DVWSimonsVignette}. It led to the development of the method of spectral filtering~\cite{diakonikolas2023algorithmic} and the sum-of-squares method for robust statistics~\cite{kothari_outlier-robust_2017, kothari_better_2017,hopkins2018mixture} that eventually led to the full resolution of the question~\cite{bakshi2020outlier,bakshi2022robustly} via connections to algorithmic properties related to verifying concentration and anti-concentration~\cite{karmalkar_list-decodable_2019,raghavendra_list_2019} of high-dimensional probability distributions. 

The main goal of this work is to introduce a new class of methods that apply to learning spherical (i.e., covariance $\propto I$) mixture models. To show the contrast with previous work and motivate our methods, we summarize three high-level conclusions that emerge from the above line of work:
\begin{enumerate}[leftmargin=15pt]
  \item \textbf{Minimum Separation.} For learning the parameters of a mixture with $k$ components, all polynomial time/sample algorithms need a \textit{minimum Euclidean separation} between the cluster means of \(\gamma = \Omega(\sqrt{\log k})\) and this is provably necessary~\cite{regev_learning_2017}.
  
  \item \textbf{Moment-Based Methods.} Virtually all known algorithms for parameter estimation in GMMs (from the work of \cite{pearson1894contributions} to recent advances, \emph{e.g.}, \cite{hopkins2018mixture,kothari2018robust,liu2022clustering}) fundamentally rely on algorithms that try to find clusters with low-degree empirical \textit{moments} matching/behaving similarly to that of a Gaussian. 
  
  \item \textbf{Certifiably Bounded Distributions.} Even the most general methods developed for learning mixture models only apply when the cluster distributions have sufficiently light tails and there is an efficiently verifiable certificate~\cite{kothari_outlier-robust_2017} of this property. We know this is true for all sub-Gaussian distributions thanks to recent advances~\cite{diakonikolas2024sos,kothari_better_2017}. But this is a rather strong condition on tails. In particular, no distribution family with even mildly heavy tails (\emph{e.g.,} sub-exponential distributions!) is known to satisfy it so far (as pointed out in~\cite{diakonikolas2024sos}). 
\end{enumerate}

In this work, we develop a new method for estimating the parameters of a mixture model based on the \emph{Fourier} transform of the mixture. Our methods go beyond the method of moments (indeed, our results {apply to mixture models which \emph{provably} cannot be learned via just low-degree moment information} (see \cref{thm:intro:moment})) and apply to various distributions with heavy tails (indeed, as we discuss in \cref{rem:sfd}, even with infinite variance). Surprisingly, in sharp contrast to the case of (sub)-Gaussian distributions, our methods, {whenever applicable,} runs in polynomial (in both the dimension $d$ and the number of clusters $k$) samples and time to learn the spherical mixture \emph{without any {minimum} separation requirement}. Our methods apply to a broad class of distributions --- we now take a short detour to introduce this family before describing our results.

\medskip

\noindent \textbf{Slow Fourier Decay.} Our methods apply whenever the component distributions $D$ satisfy a certain \emph{Fourier decay} property. Recall that the Fourier transform {(characteristic function)} of a distribution $D$ on $\R^d$ is defined by: 
\[
\phi_D(t) = \E_{X \sim D}[e^{i \<t, X\>}].
\]
where $t \in \R^d$.

\begin{definition}
[Slow/Fast Fourier Decay]
\label{def:sfd-ffd}
Let $D$ be a probability distribution over $\R^d$. We say that $D$ satisfies \emph{Slow Fourier Decay (SFD)} with parameters $c_1, c_2 \ge 0$ if it holds that
\[  \inf_{t : \|t\| \leq T} |{\phi_D(t)}| \gtrsim 
 d^{-c_1} T^{-c_2} \,.
  \]
In contrast, $D$ satisfies \emph{Fast Fourier Decay (FFD)} with parameters $c_1', c_2' \ge 0$ 
if it holds that
\[
\sup_{t : \|t\| \geq T} |{\phi_D(t)}| \lesssim d^{-c_1'} T^{-c_2'}\,.
  \]
\end{definition}

The SFD property requires that, the magnitude of the characteristic function inside the ball of radius $T$ decays \emph{slower} than some polynomial of $1/T$ and $1/d$, while the FFD property aims to capture the complementary behavior, i.e., that the magnitude as $t$ grows, decays at a rate faster than some polynomial of $1/T$ or $1/d$.
\begin{figure}[!ht]
    \centering 
    \includegraphics[width=0.5\linewidth]{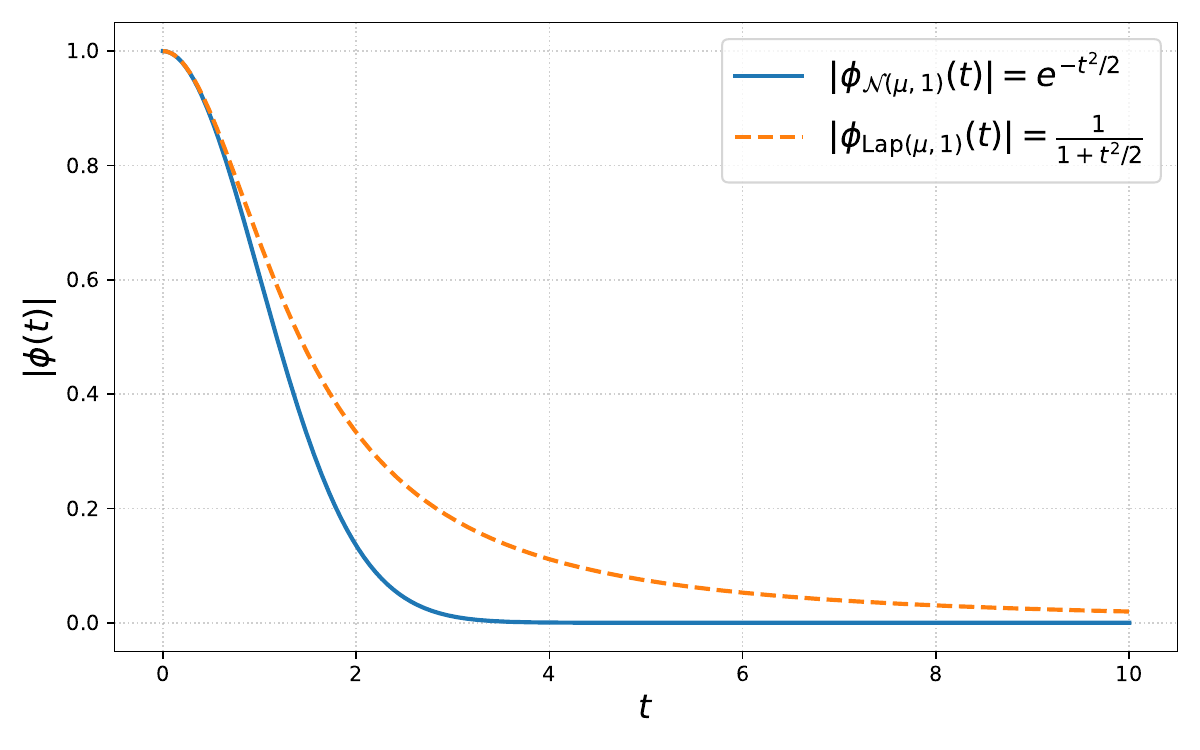}
    \caption{Fourier Decay for Gaussian and Laplace in one dimension.}
    \label{fig:placeholder}
\end{figure}

To illustrate these definitions, let us consider the case where $D = \cN(\mu, 1)$ or $D = \mathrm{Lap}(\mu, 1)$\footnote{The density of the Laplace distribution in $d$ dimensions (with mean \(\mu \in \R^d\) and covariance \(I_d\)) is $p_{\lap(\mu, I_d)}(x) =  {\frac {2}{(2\pi )^{d/2}}}\left({\frac {\|x-\mu\|_2^2 }{2}}\right)^{v/2}K_{v}\left(\sqrt{2} \|x-\mu\|_2\right),$ where \(v=(2-d)/2\) and \(K_{v}\) is the modified Bessel function of the second kind. {In particular, when \(d=1\), \(p_{\lap(\mu, 1)}(x) = \frac{1}{\sqrt{2}}\exp(-\sqrt{2}|x-\mu|)\).}} in one dimension ($d = 1$). Observe that the modulus of Gaussian characteristic function, which is equal to $t \mapsto \nicefrac{1}{e^{t^2/2}}$, vanishes exponentially faster than that of the Laplace distribution, which equals $t \mapsto \nicefrac{2}{2 + t^2}$ (\Cref{fig:placeholder}). This means that the 1-D Laplace distribution is SFD for parameter $c_2=2$ but the 1-D Gaussian is FFD for some parameter $c_2'$ (in fact for any \(c_2'\ge 0\)). 

The situation is similar for $d > 1$ dimensions, where the modulus of the characteristic function of the Laplace distribution is $t \mapsto \nicefrac{2}{2+\|t\|^2_2}$. Beyond Laplace, more examples of distributions satisfying SFD are: chi-squared (with constant degrees of freedom), gamma (with constant shape parameter), and the exponential distribution\footnote{For every real-valued, even, continuous function $\phi$ with \(\phi(0) = 1\) and \(\phi(\infty)=0\) that is convex on \((0, +\infty)\), P\'olya's theorem implies the existence of a distribution with characteristic function $\phi.$}.

\begin{remark}
[Comparing SFD with Tail Behavior] 
\label{rem:sfd}
The following examples indicate that the decay of the characteristic function is quite different from the tail behavior:
\begin{enumerate}
    \item There is an SFD distribution with sub-Gaussian tails (e.g., \sledit{the uniform mixture of a Dirac measure at \(0\) and a uniform distribution over \([-1,1]\)}).
    
    \item There is an SFD distribution with tails that are sub-exponential but not sub-Gaussian (e.g., Laplace distribution and chi-squared distribution with constant degrees of freedom).

    \item There is an SFD distribution with infinite variance (e.g., Linnik distribution \cite{anderson1993linnik}). 
\end{enumerate}
\end{remark}

\noindent \textbf{Our Results.} In this work we focus on learning mixtures $\mathcal M$ of SFD and FFD distributions in high dimensions.  As a driving example through the paper, the reader should think of $\mathcal M$ as a mixture of Laplace (SFD part) and Gaussian (FFD part) components.

Our first result is an efficient algorithm for recovering the means and weights of $\mathcal M$ {when the mixture model only consists of SFD components (e.g., mixture of Laplace distributions).}

\begin{theorem}
[Informal, Learning SFD Mixtures, see \Cref{thm:learning-SFD-mixtures}]
\label{thm:sfd}
Consider a mixture model $\mathcal M$ consisting of $k$ translations of an SFD distribution $D$ with parameters $c_1,c_2 = O(1)$ and with means $\mu_1,...,\mu_k$, weights $w_1,...,w_k = \Omega(1/k)$, and separation $\gamma = \min_{i \neq j} \|\mu_i - \mu_j\|$.
There exists an algorithm that uses $n = {\poly(d, k, 1/\gamma, 1/\varepsilon )}$ samples from $\mathcal M$, runs in time $\poly(n)$ and computes $\{\wh w_i, \wh \mu_i\}_{i \in [k]}$ such that with probability $99\%$ for all $j \in [k]$, $\min_i \|\wh \mu_i - \mu_j\| \leq \epsilon$, and $\min_i |\wh w_i -  w_j| \leq \epsilon$.
\end{theorem}

Below, we outline why this result represents a departure from the high-level takeaways of previous studies on learning mixture models.
\begin{enumerate}[leftmargin=15pt]
    \item \textbf{No minimum separation.} The key idea of \cite{regev_learning_2017} is that when $\gamma = o(\sqrt{ \log k})$, and $d = \Omega(\log k))$, they can design two GMMs whose parameter distance is very large, but whose total variation distance is $k^{-\omega(1)}$. This implies that a separation of {$\Omega(\sqrt{\log k})$} is required to achieve sample complexity that is polynomial in $k$. 
    Somewhat surprisingly, \Cref{thm:sfd} implies that this intuition is actually wrong for distribution families that satisfy the SFD property\footnote{We note that the work of \cite{qiao2022fourier} showed that one can learn the parameters of a spherical Laplace mixture without a separation requirement in the parameter regime when $k \geq 2^{\omega(d)}$ -- that is, the number of components grows super-exponentially in the dimension. In this case, note that a polynomial bound in $k$ is exponential as a function of the dimension $d$. Indeed, the lower bound of \citet{regev_learning_2017} only applies in the regime when $k=O(\log d)$. We refer \hyperref[qiao]{here} for a more detailed comparison. In contrast, our result shows that learning SFD distributions in polynomial time does not suffer from a separation requirement in \emph{any} parameter regime, including the more standard setting where $d$ and $k$ are comparable.}.

    A corollary of our results is that in the case of a mixture of Laplace distributions, there is an algorithm that recovers the means of the mixture with sample complexity and runtime $\poly(d, k, 1/\gamma,1/\epsilon)$ without any non-trivial separability assumption on $\gamma$. 
    
    This result is interesting from the perspective of \emph{clustering} as well: for the Gaussian mixtures case \cite{liu2022clustering}, $\poly(d,k)$-time parameter estimation {is possible} only in the regime when the clusters are non-overlapping (i.e., total variation distance $\rightarrow 1$ as $d,k \rightarrow \infty$). In contrast, we can achieve statistically and computationally efficient parameter estimation for mixtures of Laplace distributions even when the mixture is not clusterable (i.e, arbitrarily small total variation distance between distinct clusters).

    \item \textbf{Beyond Moment-Based Methods.} Another interesting aspect of \Cref{thm:sfd} is that in this mixture problem the method of moments is provably inefficient. In fact, in order to efficiently estimate Laplace mixtures, it is \emph{necessary} to depart from the standard moment-based methods as our following theorem implies.

    \begin{theorem}[Moment-Matching Lower Bound] \label{thm:intro:moment}
      \label{thm:Moments}
        There exist two uniform $k$-mixtures of SFD distributions with parameters $c_1,c_2 \in \Theta(1)$ in $d = \log k$ dimensions such that: (i) their parameters are $\sqrt{\log k}$ separated \footnote{{If $\{\mu_1,...,\mu_k\}$ and $\{\mu_1',...,\mu_k'\}$ are the two sets of parameters, then they are separated both within the mixture (i.e., $\min_{i\ne j}\|\mu_i - \mu_j\| \geq \sqrt{\log k},$~~$\min_{i\ne j}\|\mu_i' - \mu_j'\| \geq \sqrt{\log k}
        $) and across mixtures (i.e., $\min_{\pi} \sum_{j}\|\mu_j - \mu_{\pi(j)}'\| \geq \sqrt{\log k}
        $) (that is the two mixtures have large parameter distance \cite{regev_learning_2017}).}} but (ii) their first $\log k$ moments match up to $1/k^{\log \log k}$ error in Frobenius norm.
    \end{theorem}
    
    The fact that we show moment-matching in Frobenius norm is crucial since it implies a $d^{\log k} = k^{\log \log k}$ \emph{sample complexity} lower bound for any moment-based algorithm. In particular, the above result implies that there is an SFD mixture estimation problem that is solvable with sample and computational complexity $\poly(d, k)$ but any moment-based method requires number of samples that are super-polynomial in $k$. We refer to \Cref{sec:sketck_moments} for more discussion. 

    \item \textbf{No Tail Requirement.} As we mentioned in \Cref{rem:sfd}, our SFD condition is essentially incomparable to the tail-behavior of the mixture components. This allows us to learn mixtures of even heavy-tailed distributions, e.g., even distributions with infinite variance (see \Cref{rem:sfd}), using our Fourier-based method as long as their characteristic function decays sufficiently slow. This opens a new avenue for learning mixtures of heavy tailed distributions and bypasses the difficulties faced by Sum-of-Squares based methods.
\end{enumerate}

\noindent \textbf{Composing our result with SoS.} An additional advantage of our Fourier-based tool is that it composes well with the existing sum-of-squares framework for learning mixture models (that currently applies to the widest known cluster distributions). This allows to learn mixture models that have both SFD and FFD components as our next theorem shows.

\begin{theorem}
[Informal, see \Cref{thm:recover-sfd-ffd}]\label{thm:sfd-ffd}

Consider a mixture model $\mathcal M$ in $d$ dimensions that consists of $k + k'$ components of the following form:
\begin{enumerate}
    \item (SFD part) $k$ translations of a sub-Weibull and SFD distribution $D$ with parameters $c_1,c_2 = O(1)$ and with means $\mu_1,...,\mu_k$, and,

    \item (FFD part) 
    $k'$ distributions $D_1,...,D_{k'}$  which are all FFD with parameters $2c_1, 2c_2$, certifiably bounded (\Cref{def:cb}), sub-exponential, and with means $\mu_1',...,\mu_{k'}'$.
\end{enumerate}
Furthermore, we assume that the minimum weight is at least $\Omega(1/(k + k'))$, the separation between the SFD components is $\gamma_S > 0$, the separation between the FFD components is $\gamma_F = k'^{O(1/t)}$, and the separation between SFD and FFD components is $\gamma_{\text{SF}}=k'^{O(1/t)}$ for some $t > 0$. Then there exists an algorithm that uses 
\[ n = \underbrace{\poly(d,k,1/\gamma_S,1/\epsilon)}_{\mathrm{SFD~estimation}} + \underbrace{
\poly(d^{t}, k')}_{\mathrm{FFD~estimation}} \]
samples from $\mathcal M$, runs in time $n^{O(t)}$, and computes 
$\{\wh \mu_i\}_{i \in [k]},
\{\wh \mu_i'\}_{i \in [k']}$
such that
\begin{enumerate}
    \item (SFD estimation) for all $j \in [k]$, $\min_i \|\wh \mu_i - \mu_j\| \leq \epsilon,$ and,
    \item (FFD estimation) for all $j \in [k']$, $\min_i \|\wh \mu_i' - \mu_j'\|  \leq \poly(1/k')$
\end{enumerate}
with probability $99\%$.
\end{theorem}

This result employs the structure of the SFD distributions to perform the Fourier-based algorithm, and then uses these estimations combined with the SoS framework to learn the FFD part. The additional assumptions in the SFD and FFD parts are requires to make use of the SoS toolbox:
\begin{enumerate}
    \item For the SFD components, we need a resilience property (see \Cref{def:resilience}), which is true, e.g., if the components are sub-Weibull (a property strictly weaker than sub-exponential tails).

    \item For the FFD components, we need the components to be certifiably bounded \cite{hopkins2018mixture,kothari_better_2017,kothari2018robust}
    and this property is satisfied by all sub-Gaussian distributions~\cite{kothari_better_2017,diakonikolas2024sos}. We do not need to make any specific parametric assumptions such as Gaussianity.
\end{enumerate}

We note that the question of finding efficient learning algorithms for mixture models beyond sub-Gaussian clusters was recently explicitly stated in the work of \citet{diakonikolas2024sos}. Their work implies such algorithms for all sub-Gaussian distributions by showing low-degree sum-of-squares certificates of sub-Gaussian moments. They specifically pose the question of tackling sub-exponential distribution families (that include, e.g., all log-concave distributions, but is more general).  
Currently, we do not know how to find such certificates {for the class of sub-exponential distributions.}
Our results nevertheless show a polynomial time algorithm for learning mixtures of Laplace distributions (surprisingly, without any need for Euclidean mean separation). Our work also makes progress on the research direction (suggested in \citet{diakonikolas2024sos}) of finding algorithms for high dimensional tasks that work for broad distribution families without solving large convex programs. 
\smallskip

\noindent \textbf{Comparison with \cite{qiao2022fourier}.} \label{qiao}
The work of \citet*{qiao2022fourier} provides an algorithm that learns the means of a uniform mixture model, where each component is a shift of some distribution $D$ and whose sample/time complexity depends on the characteristic function of $D$. In particular, their algorithm requires samples and time $\poly(k) \cdot 2^d \cdot (1/\min_{\|t\| \leq T} \norm{\phi_D(t)})$ where $T \lesssim \gamma^{-1}\sqrt{d \log k}$ for $\gamma$-separated means\footnote{The sample complexity of \cite{qiao2022fourier} is inherently exponential in $d$ since it uses a tournament-based technique which relies on the realization of an event that has probability $2^{-d}$.}. Hence, the results of \cite{qiao2022fourier} are sample-efficient only in the regime where $d = O(\log k)$. This is in contrast to our algorithm from \Cref{thm:sfd} which has polynomial sample complexity and running time and applies to mixtures with arbitrary weights (see \Cref{thm:learning-SFD-mixtures}).
\medskip

\subsection{Fast High-Dimensional Sparse Fourier Transforms} \label{sec:intro:sft}

{In this section we describe one major component of our estimation algorithms for mixture models that we believe can be of independent interest. The key idea for our new algorithmic tool can be quickly described as follows: let $\mathcal M = \sum_{i\in[k]} w_i D(\mu_i)$ be a mixture of translation $\mu_1, \dots, \mu_k$ of a known probability distribution $D$. Then the characteristic function of the mixture becomes
\[ \E_{Y \sim \mathcal{M}}[e^{i \<t,Y\>}] = \sum_{j \in [k]} w_j e^{i \<t, \mu_j\>} \phi_D(t). \]
But since $D$ is known, we can divide both sides with $\phi_D(t)$ and we get that $x^{\star}(t) = \E[e^{i \<t,Y\>}]/\phi_D(t)$ is a signal which in the Fourier domain has $k$ active frequencies. Furthermore, these frequencies correspond to the translations $\mu_1, \dots, \mu_k$ that we want to estimate, so we can write our problem as a Fourier estimation problem. Of course, we do not have access to the signal $x^{\star}(t)$ and this requires to utilize the literature on computing sparse Fourier transforms. In fact, we need to develop our own algorithm that is suitable for our application in statistics and learning theory. We now briefly discuss the background for this problem.}

\noindent \textbf{Problem Formulation.}
For {fixed \(T > 0\) and $t \in B_T^d(0):=\{\tau\in \R^d: \|\tau\|_2 \le T\}$} let $x^\star(t) = \sum_{j=1}^k w_j e^{i \<\mu_j, t\>}$ be a $k$-sparse signal with weights $w_j\in \C$ and frequencies $\mu_j \in \R^d$ for $j \in [k]$. Assume that the learner has query access to the noisy signal {over \(t \in B_T^d(0)\)}, 
\begin{equation}
\label{eq:signal}
    x(t) = x^\star(t) + g(t)\,,
\end{equation}
where $g : B^d_T(0) \to \C$ is some (potentially adversarial) noise function with bounded magnitude. The key question then is the following:
\textit{what is the number of queries and the computation time needed to recover the weights and frequencies of the $k$-sparse signal $x^\star$?} In this context, query access to $x(t)$ means that there exists an oracle such that given a time $t$ returns the value $x(t)$.

The work of \citet{PS15-SFT} answered this question in the one-dimensional setting ($d = 1$). \cite{PS15-SFT} developed an algorithm that recovers
the frequencies of the signal $x^\star(t)$ with error $O(\nicefrac{\mathcal N}{T})$ from $\wt{O}(k \log(T))$ queries on the signal \(x(t)\) and runs in time $\poly(k \log(T))$, where we can think of $\mathcal N$ as the $L_2$ norm of the noise signal $g$.
Follow-up work by \citet{jin2022superresolutionrobustsparsecontinuous}
studied the extension of this problem to high dimensions ($d > 1$). Their algorithm recovers the frequencies with error 
$\poly(d) \cdot \nicefrac{\mathcal N}{T}$ but requires $\wt{O}(k) \cdot \exp(d)$ time {and queries}. 

For our applications to statistics and learning, we need to prove the following result, whose proof relies on a careful adaptation of the one-dimensional method of \citet{PS15-SFT} together with standard techniques of low-dimensional projections, from the work of \citet{moitra2010settling}.

\begin{theorem}
[Informal, see \Cref{thm:d-sft}]
{For any fixed \(T>0\),}
consider any signal \(x(t) = x^\star(t) + g(t)\in \C\) over \(t\in B_T^d(0)\), where \(g(t)\) is adversarial noise and $x^\star$ is \(k\)-sparse, as in \eqref{eq:signal}, with frequency separation \(\gamma = \min_{j' \ne j} \|\mu_{j'}-\mu_j\|_2\). {If \(T \ge \Omega(d^{5/2}\log (k)/\gamma)\), then} there is an algorithm that queries the signal $x(t)$ on the points $t_1, \dots, t_m \in B_T^d(0)$ with $m = \wt{O}(k \cdot d \cdot \log(T))$, runs in time $\wt{O}(m)$, and computes parameters $\{(\wh w_i, \wh \mu_i)\}_{i\in [k]}$ such that, with probability at least \(99\%\), for any $j \in [k]$ with \(|w_j| =\Omega(\mathcal N)\), 
\[
\min_{i \in [k]}\|\mu_j - \wh \mu_{i}\|_2 \le O\left(\frac {d^3 \cdot  \mathcal{N}}{\gamma \cdot T \cdot |w_j|} \right), \quad
{\min_{i \in [k]} |w_j - \wh{w}_{i}| \leq O(\mathcal{N})}\,,
\] 
where $\mathcal N \approx {\max_{j \in [m]} |g(t_j)|} + \theta \| w\|_2$ for some appropriately chosen parameter $\theta$. 
\label{thm:main}
\end{theorem}
{We mention that the stated estimation guarantees hold with high probability over the randomness of the algorithm ({including the choice of }$t_1,...,t_m$).}
Note that both the query complexity and the runtime of the algorithm are nearly linear in $d$ and $k$. 
As mentioned in \cite{PS15-SFT}, the requirement for the lower bound on the weight $w_j$ is necessary since otherwise the noise $g$ could cancel completely this signal. Moreover, for the tones of high magnitude, the error converges to 0 as the noise level $\mathcal N$ decreases, a phenomenon known as
super-resolution \cite{donoho1992superresolution,candes2014towards,moitra_super-resolution_2015,huang2015super,chen2021algorithmic}. 

While our result seems to combine a few known techniques in literature, we have not found an already existing result that suffices for our applications. Indeed, we believe that our formulation here will likely be useful in applying sparse Fourier transforms in statistical estimation because it combines several properties that, to the best of our knowledge, are not satisfied by existing methods: (1) it uses a polynomial number of queries, (2) it runs in polynomial time in high-dimensions, and (3) the error parameter $\mathcal{N}$ depends on the modulus of $g$ evaluated only on the queried points $t_1, \dots, t_m$, instead of, e.g., the $L_2$ norm of the signal $g$ which in this context is $\frac{1}{T} \int_0^T |g(t)|^2 \mathrm{d} t$ {when \(d=1\)}. This last property is crucial to our applications. This is because in our setting, the noise $g$ captures the statistical error incurred in estimating the characteristic function from samples ({for any $t$)} and so it is not clear how to argue about its value outside the queried points.

\subsection{Further Applications: Oblivious Robust Statistics}
\label{sec:intro:robust}

Beyond the fundamental problem of parameter estimation in mixture models, our method can be applied to robust statistics \cite{huber2011robust,diakonikolas2023algorithmic,diakonikolas2019robust} to handle contamination models that assume less powerful adversaries than Huber's contamination model \cite{huber2011robust} and hence lead to better estimation guarantees. Following the nomenclature of \cite{li2023thesis},  where this model was introduced for the first time, we call this  model \textit{noise-oblivious contamination}.

\begin{definition}
[\textsc{Noise-Oblivious Contamination}]
\label{def:subset-of-signals}
Let $D$ be a distribution and $D(\mu)$ the translation of $D$ that has mean $\mu \in \R^d$. Fix also $\alpha \in [0,1]$ to be the contamination level and $n$ to be the number of samples. The noise-oblivious contamination procedure can be described as follows:
\begin{enumerate}[leftmargin=15pt]
    \item An adversary chooses $\mu_1,...,\mu_n$ with the restriction that for $(1 - \alpha)$-fraction of $\mu_i$'s satisfy $\mu_i = \mu$.
    \item Then, for each $i$, the sample $x_i$ is drawn independently from $D(\mu_i)$.
\end{enumerate}
The dataset $\{x_1,...,x_n\}$ is called \(\alpha\)-corrupted and our goal is to estimate $\mu$.
\end{definition}

There are multiple ways to motivate this problem: (1) in many settings the contamination happens before some noise is added to the data, e.g., the max-affine regression problem as it is described in \cite{li2023thesis}, and (2) in large-scale multiple testing most samples follow a null distribution centered at an unknown mean, and a minority arise from shifted alternatives. This setting, studied in \cite{carpentier2021estimating, kotekal2025optimal,diakonikolas2025efficient} and it is related to empirical Bayes' models of \cite{efron2004large}. Finally, the noise-oblivious contamination model is a classical instance of learning from \emph{heterogeneous} data \cite{compton2024near}, where samples are drawn independently, but from non-identical distributions. We refer to \Cref{sec:related} for a more detailed comparison with previous work. 
\medskip

\noindent \textbf{Our results.} Our final result is to show that, under the Slow Fourier Decay condition, our Fourier-based technique implies an efficient algorithm with polynomial sample complexity to solve the mean estimation problem with noise-oblivious contamination. One important aspect of this result is that even when the contamination level $\alpha$ is constant we can still recover the mean $\mu$ with a rate that goes to $0$ as $n$ goes to $\infty$. 

\begin{theorem} 
[Consistent Estimation for Noise-Oblivious Contamination; Informal, see \Cref{thm:noise-obi-mean-estimation-full}]
\label{thm:intro:robust}
Consider the \(d\)-dimensional mean estimation problem in the setting of \Cref{def:subset-of-signals} with distribution $D(\mu)$ with  true mean $\mu \in \R^d$ such that \(\|\mu\|_2 \le B\) for some \(B > 0\)\footnote{If $D$ has some additional properties, e.g., bounded covariance, we can get rid of the dependence on $B$ (see \Cref{sec:noise-obl} for details)}.
Define 
\[R(T) := \sup_{t:\|t\|_2 \le T}|\phi_{D}(t)|^{-1}\] 
for any $T>0$. 
If the corruption rate \(\alpha \le \alpha_0\) for some absolute constant \(\alpha_0>0\), then there is an algorithm that computes an estimate $\wh \mu \in \R^d$ such that $\| \mu - \wh \mu\|_2 < \eps$ with probability $99\%$. The algorithm uses \(n = \wt{O}\left(R( d^3 B/\varepsilon)^2\right)\) samples and runs in time $\poly(n)$.
\label{thm:NoiseObl}
\end{theorem}

This result indicates that the sample complexity of the noise-oblivious contamination model is also controlled by the SFD property. As a corollary we get that if $D$ is a Laplace distribution then the mean estimation problem with noise-oblivious contamination is solvable in polynomial samples and running time whereas if $D$ is a Gaussian then the sample complexity that is needed is exponentially large in $1/\epsilon$ (even in one dimension). 

\begin{corollary} In the setting of \Cref{thm:NoiseObl}:
    \begin{enumerate}
    \item If $D$ is the Laplace distribution, there is an algorithm that computes an estimate $\wh \mu \in \R^d$ such that $\|\mu - \wh \mu\| < \eps$ with probability $1-\delta$. The algorithm uses $n = {\widetilde O(\poly(d/\varepsilon))} \log(1/\delta)$ samples and runs in time $\poly(n)$.  
    
    \item If $D$ is the single-dimensional standard Gaussian distribution, there is an algorithm that computes an estimate $\wh \mu \in \R$ such that $|\mu - \wh \mu| < \eps$ with probability $1-\delta$. The algorithm uses $n = 2^{O(1/\varepsilon^2)} \log(1/\delta)$ i.i.d. samples and runs in time $\poly(n)$.  
\end{enumerate}
\label{cor:gaus_laplace}
\end{corollary}

The first observation is that the designed estimators are \emph{consistent}, i.e., its error goes to 0 with the number of samples. This is in contrast to the standard Huber's contamination model where the information-theoretic estimation limit is the corruption rate \cite{huber2011robust}.

The two guarantees have a gap in their sample complexity. This is again due to the fact that Laplace is an SFD distribution whereas Gaussian is an FFD distribution. The sample complexity for the Gaussian case is exponential in $1/\epsilon$. This is surprisingly tight based on existing information-theoretic lower bounds \cite{kotekal2025optimal}. On the other hand, for Laplace distributions (and any distribution satisfying SFD), the estimator has polynomial sample complexity. Both estimators have sample polynomial running time.
\medskip

\noindent \textbf{Comparison with \cite{diakonikolas2025efficient}.} The work of \citet{diakonikolas2025efficient} resolves the high-dimensional version of the above Gaussian mean estimation problem with noise-oblivious adversaries using a preliminary version of our result (appearing in~\cite{li2023thesis}) as a black-box component. Their algorithm first carefully projects the observations in a low-dimensional data-dependent subspace and then applies our Fourier-based estimator as a black-box \cite[Proposition 2.1, Fact 2.2]{diakonikolas2025efficient}\footnote{To be precise, \cite{diakonikolas2025efficient} cites (i) the one-dimensional algorithm of \Cref{cor:gaus_laplace} as it appeared in the Master's thesis \cite{li2023thesis} (results of which are presented for publication for the first time in this paper) and (ii) the concurrent and independent work of \cite{kotekal2025optimal}.} (whose sample and time complexity becomes exponential in {\(1/ \varepsilon\)} due to the fact that Gaussians are FFD, i.e., they have very fast Fourier decay). Their estimation algorithm uses $\sim d/\epsilon^{2+o(1)} + 2^{O(1/\epsilon^2)}$ samples and runs in sample-polynomial time.

\subsection{Technical Overview} \label{sec:intro:techniques}

In this section, we give an overview of the techniques that we use to prove our main results.

\subsubsection{Efficient Sparse Fourier Transforms -- \Cref{thm:main}}
We start with a sketch of the SFT algorithm that we provide, which will be the main tool for our applications later on. Let us recall the problem of interest. Our goal is to query the noisy signal $x(t) = x^\star(t) + g(t)$ in linearly in $k$ many points $t \in \R^d$ and efficiently recover the $k$-sparse signal $x^\star(t) = \sum_{i \in [k]} w_i e^{i \<\mu_i,t\>}$.
As we have already mentioned, in dimension $d=1$, the work of \citet{PS15-SFT} manages to solve this problem. However, it cannot be directly extended in high-dimensions. The follow-up work of \citet{jin2022superresolutionrobustsparsecontinuous} studies the high-dimensional version of the sparse recovery problem and gives an algorithm that efficiently recovers $x^\star$ for any \emph{constant} dimension with $\wt{O}(k)$ queries; however, in general, the query and time complexity scale as $2^d$. Hence, the first obstacle that we have to avoid is the exponential dependence on the dimension. 

Our approach is inspired by works in mixture models (e.g., \cite{moitra2010settling}) that deal with the high-dimensionality of the data by studying low-dimensional projections. Such a connection between Fourier transforms and low-dimensional projections appears in the work of \citet{chen2021algorithmic} in the study of two-dimensional Airy disks. At a conceptual level, we follow a similar approach: given query access to the signal $x(t)$ for $t \in \R^d$ with \(\|t\|_2 \le T\), we first project the time variable $t$ in various directions $v_1,...,v_m$, and then study the one-dimensional signals
\[
x^{v_\ell}(t) := x(t \cdot v_\ell) = \sum_{j\in [k]} w_j e^{i t \<\mu_j, v_\ell\>} + g(t \cdot v_\ell), ~~ t \in [-T,T]\,.
\]
Observe that the weights of the projected signal $x^{v_\ell}$ are preserved for all $\ell $ and the means are projected in direction $v_\ell$. Our goal is to apply the one-dimensional algorithm of \citet{PS15-SFT} in each one of these signals $x^{v_\ell}(t)$, which will allow us to recover the frequencies parameters of the projections of the true signal
 \[
 x^\star(t \cdot v_1), \dots ,x^\star(t \cdot v_m)\,. 
 \]
{Unfortunately, the analysis of \cite{PS15-SFT} has error that relies on the $L_2$ norm of $g$ which is not suitable for statistics applications. For this reason we have to analyze the algorithm of \cite{PS15-SFT} in a different way to make sure that our error only depends on the modulus of $g$ on the queried points. We give more details about this in \Cref{sec:sft:1d}.}

Once we have recovered $x^\star(t \cdot v_1), \dots ,x^\star(t \cdot v_m)$ we have the $m \times k$ inner products between $d$-dimensional vectors 
$ \{ \<\mu_j, v_\ell\> \}_{j \in [k], \ell \in [m]} $
and we can extract the true means $\mu_1,...,\mu_k \in \R^{d}$ by solving a linear system.

The remaining part is to determine how to pick the projections $v_1,...,v_m$. 
For this we use the idea of \citet{kalai2010efficiently} for learning mixture models, where they project along directions that are \emph{sufficiently close} to each other. 
The benefit of projecting along close-by directions is that the ordering of the means can be preserved with high probability.
In other words, if there is an ordering $\pi$ such that
\[
\<\mu_{\pi(1)}, r\> 
\leq ...
\leq 
\<\mu_{\pi(k)}, r\> 
\]
in the random direction, there the same ordering is preserved in any projection $v_\ell$, $\ell = 1,...,d$. Having preserved the order, one can recover the true means $\mu_1,...,\mu_k$ by solving $k$ linear systems of the form
\[
V [\mu_{j,1},...,\mu_{j,d}]^\top = [\<v_1, \mu_j\>,...,\<v_d,\mu_j\>]^\top
\]
where the matrix $V \in \R^{d\times d}$ contains $v_1,...,v_d$.
By picking $\epsilon_1$ appropriately, the condition number of these linear systems will be polynomially bounded. Solving this system results in an estimation of the means that implies the bound of \Cref{thm:main}.

\subsubsection{Learning Mixture Models -- \Cref{thm:sfd,thm:sfd-ffd}}

Our goal is to learn the parameters of a mixture model that can be written as
\begin{equation}
    \mathcal M = \underbrace{\sum_{i\in[k]} w_i D(\mu_i)}_{\mathrm{SFD~part}}
    +
    \underbrace{\sum_{i\in[k']} w_i' D_i'(\mu_i')}_{\mathrm{FFD~part}}
\end{equation}
Our goal is to learn the parameters of this mixture model under some mild assumptions on the distributions and the separation of the parameters. At a high-level, our algorithm works in two stages, that we explain below.

\paragraph{Step I: Recovering the SFD part} For the SFD part (e.g., Laplace distributions), the only assumption that we need is that the FFD components (e.g., Gaussian distributions) have a faster Fourier decay. Other than that, we place no minimum separation assumptions for the means $\mu_1,...,\mu_k$. The algorithm that recovers the SFD means is using our robust sparse Fourier transform (\Cref{thm:main}).

Let us assume that the SFD part consists of translations of a distribution $D$ which is SFD with parameters $c_1,c_2$ and assume that the FFD components (with parameters $c_1',c_2')$ decay faster than that, i.e., $c_2' > c_2$. Then our algorithm works as follows. For a sample $Y \sim \mathcal M$, we can write
\[
\E_{Y}[e^{i \<t,Y\>}] = \sum_{j \in [k]} w_j e^{i \<t, \mu_j\>} \phi_D(t) + \sum_{j \in [k']} w_j' e^{i \<t, \mu_j'\>} \phi_{D'_j}(t)
\]
which can be equivalently written as
\[
\phi_D(t)^{-1}\E_{Y}[e^{i \<t,Y\>}] = \sum_{j \in [k]} w_j e^{i \<t, \mu_j\>} + \sum_{j \in [k']} w_j' e^{i \<t, \mu_j'\>} \frac{\phi_{D'_j}(t)}{\phi_D(t)}\,.
\]
The first observation is that for a fixed $t$, the left-hand side can be estimate with sample from $\mathcal M$ using standard concentration tools. Now, in the right-hand side, the first term is corresponds to a $k$-sparse signal, whose tones we want to estimate (this is the SFD part). The second term consists of the FFD components and the key observation is that this term vanished as $t$ increases, thanks to the behavior of the characteristic functions. 
 
Hence, in short, our idea is to employ the sparse Fourier algorithm of \Cref{thm:main} with true signal $x^\star(t)$ corresponding to the SFD components and noise $g(t)$ that contains (i) the vanishing term coming from the FFD part and (ii) the estimation error of the left-hand side. The algorithm has to carefully tune the duration $T$, the number of samples $n$ from $\mathcal M$, and the number of queries to the noisy signal $x(t) = x^\star(t) + g(t)$ in order to bound the noise level $\mathcal N^2$ of \Cref{thm:main}. The details appear in \Cref{sec:sfd-proof}. 

\paragraph{Step II: Recovering the FFD Part}  The second step of the algorithm is using the SoS framework to recover the FFD part. To do that, we need to put some constraints in both the SFD and the FFD distributions. For this step, we have to make use of the means estimated in Step I. Let us explain the assumptions that we need.
    \begin{itemize}
        \item For the FFD part (e.g., Gaussian distributions), we need some minimum parameter separation of order $\poly(k)$. This is in general unavoidable since we want polynomial sample complexity. More to that, we need to bound the tails of the FFD part. To this end we will assume that the FFD components are $(2t,B)$-certifiably bounded and sub-exponential. Both assumptions are standard and are already needed from prior work  \cite{kothari2018robust,kothari_better_2017}.

        \item For the SFD part (e.g., Laplace distributions), we will still not require any non-trivial separation between the SFD means but we will require some non-trivial separation $\gamma_{\text{SF}}$ between the SFD and the FFD means. This is expected and the order of the separation is controlled by \emph{resilience} property of the SFD components (see \Cref{def:resilience})\footnote{Resilience of a distribution $D$ is a key concept in robust statistics that guarantees (roughly speaking) that the empirical mean of any $\alpha n$ subset of a sample from $D^n$ will be close to the true mean with high probability. See \Cref{def:resilience} for details. }. For instance, for Laplace components, the deviation of the mean of an \(\alpha\) fraction of the sample will be at most \(O(\log(1/\alpha))\), and so $\gamma_{\text{SF}} \approx \poly(k)$. In general, this is expected since we do not put any tail requirement on the SFD part. 
    \end{itemize}

    Under the above conditions, there is a natural SoS-based algorithm that will recover the FFD components. Our algorithm combines classical tools from robust statistics such as robust mean estimation and list-decodable mean estimation procedures that use SoS \cite{kothari_better_2017}. 
    Assume that we run the SFD algorithm from Step I and we have a list of predictions for the SFD means $\mu_1,...,\mu_k.$
    Our algorithm, apart from this list, has access to i.i.d. samples from the mixture $\mathcal M$. The idea is that when the number of samples is sufficiently large, we can run a list-decodable mean estimation algorithm for each FFD distribution $D_i'(\mu_i')$ with $i \in [k'].$ This algorithm treats samples from all the remaining $k + k'-1$ components as ``corruptions''. The guarantee of this algorithm (see \Cref{thm:list-decode})
    is a sequence of subsets $S_1,..,S_m \subseteq [n]$ with $m \approx \poly(k)$ with the guarantee that the empirical mean $\frac{1}{|S_j|} \sum_{i \in S_j} x_i$ for some $j \in [m]$ is close to the target mean (here $\{x_i\}$ are the given training samples).

    Now, given this list of sets, we have to reject the subsets that correspond to SFD clusters. In particular, we use the list of SFD mean estimates given to the learner $\wh \mu_1,...,\wh \mu_k$ (which is generated in Step I before) to remove all the sets $S_j$ with empirical mean $\gamma_{\text{SF}}$-close to one of these points. For the removal, we make use of the observation that the SFD and the FFD components are well-separated but also that the SFD means are estimated with accuracy smaller than this separation.

    Next, we have to deal with the survival sets. Our algorithm merges all the sets whose empirical means are closer than $\gamma_F/2$, where $\gamma_F$
    is the minimum separation between FFD components. Using the separation assumption, this merging will result in a collection of $k'$ sets $S_1',...,S'_{k'}$ and in each one of those sets we can prove that, apart from a constant fraction $\alpha = c^{-2t}$ (for some $c)$ of the points, all the remaining observations are drawn from the same FFD distribution $D'_j$ for some $j \in [k'].$ This implies that we can use a standard robust mean estimation algorithm to estimate the true FFD means up to accuracy $B \alpha^{1-1/2t}$ (see \Cref{thm:sos-robust-mean-est}).

    For the specific case, where the FFD part is Gaussian, we can get arbitrarily close to the true means, by modifying the local convergence method of \citet{regev_learning_2017} (see \Cref{app:local-convergence}).

\subsubsection{Moment-Matching for SFD Mixtures -- \Cref{thm:sfd-ffd}}
\label{sec:sketck_moments}

We mentioned in \cref{sec:intro:mixture} that moment-matching of the first \(r\) moments in Frobenius norm implies a \(d^r\) sample complexity lower bound for any moment-based algorithm. The reason is as follows. The empirical \(r\)-th order moment tensor \(\wh T\) will have \(\E[\|\wh T - \E[\wh T]\|_{\mathrm F}^2] = \Omega(d^r)\), since the variance of every entry of \(\wh T\) is \(\Omega(1)\). Thus, if we are estimating the moment tensor with \(n\) samples, the expected error will be \(\Omega(d^r / n)\). Using \Cref{thm:Moments}, considering moments of order at least $\log k$ is needed. Thus, the sample complexity of any moment-based method (using the standard empirical estimators) would be \(n = \Omega(d^{\log k}) = \Omega(2^{\log k \log\log k})\), while our Fourier-based algorithm has $\poly(k)$ sample complexity.

As a consequence, this implies that the sample complexity of moment-based methods scales at least super-polynomially with the number of components $k$, while our algorithm of \Cref{thm:sfd} achieves a polynomial dependence on $k$. Hence, our Fourier-based tool is a method that provably bypasses the limitations of the method of moments. 
The proof is inspired by the pigeonhole argument of \citet{regev_learning_2017} and appears in \Cref{sec:moments}. 
\smallskip

Next, we discuss the technical overview of \Cref{thm:Moments} which shows that moment-based methods are not useful for {learning mixture models} when the distribution satisfies the SFD condition. To do that, we show that there exist two mixtures of $k$ Laplace distributions whose parameters are very far but their first $\log k$ moments are very close. To do that, we adapt the techniques of \citet{regev_learning_2017}. First, it is important to explain what we mean by moment-matching. Closeness in moments will be measured using the Frobenius norm, which is defined as $\fnorm{T} = \left( \sum_{i_1,i_2,\dots, i_\ell}  T_{i_1,i_2,\dots, i_\ell}^2 \right)^{1/2}$ for some order-$\ell$ tensor $T$.

Our result is as follows: There exist  two uniform mixtures of Laplace distributions \(Y\) and \(\tilde Y\) in $\Theta(\log k)$ dimensions, {consisting of Laplace components with means \(\mu_1,\dots, \mu_k\) and \(\tilde\mu_1,\dots, \tilde\mu_k\), respectively, } such that
\begin{enumerate}
    \item (Moment matching) 
    Their moments are close in the Frobenius norm: For any order $r =1,2,\dots, \Theta(\log k),$ it holds that
    $\| \E Y^{\otimes r} - \E \tilde Y^{\otimes r} \|_{\mathrm{F}} \le k^{-\Omega(\log\log k)}$. 

    \item (Parameters are far) Their parameter distance (i.e., $\min_{\pi \in S_k} \sum_j \|\mu_j - \tilde \mu_{\pi(j)} \|_2 $ ) is at least \(\Omega(\sqrt{\log k})\).
\end{enumerate}

To show the moment-matching guarantee we use a packing argument, as in \citet{regev_learning_2017}. In more detail, one can use the pigeonhole principle to show (see \Cref{lem:dirac-moment-matching-rv17}) that for any large enough collection (roughly $\exp((R/d)^d)$ of Laplace mixtures, for most Laplace mixtures
in the collection one can find other mixtures which approximately match in their first $R$ mean
moments in Frobenius norm with error $d^{-2R}.$ To show the gap in the parameter distance, one can construct the above collection by selecting means uniformly at random from the ball of radius $\sqrt{d}$ (see \Cref{lem:separation}). Then it is standard that the pairwise distance between the means is large. Combining the two arguments, we get the desired result. For details, we refer to \Cref{sec:moments}.

\begin{remark}
[Connection between SFD and Moment-Matching]
{\citet{regev_learning_2017} used a weaker notion of closeness, i.e., the symmetric injective norm. It is important to note that this notion of closeness allows them to translate moment-matching to p.d.f. closeness for Gaussians. However, for Laplace distributions and other distributions with heavy-tailed characteristic function (i.e., which satisfy SFD), Lemma 3.7 in \cite{regev_learning_2017} does not hold. This is exactly why we can bypass the moment-based methods using our Fourier analytic tools.}
\end{remark}

{
\subsubsection{Mean Estimation with Noise-Oblivious Adversaries -- \Cref{thm:intro:robust}}
Recall that, under the setting of \cref{def:subset-of-signals}, the input of the algorithm can be viewed as \(n\) independent random variables, with a \((1-\alpha)\) fraction being sampled from \(D(\mu)\), and the rest \(\alpha\) fraction being sampled from \(D(z_k)\), where \(z_k\) is chosen by the adversary, for \(k=1,2,\dots, \alpha n\). Our goal is to recover the true mean \(\mu\). Note that the input distribution can also be viewed as a mixture model, but now we only care about the major component (i.e., \(D(\mu)\)). Thus, the analysis will be very similar to that of the mixture learner. 

Given sample \(\{Y_j\}_{j\in [n]}\) generated according to \cref{def:subset-of-signals}, we have \[\frac 1 n\sum_{j=1}^n \E[e^{i\langle t, Y_j \rangle}] = (1-\alpha) e^{i\langle t, \mu \rangle} \phi_{D}(t) + \frac 1 n \sum_{k=1}^{\alpha n} e^{i\langle t, z_k \rangle} \phi_{D}(t),\]
which is equivalent to \[\phi_{D}(t)^{-1}\cdot \frac 1 n\sum_{j=1}^n \E[e^{i\langle t, Y_j \rangle}] = (1-\alpha) e^{i\langle t, \mu \rangle} + \frac 1 n \sum_{k=1}^{\alpha n} e^{i\langle t, z_k \rangle }.\]
This time, in the right-hand side, we will view the first term as a \(1\)-sparse signal, and the second term as noise. 
Note that, by triangle inequality, the modulus of the second term is upper bounded by \(\alpha\). 
Therefore, as long as \(\alpha\) is at most some absolute constant, we can again apply the sparse Fourier transform algorithm of \cref{thm:main}, with true signal \(x^\star(t)= (1-\alpha) e^{i \langle t, \mu \rangle}\) and noise \(g(t)\) that contains (i) the term from the adversarial corruption and (ii) the estimation error of the left-hand side. The details appear in \cref{sec:noise-obl}. 
}

\subsection{Open Questions} 
We believe that our work opens some new algorithmic directions in {learning mixtures} and {distribution learning under sample contamination}. To this end, we identify and leave some immediate open problems.

\smallskip

\noindent \textbf{Open Problem \#1.} Is there an efficient algorithm for \emph{robust} learning {mixtures of} SFD distributions? 

At a technical level, it is not clear how to apply the connection to sparse Fourier transforms, when there are outliers in the sample.

\smallskip

\noindent \textbf{Open Problem \#2.} 
Is there an efficient algorithm for learning {mixtures of} SFD distributions (or learning an SFD distribution in the noise-oblivious model) with \emph{unknown} covariance? 

\smallskip
Our algorithms rely on decomposing the periodic part of the characteristic function (i.e., $e^{i t \mu}$ which contains the unknown mean) from the ``tail'' of the characteristic function, which is associated with the SFD property. However, when the variance is also unknown, such a decomposition is no longer possible.

\subsection{Related work}
\label{sec:related}

In this section, we discuss works that are related to our paper.
\paragraph{Gaussian Mixture Models} Gaussian mixture models (GMMs) are one of the most well-studied {parametric distribution families for density estimation in particular and, also} in statistics {more broadly}, with a history going back to the work of Pearson \cite{pearson1894contributions} {(also see the survey by \citet{titterington1985statistical} for applications for GMMs in the sciences).}
The study of {statistically and computationally} efficient algorithms for {estimating} GMMs goes back to the seminal {works} of \citet{redner1984mixtures,dasgupta1999learning,lindsay1995mixtures} and has {since} attracted significant interest from theoretical computer scientists, \emph{e.g.,} \cite{vempala2002spectral,kannan2005spectral,brubaker2008isotropic,feldman2006pac,kalai2010efficiently,moitra2010settling,hopkins2018mixture,bakshi2020outlier,diakonikolas2020robustly,liu2022clustering,liu2022learning,buhai2023beyond,anderson2024dimension,li2017robust,diakonikolas2020small,bakshi2022robustly,bruna2021continuous,gupte2022continuous,diakonikolas2024implicit}. The closest to our work is probably the work of \citet{regev_learning_2017} which shows a tight lower bound for the minimum separation required for {parameter estimation} in GMMs{: the class of Gaussian mixture models with $k$ components in $d$ dimensions requires super-polynomially many samples when the minimum distance $\gamma$ between the parameters of the different components is of the order $\gamma = o(\sqrt{\log k})$ in $d = O(\log k)$ dimensions, even for the class of \emph{spherical} GMMs. In contrast, if $\gamma = \Omega(\sqrt{ \log k})$, then $\poly(d,k)$ samples are sufficient \cite{regev_learning_2017}.}
{At a technical level, our work is also inspired by the projection technique of \citet{kalai2010efficiently} in order to speed-up the high-dimensional robust sparse Fourier transform algorithm.}

\paragraph{Non-Gaussian Mixtures} 
The study of {mixture models} extends to mixtures with non-Gaussian components. Unlike the SFD property we are considering, most of the works study mixtures with components that are somehow concentrated \cite{achlioptas2005spectral}, e.g., (SoS-certifiable) sub-Gaussian \cite{mixon2017clustering,hopkins2018mixture,kothari2018robust,dmitriev2024robust} or have only bounded covariance but satisfy some separation assumption \cite{diakonikolas2025clustering}. Moreover, recent works study algorithms for non-parametric generalizations of spherical Gaussians, and, in particular, the class of Gaussian location mixtures \cite{chewi2025ddpm,gatmiry2024learning}; these works also aim to go beyond moment-based methods, by using algorithms based on diffusion models \cite{chen2024learninggeneralgaussianmixtures}. \akedit{Finally, the work of \citet{dasgupta2005learning} studied mixtures of heavy-tailed distributions but their motivation was different: their goal was to understand separation conditions that allowed for polynomial sample complexity and hence their algorithms are not computationally efficient.}

{\paragraph{Moment-Based Methods} Virtually all known algorithms for parameter estimation in Gaussian Mixture Models (GMMs)—from the foundational work of \cite{pearson1894contributions} to recent advances in theoretical computer science (\emph{e.g.}, \cite{hopkins2018mixture,kothari2018robust,liu2022clustering})—are fundamentally \textit{moment-based}. 
A spectacular use of the method of moments was the sequence of classical works that settled the efficient learnability of a high-dimensional mixture of Gaussians \cite{kalai2010efficiently,moitra2010settling,belkin2010polynomial,hardt2015tight} under minimal
information-theoretic separation assumptions. The running time of these algorithms however turns out to be {$(d/\epsilon)^k (1/\epsilon)^{k^{k^2}}$} for accuracy $\epsilon$ and at least a $d^{\Omega(k)}$ cost appears necessary~\cite{diakonikolas2017statistical}. Subsequent work, including \cite{hsu2013learning,bhaskara2014uniqueness,anderson2014more,ge2015learning,bhaskara2014smoothed}, leveraged tensor decomposition techniques, focusing on extracting low-rank structure from empirical third- and fourth-order moment tensors—structures that are especially tractable in the case of Gaussian mixtures. Motivated by applications to robust statistics, recent works \cite{hopkins2018mixture,kothari2018robust,diakonikolas2018list} introduced the use of higher moments to enable {parameter estimation} with separation as small as $k^\epsilon$ for any $\epsilon > 0$, even beyond the Gaussian setting. Building on this moment-based framework, numerous follow-up works extended these techniques to more general {statistical} problems on {mixture models} across multiple directions (\emph{e.g.}, \cite{bakshi2020outlier,diakonikolas2020robustly,kane2021robust,liu2021settling,bakshi2022robustly,liu2022clustering,liu2022learning,buhai2023beyond,anderson2024dimension}). As a result, moment-based methods have become the dominant algorithmic paradigm for {parameter recovery in GMMs}. {Regarding the fundamental class of spherical GMMs, the runtime of moment-based algorithms has been recently improved from $\poly(d,  k^{\polylog(k)})$  \cite{hopkins2018mixture,kothari2018robust,diakonikolas2018list,diakonikolas2020small} to $\poly(d,k)$ under either a slightly stronger separation  \cite{liu2022clustering} or under  the assumption
that the largest pairwise distance is comparable to the smallest one \cite{diakonikolas2024implicit}.}
}
\paragraph{Fourier-Based Methods} In this section, we add some related TCS works that make use of Fourier transforms. 
The main algorithmic tools we use rely on robust sparse Fourier transforms. \citet{PS15-SFT} gave the first robust sparse Fourier transform algorithm in the continuous setting in one dimension. Later, \citet{jin2022superresolutionrobustsparsecontinuous} generalized it to \(d\) dimensions. For \(k\)-sparse signal with \(\gamma\)-separated frequencies, the sample duration needed in \cite{jin2022superresolutionrobustsparsecontinuous} is \(T = O(\log (k) / \gamma)\). 
On the negative side, \citet{moitra_super-resolution_2015} shows a lower bound of \(T = \Omega(1/\gamma)\) by  determining the threshold at which noisy \emph{super-resolution} is possible.
Moreover, the works of \citet{chen2016fourier,song2023quartic} study the problem of interpolating a noisy Fourier-sparse signal even when tone estimation is not possible. As we mentioned the main focus of the SFT problem is to achieve fast estimation using nearly linearly many queries; the work of \citet{huang2015super} achieves an efficient algorithm in both $k$ and $d$ with samples that scale quadratically with $k$ and $d$. We mention that for our purposes we could not use off-the-shelf this algorithm since in \Cref{thm:d-sft} we only assume bounded $L_{\infty}$ at the queried points {as we discussed in the last paragraph of \Cref{sec:intro:sft}}.

Fourier-based methods have found applications to algorithmic statistics. \citet{diakonikolas2016efficient,diakonikolas2016properly,diakonikolas2016fourier} have used discrete Fourier transforms for learning sums of integer-valued random variables.
\citet{chakraborty2020learning} give an algorithm for learning mixtures of spherical
Gaussians in dimensions $\omega(1) \leq d \leq O(\log k)$ via deconvolving the mixture using Fourier transforms.
\citet{chen2020learning}
study the problem of learning mixtures of linear regressions (MLRs), which
can be reduced to estimating the minimum variance in a mixture of zero-mean Gaussians. They
solved this problem by estimating the Fourier moments – the moments of the Fourier transform,
and gave the first sub-exponential time algorithm for learning MLRs. \citet{chen2021algorithmic} study learning
mixtures of Airy disks, a problem that is motivated by the physics of diffraction. Their algorithm
also proceeds by first estimating the Fourier transform of the mixture, and then dividing it pointwise
by the Fourier spectrum of the “base” distribution. Finally, we have already discussed in the introduction the work of \cite{qiao2022fourier}. 

\paragraph{Noise-Oblivious Contamination}
{The content of \Cref{sec:noise-obl} contains the results appearing in the recent Master's thesis} \cite{li2023thesis}. 
\cite{kotekal2025optimal} independently study the model of  in one dimension and derive matching
information theoretic upper and lower bounds for Gaussian mean estimation. They also consider 
the unknown variance case. Before these works, \cite{carpentier2021estimating} studied the
sample complexity of noise-oblivious robust Gaussian mean estimation in the special case where the corruption points $z_i$ satisfy $z_i - \mu > 0$.
\cite{diakonikolas2025efficient} studies Gaussian mean estimation in the multivariate case.
\cite{diakonikolas2025efficient} builds on the single-dimensional algorithm of \cite{li2023thesis} and provides a high-dimensional algorithm for the noise-oblivious contamination model (which they refer to as mean-shift contamination). The analogous problem where the adversary instead of the mean corrupts the variance in one dimensions is studied by \cite{chierichetti2014learning,liang2020learning,compton2024near} and the high-dimensional variant by \cite{diakonikolas2025entangled}.

\section{Algorithms for Sparse Fourier Transforms }
\label{sec:fourier}
{
In this section, we introduce the main algorithmic tools that we will use. 
First, we provide a modified version of the one-dimensional robust sparse Fourier transforms, studied by \cite{PS15-SFT}. Next, we introduce a high-dimensional extension of this algorithm. We remark that while a similar high-dimensional algorithm has appeared in prior work \cite{jin2022superresolutionrobustsparsecontinuous}, our algorithm is computationally efficient while the one presented in prior work runs in time exponential in the dimension.}

\subsection{Robust Sparse Fourier in One Dimension} \label{sec:sft:1d}

The first key result that we will use is a modification of the algorithm of \citet{PS15-SFT} for robustly computing sparse Fourier transforms in the continuous
setting.
To state the result intuitively, let $x(t) = x^\star(t) + g(t)$, where $x^\star$ has a $k$-sparse Fourier transform and $g$ is an arbitrary
noise term. Given query access to $x(t)$ for times $t \in [0,T]$, the algorithm is able to estimate the frequencies and the weights, i.e., the \emph{tones} of the true signal $x^\star$, with an estimation error that depends on the \emph{noise level,} i.e., on how large $g$ is in some average sense. In particular, the algorithm queries the signal roughly $k \log T$ times and outputs estimates $\wh f_1,...,\wh f_k$ of the true frequencies such that there is a permutation $\pi$ with
\[
\max_{j \in [k]} |f_j - \wh f_{\pi(j)}| \lesssim \frac{\mathcal N}{T |w_j|}
\]
where $\mathcal N^2 \approx \frac{1}{T} \int_0^T |g(t)|^2 \mathrm{d} t$ and $w_j$ is the $j$-th weight, \akedit{whenever the tone has large magnitude, i.e., $|w_j| = \Omega(\mathcal N)$}. Notably, \sledit{the error goes to 0 as the noise level decreases to 0.}
This phenomenon is known as
super-resolution \cite{moitra_super-resolution_2015} --- one can achieve very high frequency resolution in sparse, nearly noiseless settings. Moreover, the rate of estimation is optimal \cite{PS15-SFT}.

For our purposes, we need to modify the statement of \citet{PS15-SFT} and adapt its proof. We provide the modification below.

\begin{theorem}
\label{thm:sparse-Fourier-transform-Modified}
\sledit{For any fixed \(T>0\)}, consider any signal \(x(t) = x^\star(t) + g(t)\in \C\) over \(t\in [0, T]\), for arbitrary noise \(g(t)\) and exactly \(k\)-sparse \(x^\star(t) = \sum_{j=1}^k w_j e^{ i f_j t}\) with \(f_j \in [-B, B]\) and frequency separation \(\gamma = \min_{j' \ne j} |f_{j
'}-f_j|\). Let \(\theta>0\) and \(\delta>0\) be some parameter. \sledit{If \(T\ge\Omega(\log(k/\theta)/\gamma)\),
then} there is an algorithm $\mathsf{SFT}_1(x,k, T, B, \gamma, \theta,\delta)$ that (i) randomly draws times $t_1,...,t_N$ with
\[
N = O(k\log(B T)\log(k/\theta)\log (k/\delta)),
\] 
(ii) queries the signal \(x(t)\) at $t \in \{t_1,t_2,\dots, t_N\}$, and (iii) computes \(\{(\wh w_j, \wh f_j)\}\) in 
\[
O(k\log(BT)\log(BT/\theta)\log (k/\delta))\]
running time, such that the following holds.

Define the event $\mathcal E$ to be the set of the random strings \(r\) used by the algorithm, and $t_1 = t_1(r),...,t_N = t_N(r)$ to be the times picked by the algorithm such that, the algorithm running with randomness \(r\) outputs \(\{(\wh w_j, \wh f_j)\}_{j \in [k]}\) with the property that there is a permutation $\pi$ such that for any \(w_j\) with \(|w_j| =\Omega(\mathcal N)\), 
\[
|f_j - \wh f_{\pi(j)}| \le O \left(\frac {\mathcal N}{T|w_j|} \right),\qquad {|w_j - \wh w_{\pi(j)}| \le O(\mathcal{N}),}
\] 
where \[
\mathcal N^2 = \max_{j \in [N]} |g(t_j)|^2 +\theta \sum_{\ell=1}^k |w_\ell|^2. 
\]
Then \(\Pr_{r}\left[r\in \mathcal E\right] > 1-\delta\,.\) 
\end{theorem}

\begin{remark}
[Comparison of \Cref{thm:sparse-Fourier-transform-Modified} with \citet{PS15-SFT}]
{The above statement is an adaptation of Theorem 1.1 of \citet{PS15-SFT}. We need to adapt the steps of the analysis for our application to recovering the parameters of SFD mixtures. In our proof, we explain how the algorithm of \cite{PS15-SFT} works and what we need to modify for our purposes. The main modification (which is implicit in the analysis of \citet{PS15-SFT}) is that the noise level $\mathcal N^2$ scales with the maximum of the noise function $|g(t)|^2$ on the times $\{t_1,..,t_N\}$ picked by the learning algorithm, while in the analysis of \cite{PS15-SFT}, the noise level scales with the integral $\int_0^T |g(t)|^2 \mathrm{d} t.$ We need the former, because in our application we only have control of $g$ on the queried points.  } 
\end{remark}
\begin{proof}
{In each part of the analysis, we will explain how the original algorithm of \citet{PS15-SFT} works and further discuss our modification.

\paragraph{Hashing} The algorithm of \citet{PS15-SFT} proceeds in stages, each of which hashes the frequencies to $\mathcal B$ bins. The hash
function depends on two parameters $\sigma$ and $b$, and so we define it as $h_{\sigma, b} : [-F,F] \to [\mathcal B]$.
A tone with a given frequency $f$ can have two “bad events” : (1) colliding with another frequency of $x^\star$ or (2) landing near the boundary of the bin; they each will occur with small constant probability.

The algorithm $\mathsf{HashToBins}$ hashes frequencies into different bins in order to reduce the \(k\)-sparse recovery to \(1\)-sparse recovery.
More precisely, define \(P_{\sigma, a, b}\) as an operator on the signal such that \((P_{\sigma, a, b}x)(t) = x(\sigma(t-a)) e^{-2\pi i \sigma b t}\).
The algorithm gets as input $x, P_{\sigma, a,b}$ and $\mathcal B$ and returns a vector $\wh u \gets \mathsf{HashToBins}(x, P_{\sigma, a, b}, \mathcal{B})$. 

We now explain how to compute $\wh u$ and what is its meaning. 
Let us start with the computation.
Let \(\wh G(f)\) approximate \(\mathbf{1}[|f|\le \frac{\pi}{\mathcal{B}}]\), where \(\wh G(f) = \sum_{j=1}^M G_j e^{2\pi i jf/M}\) is sparse, and \(M = O(\mathcal{B}\log(k/\theta))\). For input signal \(x(t)\), set \(y = G\cdot P_{\sigma, a, b}x\). Its Fourier transform will be \(\wh y = \wh G * \wh{P_{\sigma, a, b}x}\). Finally, let \(\wh u_j = \wh y_{jB/\mathcal{B}}\). The key property of the algorithm is that, if neither “bad” event holds for a frequency $f$, then for the bin $j = h_{\sigma, b}(f)$, we have that $|\wh u_j| \approx |\widehat{x^\star}(f)|$ with a phase depending on $a.$ In other words, the observation of \cite{PS15-SFT} is that $|\wh u_j|$  will be approximately the sum of all the tones hashed into the \(j\)-th bin, up to a phase shift, where the hash function \(h_{\sigma, b}(f)\) only depends on \(\sigma\) and \(b\). They show that, if \(\sigma\) and \(b\) are chosen uniformly at random from some intervals, with high probability there will be no collision and every frequency will not be too far away from the center of each bin. For the reduction, it remains to show how the noise is distributed across all the bins. 
}
They show that the total noise in all the bins is bounded by the noise rate \cite[Lemma~3.2]{PS15-SFT}: 
\begin{equation}
\label{eq:original-3.2}
\E_{\sigma, a, b} \left[ \sum_{f\in H} \left| \wh u_{h_{\sigma, b}(f)} - \wh{x^\star}(f) e^{2\pi i a \sigma f} \right| + \sum_{j\in I} \wh u_j^2 \right] \lesssim \mathcal{N}^2 := \frac{1}{T}\int_0^T |g(t)|^2 \mathrm{d} t + \theta \sum_{\ell=1}^k |w_\ell|^2.
\end{equation}

Before proceeding with our modification, let us summarize the notation that we will need for our restatement of \cite[Lemma 3.2]{PS15-SFT}.
\begin{itemize}
    \item \(\sigma, a, b\) are the ({bounded real-valued and chosen uniformly at random}) parameters of some permutation \(P_{\sigma, a, b}\) on the signal \(x\), defined as \((P_{\sigma, a, b}x)(t) = x(\sigma(t-a)) e^{-2\pi i \sigma b t}\). Importantly, {the signal $x(t)$ is} queried {at times that} depend only on {the first and the second parameters \(\sigma\) and \(a\)}, not \(b\); and the indices of the bins for the frequencies only depend on \(\sigma, b\).
    \item \(h_{\sigma, b}(f)\) is the index of the bin that \(f\) is hashed into, and \(\wh u_j\) is the total ``mass'' of signal hashed into the \(j\)-th bin with a phase depending on \(a\). 
    \item \(H\) is the set of true frequencies that are hashed without collisions and large offsets from the centers of the bins. \(I\) is the set of the indices of the bins with no true frequencies hashed into it. 
\end{itemize}

{
Lastly, we mention that the algorithm $\mathsf{HashToBins}$ is called 3 times where the second argument is $P_{\sigma, \xi, b}$ with $\xi = \{a, \gamma, \gamma+ \beta\}$. This essentially implies 3 variants of \Cref{eq:original-3.2} (one for each value of $\xi$, \Cref{eq:original-3.2} corresponds to $\xi = a)$.

{We now provide our modified version of the inequality.} {We will remove the expectation over \(\sigma\) and \(a\) in \cref{eq:original-3.2}, and instead state another inequality which holds for any fixed randomness \(r\) used by the algorithm to determine the value of all the variables but \(b\). In particular, the inequality will consist of an expectation over $b$ and will be true for all the values of  \(\sigma\) and $\xi$ (i.e., \(a, \gamma,\) and \( \gamma + \beta\) in \cite[Algorithm 2]{PS15-SFT}) of \(P_{\sigma,\xi,b}\) that are passed as the argument of \(\mathsf
{HashToBins}(x, P_{\sigma,\xi,b}, \mathcal{B})\) in \cite[Algorithm 2, lines 8, 26, 27]{PS15-SFT}).  
}
\begin{lemma}
[{\cite[Lemma 3.2 (Modified)]{PS15-SFT}}]  
Fix a random string \(r\) that determines all the variables but \(b\). For all values of \(\sigma, a, \gamma, \beta\) used by the algorithm running with randomness \(r\), 
\[\E_{b} \left[ \sum_{f\in H} \left| \wh u_{h_{\sigma, b}(f)} - \wh{x^\star}(f) e^{2\pi i \xi \sigma f} \right| + \sum_{j\in I} \wh u_j^2 \right] \lesssim \mathcal{N}^2 = \max_{j \in [N]} |g(t_j)|^2 +\theta \sum_{\ell=1}^k |w_\ell|^2,\]
where \(\wh u = \mathsf{HashToBins}(x, P_{\sigma, \xi,b}, \mathcal{B})\), \(h_{\sigma, b}\), \(H\), \(I\) are defined as above, {for \(\xi = \{ a, \gamma, \gamma + \beta \}\)}. 
\end{lemma}

{The above Lemma controls the quality of the approximation of $\mathsf{HashToBins}$ and shows that the total error over all tones is bounded by $\mathcal{N}^2$. Note that since the randomness $r$ across the whole execution of the algorithm is fixed (except of $b)$, }{the values of the first and the second parameters of \(P_{\sigma,\xi,b}\) are determined and hence the values $t_1,...,t_N$ are fixed.} Finally, our definition of the noise rate $\mathcal N$ is the main difference compared to \cref{eq:original-3.2} (we pay the worst choice of the algorithm given the random string $r$ instead of the ``average cost'' of \cref{eq:original-3.2}). 

We proceed with the proof of the modified Lemma. \citet{PS15-SFT} prove their version of the  inequality (i.e., \Cref{eq:original-3.2}) by considering two cases, \(x^\star(t) = 0\) (see \cite[Lemma 3.3]{PS15-SFT} and \(g(t) = 0\) (see \cite[Lemma 3.4]{PS15-SFT}, separately, and then combining them together by linearity.
Due to our change on the definition of $\mathcal N$, we need to modify only the statement and the proof of the first case \cite[Lemma~3.3]{PS15-SFT} ({i.e., when $x^\star(t) = 0)$}. {We now provide the proof of our modified version of \cite[Lemma~3.3]{PS15-SFT}, }{when the second parameter of \(P_{\sigma, \xi, b}\) is \(\xi = a\). The proof for \(\xi = \{\gamma, \gamma + \beta\}\) is the same. }
}
\begin{lemma}
[{\cite[Lemma 3.3 (Modified)]{PS15-SFT}}]
\label{lem:3.3mod}
Assume that $x^\star(t) = 0$ for all $t \in [0,T].$ Fix a random string \(r\) that determines all the variables but \(b\). For all values of \(\sigma, a\) used by the algorithm running with randomness \(r\),  
\begin{align}
\E_{b} \biggl[ \sum_{j=1}^{\mathcal{B}} \left| \wh u_j \right|^2 \biggr] \lesssim \max_{j\in [N]} |g(t_j)|^2. \label{eq:maxgti}
\end{align}
\label{lemma:Lemma 3.3}
\end{lemma}

{Recall that since the randomness $r$ across the whole execution of the algorithm is fixed (except of $b)$, the values of $\sigma, a$ are determined and hence the values $t_1,...,t_N$ (appearing in the above right-hand side) are fixed.}
\begin{proof}
[Proof of \Cref{lemma:Lemma 3.3}] 
{Let us now see how we can derive inequality \eqref{eq:maxgti}.
The proof is exactly the same as the proof of \cite[Lemma~3.3]{PS15-SFT} until reaching the point where it is shown that for any \(\sigma, a\), \begin{align*}
\E_{b} \biggl[ \sum_{j=1}^{\mathcal{B}} \left| \wh u_j \right|^2 \biggr] = \mathcal{B}\cdot \sum_{j=1}^{\mathcal{B}\cdot \log(k/\theta)} |G_j|^2 |g(\sigma(j-a))|^2,
\end{align*}
where \(G_i\) satisfies \(\sum_j |G_j|^2 \asymp \frac{1}{\mathcal{B}}\). Then the original proof goes through by taking the expectation over \(a\), and by noting that \(\E_a |g(\sigma(j-a))|^2 \lesssim \frac{1}{T} \int_0^T |g(t)|^2 \mathrm{d} t\). In our modified proof,} we can bound \(|g(\sigma(j-a))|^2 \le \max_{j\in [N]}|g(t_j)|^2\), {since every \(\sigma(j-a)\) is one of the times queried \(t_1,\dots, t_N\). Therefore,} \(\E_{b} \bigl[ \sum_{j=1}^{\mathcal{B}} \left| \wh u_j \right|^2 \bigr] \lesssim \frac{\mathcal{B}}{\mathcal{B}} \max_{j\in [N]} |g(t_j)|^2 = \max_{j\in [N]} |g(t_j)|^2\). 

\end{proof}
The above provides the modification of \cite[Lemma~3.3]{PS15-SFT} and completes our sketch for the modification of the hashing step, where instead of the ``average cost'' of the noise $g$, the algorithm pays the maximum of $g$ at the times it queries.

\paragraph{One Stage of Recovery} 
{
Given the hashing step, we have reduced the problem to a 1-sparse recovery problem. Regarding recovery of the frequencies, the main tool of \citet{PS15-SFT} is \cite[Lemma 3.6]{PS15-SFT}, which relies on \cite[Lemma C.1]{PS15-SFT} and provides the guarantees for the algorithm $\mathsf{LocateInner}$. This algorithm, roughly speaking, splits the frequency domain into regions and uses the hashing mappings of the previous part and queries to the signal $x(t)$ to assign votes to different regions for the location of the target frequency. Then \cite[Lemma 3.7]{PS15-SFT} provides the more general $\mathsf{LocateKSignal}$, that calls $\mathsf{LocateInner}$ multiple times. 

Roughly speaking, in one step of the algorithm, the region that contains the true frequency will get the vote, and the regions that are far away from the true frequency will not get the vote, with high probability under the condition \[\E_{\gamma}[|\wh u_{h_{\sigma, b}(f)} - e^{2\pi i \gamma \sigma f} \wh x (f)|^2] \le \frac{1}{\rho^2} |\wh x(f)|^2.\]

 Let us see how  \cite[Lemma~C.1]{PS15-SFT} should be modified. In our modification, the above condition will be changed accordingly by replacing the expectation with a maximum.

\begin{lemma}
[{\cite[Lemma C.1 (Modified)]{PS15-SFT}}]
\label{lem:C.1mod}
Let \(r_\beta\) and \(r_{-\beta}\) denote the randomness used by the algorithm to determine the value of \(\beta\) and the all the other variables, respectively. 
For any \(s\in (0, 1)\), with probability at least \(1 - 15s\) over \(r_\beta\), the following holds for any fixed \(r_{-\beta}\). 
Suppose that the frequency \(f\) is in the \(q'\)-th region, and \begin{align*}
    \max_{\gamma}\{|\wh u_{h_{\sigma, b}(f)} - e^{2\pi i \gamma \sigma f} \wh x (f)|^2\} &\le \frac{1}{\rho^2} |\wh x(f)|^2, \\
    \max_{\gamma, \beta}\{|\wh u'_{h_{\sigma, b}(f)} - e^{2\pi i (\gamma+\beta) \sigma f} \wh x (f)|^2\} &\le \frac{1}{\rho^2} |\wh x(f)|^2,
\end{align*} 
where \(\wh u = \mathsf{HashToBins}(x, P_{\sigma, \gamma, b}, \mathcal{B})\), \(\wh u' = \mathsf{HashToBins}(x, P_{\sigma, \gamma+\beta, b}, \mathcal{B})\), and the \(\max\) is taken over all the values of \(\gamma\) and \(\beta\) used by the algorithm running with randomness \(r=(r_\beta, r_{-\beta})\). Then for one round of voting \cite[Algorithm 2, lines 24--35]{PS15-SFT}, where \(\gamma \in [\frac{1}{2}, 1]\), \(\beta \in [\frac{st}{4\sigma \Delta l}, \frac{st}{2\sigma \Delta l}]\), we have \begin{enumerate}
    \item the vote \(v_{{h_{\sigma,b}(f)}, q'}\) of the true region \(q'\) will increase by one.  
    \item for any \(q\) such that \(|q - q'| > 3\), \(v_{{h_{\sigma,b}(f)}, q}\) will not increase. 
\end{enumerate}
\end{lemma}
\begin{proof}
The proof is the same as \cite{PS15-SFT}, except the first step, where they use the condition \[\E_{\gamma}[|\wh u_{h_{\sigma, b}(f)} - e^{2\pi i \gamma \sigma f} \wh x (f)|^2] \le \frac{1}{\rho^2} |\wh x(f)|^2\] to derive via Markov's inequality that \[|\wh u_{h_{\sigma, b}(f)} - e^{2\pi i \gamma \sigma f} \wh x (f)| \le \frac{1}{\sqrt{\delta_0}\rho} |\wh x(f)|\] with probability \(1 - \delta_0\) for any \(\delta_0 > 0\). However, in our modification, we have that \[|\wh u_{h_{\sigma, b}(f)} - e^{2\pi i \gamma \sigma f} \wh x (f)|\le \frac{1}{\rho} |\wh x(f)|\] holds for all the values of \(\gamma\) used by the algorithm (and the same for \(\wh u'\) with \(\gamma + \beta\)), which fits in the rest of the proof. 
Therefore, the failure probability only comes from an event over the draw of \(\beta\) that relates to the true frequency \(f\) and is independent of the noise, which is \(15s\) \cite[second to last paragraph on page 27]{PS15-SFT}. 
\end{proof}

Since we have changed the condition in \cite[Lemma~C.1]{PS15-SFT}, we need to check how the rest of the proof adapts to this new condition. 
Based on \cite[Lemmas~3.6, 3.7]{PS15-SFT}, one can define \[\mu^2(f)=\E_a[|\wh u_{h_{\sigma, b}(f)} - e^{2\pi i a \sigma f} \wh x^\star(f)|^2],\] which is roughly the amount of noise in the bin that contains \(f\), and set \(\rho = |\wh x^\star(f)|/\mu(f)\). We will change it to \[\mu^2(f)=\max_\xi\{|\wh u_{h_{\sigma, b}(f)} - e^{2\pi i \xi \sigma f} \wh x^\star(f)|^2\},\] where \(\wh u = \mathsf{HashToBins}(x, P_{\sigma, \xi, b}, \mathcal{B})\), and the \(\max\) is taken over all the values of \(\xi = \{a, \gamma, \gamma+\beta\}\) used by the algorithm running with any fixed randomness. This modification matches our new condition in \cref{lem:C.1mod}. 
From \cite[Lemma~3.7]{PS15-SFT}, the subroutine \(\mathsf{LocateKSignal}\) outputs a list \(L\) that, if \(|\wh x^\star(f)| \gtrsim \mu(f)\), then there is a frequency \(\wh f\in L\) such that \(|f-\wh f| \lesssim \frac{\mu(f)}{T|\wh x^\star(f)|}\). 
Then \cite[Lemma~3.8]{PS15-SFT} relates the partial noise \(\mu^2(f)\) and the total noise \(\mathcal{N}\) by summing all the \(\mu^2(f)\) for successfully recovered true frequency \(f\). This step is also valid in our modification, from our modified \cref{lem:3.3mod}.

The above discussion summarizes our modifications to \cite[Lemma 3.6, Lemma 3.7, Lemma C.1, Lemma 3.8]{PS15-SFT}. In short, the modified Lemma 3.8 is exactly the same as in \cite{PS15-SFT} with the only change being the modified definition of the noise scale $\mathcal{N}.$
}

\paragraph{Failure Probability}

{It is implicit in \cite{PS15-SFT} that the failure probability of the whole algorithm comes from the following bad events: 
\begin{enumerate}
    \item There are two bad events for the hashing: {collision and large offset} (which are controlled by the random variables $\sigma, b$).
    \item There is a bad event in \cite[Lemma C.1]{PS15-SFT} which corresponds to the voting in the regions (which is controlled by the random variable $\beta$).
    \item There is a bad event related to the noise function $g:$ The noise \(g(t_j)\) at some time \(t_j\) queried is not concentrated. 
\end{enumerate}
In our modification, the failure probability only comes from hashing and $\beta$, as our \(\mathcal{N}\) is a universal upper bound on \(g(t_j)\) for all queried times \(t_j\). 

The above arguments imply that, one can split the randomness \(r\) into two parts, \(r_1\) (which {controls the choices of $\sigma,b$ and \(\beta\)}) and \(r_2\) (which controls the rest of the randomness, namely $a$ and \(\gamma\) in the algorithm), such that \(r\in \mathcal{E}\) if \(r_1 \in \mathcal{E}_1\), for some ``good'' set \(\mathcal{E}_1\). This is because now the bad events will only come from hashing and \(\beta\) (controlled by \(r_1\)), as we have ``for all'' statements on the error from the noise \(g(t)\). 
Therefore, the success probability of the modified algorithm \[\Pr_r[r\in \mathcal{E}] \ge \Pr_{r_1} [r_1\in \mathcal{E}_1] \ge 1 - 1/\poly(k),\] where the last inequality is from the analysis of the original algorithm. %
}

\paragraph{Boosting} {In \cite[Section D]{PS15-SFT}, the authors boost the success probability of their algorithm from a constant to \(1 - 1/\poly(k)\), by repeating their subroutine \(\mathsf{OneStage}\) \(O(\log k)\) times. However, the same proof holds if one repeats it \(O(\log(k/\delta))\) times, and this will boost the success probability to \(1-\delta\). Therefore, the modification will also succeed with probability \(1 - \delta\) by paying a \(\log(k/\delta)\) factor in the sample and time complexity. 
}
\end{proof}

\subsection{Efficient Sparse Fourier Transforms in High Dimensions}

A high-dimensional extension of the robust SFT algorithm has been explored in the work of \citet{jin2022superresolutionrobustsparsecontinuous}. Unfortunately, their robust SFT algorithm has a running time that scales exponentially with the dimension. In this section, we show how to use the one-dimensional SFT algorithm of \Cref{thm:sparse-Fourier-transform-Modified} in high dimensions and get an efficient robust SFT algorithm even for $d > 1$, which we will apply later in our parameter estimation algorithms.

A natural method to reduce the high-dimensional problem to $d=1$ is to randomly project the data along different directions, and then recover the high-dimensional means by solving a linear system. However, a key obstacle in this idea is that the ordering of the means could change among different projections. To overcome this issue, our algorithm is based on the idea of \citet{kalai2010efficiently}, where they project along directions that are \emph{close} to each other. The benefit of projecting along close-by directions is that the ordering of the means can be preserved with high probability, so that one can identify the projected means among different directions. Meanwhile, the distances between the directions should not be too small, as we want the condition number of the linear system to be polynomially bounded to recover the means efficiently. 

Using the above idea, we prove the following algorithmic result.

\begin{theorem}
[Efficient High-Dimensional SFT]
\label{thm:d-sft}
{\sledit{For any fixed \(T>0\),}
let $B_T^d(0)$ be the $d$-dimensional ball centered at $0$ with radius $T$.
Consider any signal \(x(t) = x^\star(t) + g(t)\in \C\) over \(t \in B_T^d(0)\), for arbitrary noise \(g(t)\) and exactly \(k\)-sparse \(x^\star(t) = \sum_{j=1}^k w_j e^{ i \<\mu_j, t\>}\) with \(\|\mu_j\|_2 \le B\) and frequency separation \(\gamma = \min_{j' \ne j} \|\mu_{j'}-\mu_j\|_2\).
Let \(\theta>0\) be some parameter. 
\sledit{If \(T \ge \Omega\left(\frac{d^{5/2}\log(k/\theta)}{\gamma}\right)\)}, 
then there is an algorithm $\mathsf{SFT}_d$ (\Cref{alg:d-sft}) that (i) randomly draws times $t_1,...,t_N$ with
\[
N = O(kd\log(B T)\log(k/\theta)\log(kd)),
\] 
(ii) queries the signal \(x(t)\) at $t=t_1,t_2,\dots, t_N$, and (iii) computes \(\{(\wh w_j, \wh \mu_j)\}\) in 
\[
O(kd\log(BT)\log(BT/\theta)\log (kd) ) \]
running time, such that the following holds.

Define the event $\mathcal E$ to be the set of the randomness \(r\) used by the algorithm such that, the algorithm running with randomness \(r\) outputs \(\{(\wh w_j, \wh \mu_j)\}_{j \in [k]}\)
with the property that there is a permutation $\pi$ such that
that for any \(w_j\) with \(|w_j| =\Omega(\mathcal N)\), 
\[
\|\mu_j - \wh \mu_{\pi(j)}\|_2 \le O \left(\frac {d^3 B \mathcal{N}}{\gamma T|w_j|} \right), \qquad
{|w_j - \wh{w}_{\pi(j)}| \leq O(\mathcal{N})}\,,
\] 
where
\[
\mathcal N^2 = \max_{j \in [N]} |g(t_j)|^2 +\theta \sum_{\ell=1}^k |w_\ell|^2. 
\]
Then \(\Pr_{r}\left[r\in \mathcal E\right] \ge 2/3\,.\) 

Moreover, the success probability can be boosted to \(1- \delta\), with sample complexity \[N = O(kd \log(BT)\log(k/\theta)\log(kd)\log(1/\delta))\] and time complexity \[O(kd \log(BT)\log(BT/\theta)\log(kd)\log(1/\delta) + k^3d \log(1/\delta)^2) \]

}
\end{theorem}

\begin{algorithm}[!ht]
\centering
\begin{algorithmic}[1]
\Require Sample access to the $k$-sparse signal $x(t) = x^\star(t) + g(t)$ for $t \in B_T(0) \subseteq \R^d.$
\Ensure Estimation of the tones $\{(\wh w_j, \wh \mu_j)\}_{j \in [k]}. $
\State \(\delta_0 \gets 1/3\). 
\State Pick a random direction $r \sim \text{Unif}(S^{d-1}).$ 
\State Pick {an arbitrary} orthonormal basis $\{b_1,...,b_d\}$ and set $b_0 := 0.$

\For{\(\ell \gets 0,\dots, d\)} 
\State $r_\ell \gets r + \varepsilon_1 b_\ell$, where \(\varepsilon _1 = \frac{\delta_0\gamma}{8B d^{5/2}}\). 
\State Define the projected signal $x^{r_\ell}(t) := x(t\cdot r_\ell)$ for $t \in [-T/2, T/2]$.
\State $\{(\wh w_j^{r_\ell}, \wh \mu_j^{r_\ell})\}_{j \in [k]} \gets \mathsf{SFT}_1(x^{r_\ell}, k,T, 2B, \gamma_1 = \frac{\delta_0\gamma}{4d^{5/2}} , \theta, \frac{\delta_0}{2(d+1)}).$

\State Sort $\{(\wh w_j^{r_\ell}, \wh \mu_j^{r_\ell})\}_{j \in [k]}$ in decreasing ordering according to $\{\wh \mu_j^{r_\ell}\}_{j \in [k]}$.  
\EndFor

\For{$j \gets 1, \dots, k$}
\State $\wh \mu_j \gets \sum_{\ell = 1}^{d} b_j \cdot \frac{\wh \mu_j^{r_\ell} - \wh \mu_j^{r_0}}{\varepsilon_1}$ 
\State \(\wh w_j \gets \wh w_j^{r_0} \). 
\EndFor

\end{algorithmic}
    \caption{Sparse Fourier Transform in \(d\) dimensions, constant success probability}
    \label{alg:d-sft}
\end{algorithm}

Before proving \Cref{thm:d-sft}, we will need to introduce some key lemmas from \citet{kalai2010efficiently}.
We will use the following geometric lemma from \cite{kalai2010efficiently} to show that the separation between the means is preserved after the projection.

\begin{lemma}[{\cite[Lemma~12]{kalai2010efficiently}}, Separation after Projection]
\label{lem:proj-sep}
For any \(\mu\ne \mu' \in \R^d\), \(\delta>0\), and a random \(r\) uniformly over \(S^{d-1}\), \[\Pr_{r\sim {\rm Unif}(S^{d-1})}\left[ \left| \langle \mu, r \rangle - \langle \mu', r \rangle \right| \le \frac{\delta \| \mu - \mu' \|_2}{\sqrt{d}} \right] \le \delta. \]
\end{lemma}

{
Moreover, one can show the ordering of the projected means will not change among different projections, {when the directions of the projections are defined as in \Cref{alg:d-sft}}.
In \Cref{alg:d-sft}, the first projection is random along the direction $r \sim \mathrm{Unif}(S^{d-1})$ and then for $\ell \in [d]$, the algorithm projects in the direction $r_\ell := r + \eps_1 b_\ell$, as defined in \Cref{alg:d-sft}, which adds a small perturbation (of order $\eps_1$) to $r$ in the direction of the vector $b_\ell$ of some arbitrary orthonormal basis $\{b_1,...,b_d\}$.
In particular, to prove this, it is sufficient to show that, for a fixed mean $\mu$, the projection in any direction $r_{\ell}$ for $\ell \in [d]$, i.e., $\<\mu, r_\ell \>$ will not change too much compared to $\<\mu, r\>$. }
\begin{lemma}
\label{lem:projections-close}
For any \(\mu\in \R^d\) and \(r\), \(\{r_\ell\}_{\ell=1}^d\) defined in \cref{alg:d-sft}, \(\left| \langle \mu, r_\ell \rangle - \langle \mu, r \rangle \right| \le \varepsilon_1 \|\mu\|_2. \)
\end{lemma}
\begin{proof}
We have \(r_\ell = r+ \varepsilon_1 b_\ell\), where \(\{b_\ell\}_{\ell\in [d]}\) is a basis. Thus, \begin{align*}
\left| \langle \mu, r_\ell \rangle - \langle \mu, r \rangle \right| = \varepsilon_1 \left| \langle \mu, b_\ell \rangle \right| \le \varepsilon_1  \|b_\ell\|_2 \|\mu\|_2 \le \varepsilon_1 \|\mu\|_2. 
\end{align*}
\end{proof}

After projecting in these $d+1$ directions, one can run the univariate sparse Fourier transform to estimate the projections of the means. We can then recover the means from the information of the projections. 
\begin{lemma}[{\cite[Lemma~15]{kalai2010efficiently}, Solving the System}]
\label{lem:solve-mean-from-proj}
For any \(\mu\in \R^d\) and \(\varepsilon ,\varepsilon _1 > 0\), and \(\{r_\ell\}_{\ell=0}^d\) defined in \cref{alg:d-sft}, suppose \(|\langle  r_\ell,\mu \rangle - \wh\mu^{r_\ell}| \le \varepsilon \) for all \(\ell=0,1,\dots, d\). Then \(\wh\mu := \sum_{\ell=1}^d b_\ell \cdot \frac{\wh \mu^{r_\ell} - \wh \mu^{r_0}}{\varepsilon_1}\) satisfies \(\|\mu-\wh\mu\|_2 \le \frac{2\sqrt{d}  }{\varepsilon _1}\varepsilon \). 
\end{lemma}
\begin{proof}
Since \(\{b_\ell\}_{\ell\in [d]}\) is an orthonormal basis of \(\R^d\), \begin{align*}
    \|\mu-\wh\mu\|_2^2 &= \sum_{\ell=1}^d \langle b_\ell, \mu-\wh\mu \rangle^2 = \sum_{\ell=1}^d \left( \langle b_\ell,\mu \rangle - \langle b_\ell,\wh\mu \rangle \right)^2 \\
    &= \sum_{\ell=1}^d \left( \frac{\langle r_\ell,\mu \rangle - \langle r_0,\mu \rangle}{\varepsilon _1} - \frac{\wh\mu^{r_\ell} - \wh\mu^{r_0}}{\varepsilon _1} \right)^2 \\
    &\le 2\sum_{\ell=1}^d \left( \frac{\langle r_\ell,\mu \rangle - \wh\mu^{r_\ell}}{\varepsilon _1} \right)^2 + \left( \frac{\langle r_0,\mu \rangle - \wh\mu^{r_0}}{\varepsilon _1} \right)^2  \\
    &\le 2d\cdot 2 \left( \frac{\varepsilon }{\varepsilon _1} \right)^2 = \frac{4d \varepsilon ^2}{\varepsilon _1^2}.
\end{align*}
That is, \(\|\mu-\wh\mu\|_2 \le \frac{2\sqrt{d} \varepsilon }{\varepsilon _1}\). 
\end{proof}

We are now ready to prove \Cref{thm:d-sft}, which gives us an efficient algorithm for the high-dimensional sparse Fourier transform.
\begin{proof}[Proof of \cref{thm:d-sft}]
Let \(\delta_0 = 1/3\). 
First, by a union bound and \cref{lem:proj-sep}, with probability at least \(1 - \delta_0/2\), for any \(j_1\ne j_2 \in [k]\), \[ \left| \langle \mu_{j_1}, r \rangle - \langle \mu_{j_2}, r \rangle \right| > \frac{\delta_0 \| \mu_{j_1} - \mu_{j_2} \|_2}{2d^{5/2}} \ge \frac{\delta_0 \gamma}{2d^{5/2}}.\]
Suppose this happens. 
Up to relabeling, assume \(\langle \mu_1,r \rangle \ge \langle \mu_2,r \rangle \ge\dots \ge \langle \mu_k,r \rangle\) without loss of generality. 
Choose \(\varepsilon _1 = \frac{\delta_0\gamma}{8B d^{5/2}}\), so that by \cref{lem:projections-close}, for all \(j\in [k]\) and \(\ell\in [d]\), \[\left| \langle \mu_{j},r_\ell \rangle - \langle \mu_{j}, r \rangle \right| \le \varepsilon _1 \|\mu_{j}\|_2 \le \varepsilon _1B = \frac{\delta_0\gamma}{8d^{5/2}}.\] 
Thus, for \(j_1\ne j_2\in [k]\), if \(\langle \mu_{j_1},r \rangle \ge \langle \mu_{j_2},r \rangle\), then for any \(\ell\in [d]\), \begin{align*}
\langle \mu_{j_1},r_\ell \rangle \ge \langle \mu_{j_1},r \rangle - \frac{\delta_0 \gamma}{8d^{5/2}} > \langle \mu_{j_2},r \rangle + \frac{3\delta_0 \gamma}{8d^{5/2}} \ge \langle \mu_{j_2},r_\ell \rangle + \frac{\delta_0 \gamma}{4d^{5/2}}. 
\end{align*}
That is, the order of the projected means is preserved among each projection direction, as well as the separation, up to a constant. We will use \(\gamma_1 = \frac{\delta_0\gamma}{4d^{5/2}} = 2B \varepsilon _1\) to denote the separation in the projections. Let \[(x^\star)^{r_\ell}(t) = x^\star(t\cdot r_\ell) = \sum_{j=1}^k w_j e^{i \langle \mu_j, r_\ell \rangle t}, \] where \(\langle \mu_j, r_\ell \rangle \le \|\mu_j\|_2 \|r_\ell\|_2 \le (1+\varepsilon _1 )B \le 2B\), and \(g^{r_\ell}(t) = g(t\cdot r_\ell)\). 
Then by \cref{thm:sparse-Fourier-transform-Modified}, since \(T > O \left( \frac{d^{5/2}\log(k/\theta)}{\delta_0\gamma} \right) = O \left( \frac{\log(k/\theta)}{\gamma_1} \right)\), the algorithm \(\mathsf{SFT}_1(x^{r_\ell},k,T,2B,\gamma_1,\theta,\frac{\delta_0}{2(d+1)})\) performs \[N_1 = O(k\log(BT)\log(k/\theta)\log (kd))\] queries on \(x^{r_\ell}(t)\) at \(t=t_1^{r_\ell},t_2^{r_\ell},\dots, t_{N_1}^{r_\ell}\), and outputs \(\{(\wh w_j^{r_\ell}, \wh\mu_j^{r_\ell})\}_{j\in [k]}\) in running time \[O(k\log(BT)\log(BT/\theta)\log (kd))\] 
such that there is a permutation \(\pi_\ell\) that for any \(j\in [k]\) with \(|w_j| = \Omega(\mathcal{N}_1)\), \[\left| \langle r_\ell, \mu_j \rangle - \wh\mu_{\pi_\ell(j)}^{r_\ell} \right| \le O \left( \frac{\mathcal{N}_1}{T|w_j|} \right) , \qquad \left| w_j - \wh w_{\pi_\ell(j)}^{r_\ell} \right| \le O(\mathcal{N}_1), \] 
where \[\mathcal{N}_1^2 = \max_{i\in [N_1]} |g(t_i)|^2 + \theta \sum_{j=1}^{k} |w_j|^2 \le \mathcal{N}^2, \] with probability at least \(1 - \frac{\delta_0}{2(d+1)}\). 
Thus, by a union bound, for any \(j\in [k]\) with \(|w_j| = \Omega(\mathcal N)\), we have \(\left| \langle r_\ell, \mu_j \rangle - \wh\mu_{\pi_\ell(j)}^{r_\ell} \right| \le O \left( \frac{\mathcal{N}}{T|w_j|} \right) \) and \(\left| w_j - \wh w_{\pi_\ell(j)}^{r_\ell} \right| \le O(\mathcal{N})\) for all \(\ell \in [d]\), with probability \(1 - \delta_0/2\). Suppose this happens. 
Since the ordering of the means is preserved among all the projections, we can match the projected means in different directions after sorting. Therefore, by \cref{lem:solve-mean-from-proj}, there is a permutation \(\pi\) such that for any \(j\in [k]\) with \(|w_j| = \Omega(\mathcal{N})\), \[\left\| \mu_j - \wh\mu_{\pi(j)} \right\|_2 \lesssim \frac{2\sqrt{d}}{\varepsilon _1}\frac{\mathcal{N}}{T|w_j|} \lesssim \frac{2d^3 B\mathcal{N}}{\gamma T|w_j|}.\]  And for the weights, \(\left| w_j - \wh w_{\pi(j)}^{r_0} \right| \le O(\mathcal{N})\). {The above two error guarantees hold with probability \(1 - \delta_0/2 - \delta_0/2 = 2/3\). }

The number of queries is \[N = O(kd\log(BT)\log(k/\theta)\log(kd)),\] and the running time is \[O(kd\log(BT)\log(BT/\theta)\log(kd)). \]
By \cref{lem:boosting}, there is an algorithm that achieves the same error guarantees, with probability \(1-\delta\), using \[N = O(kd \log(BT)\log(k/\theta)\log(kd)\log(1/\delta))\] samples and \[O(kd \log(BT)\log(BT/\theta)\log(kd)\log(1/\delta) + k^3 d\log(1/\delta)^2)\] time. 

\end{proof}

We now give a lemma for boosting the success probability. For the proof we refer to \Cref{app:boosting}.
\begin{lemma}
[Boosting] 
\label{lem:boosting}
Assume that there are $k$ points $\mu_1,...,\mu_k \in \R^d$ with $\gamma = \min_{j' \neq j} \|\mu_{j'} - \mu_j \|_2$ and weights \(w_1,\dots, w_k\in \R\). 
For \(\epsilon', \varepsilon _w \in (0, 1)\), let $A(\epsilon',\varepsilon _w)$ be an algorithm that uses $n(\epsilon',\varepsilon _w)$ samples and runs in time $T(\epsilon',\varepsilon _w)$, and with probability \(2/3\), computes points $\{(\wh w_j, \wh \mu_j)\}_{j\in [k]}$ such that there is a permutation $\pi$ that $\max_{j \in [k]} \|\mu_j - \wh \mu_{\pi(j)}\|_2 \leq \epsilon'$ and \(\max_{j\in [k]} |w_j - \wh w_{\pi(j)}| \le \varepsilon _w\). 
Let $\epsilon,\delta \in (0,1)$ be the target accuracy and confidence.
Then there is an algorithm (\Cref{alg:boosting}) that uses $O(n(\min\{\varepsilon /3,\gamma/16\}, \varepsilon _w) \log(1/\delta))$ samples and runs in $O(T(\min\{\varepsilon /3,\gamma/16\}, \varepsilon _w)\log(1/\delta) + k^3d \log(1/\delta)^2) $ times,  
and with probability $1-\delta$, computes points $\wh \mu_1,...,\wh \mu_k$ 
such that there is a permutation $\pi$ such that $\max_{j \in [k]} \|\mu_j - \wh \mu_{\pi(j)}\|_2 \leq \epsilon$ and \(\max_{j\in [k]} |w_j - \wh w_{\pi(j)}| \le \varepsilon _w\). 
\end{lemma}

\section{Application I: Efficiently Learning Mixture Models  }
\label{sec:learningMixtures}

In this section we will study how to use our efficient sparse Fourier tool for learning mixture models.
First, we recall the definition of SFD  that we will need for our results. 

\begin{definition}
[Slow Fourier Decay]
\label{def:sfd}
Let $D$ be a probability distribution over $\R^d$. We say that $D$ satisfies the \emph{Slow Fourier Decay property (SFD)} with constants $c_1,c_2 \ge 0$ if the function \(R(T) = \inf_{t :\|t\|_2 \le T} |{\phi_D(t)}|\) satisfies that
\[ R(T) \gtrsim {d^{-c_1} T^{-c_2}}\,.
  \]
\end{definition}
In the next section, we show how to learn mixtures of SFD distributions. 

\subsection{Learning SFD Mixture Models }

In this section, we present our efficient parameter estimation algorithm for mixtures {models that satisfy the SFD property} in $d$ dimensions. The algorithms requires {no minimum separability assumptions}, except of the minimal information-theoretic ones and gets polynomial sample and time complexity. This is in stark contrast to the Gaussian case, which requires separation $\gamma = \sqrt{\log k}$ to get polynomial sample complexity \cite{regev_learning_2017}.

\begin{theorem}
\label{thm:learning-SFD-mixtures}
Let \(D\) be a distribution over \(\R^d\) satisfying SFD, that is, there exist constants \(c_1, c_2 \ge 0\) such that \(\inf_{t:\|t\|_2 \le T}|{\phi_D(t)}| \gtrsim d^{-c_1}T^{-c_2}\).
Consider a mixture $\mathcal M$ of $k$ distributions \(D(\mu_1),...,D(\mu_k)\) with means $\{\mu_j\}_{j\in [k]}$ and weights $\{w_j\}_{j\in [k]}$. Let $\gamma = \min_{j' \neq j} \|\mu_{j'} - \mu_j\|_2$, $w_{\min} = \min_{j\in [k]} w_j$, and \(B = \max_{j\in [k]}\|\mu_j\|_2\). There is an algorithm that given $\eps, \delta \in (0,1)$ and $n$ i.i.d.\ samples from $\mathcal M$, computes a list $\{(\wh w_j, \wh \mu_j)\}_{j\in [k]}$ such that there is a permutation $\pi$ with
\[
{\max_{j \in [k]}~
\| \mu_{j} - \wh \mu_{\pi(j)} \|_2 \leq \eps\,, \qquad\max_{j \in [k]} |w_j - \wh{w}_{\pi(j)}| \leq \eps}
\]
with probability at least $1-\delta$.  The sample complexity is \[n = \wt{O} \left( \frac{\poly_{c_1,c_2}(d, 1/\gamma) B^2 \log(1/\delta)}{w_{\min}^2 \varepsilon ^2} \right) \]
{and the running time is $\poly(n).$}
\end{theorem}

\begin{proof}
{The proof follows from the more general \Cref{thm:recover-sfd} of the upcoming Section by setting $k' = 0$.}

\end{proof}

As an illustration this result immediately yields an efficient algorithm for learning mixtures of Laplace distributions with sample and time complexity that scale polynomially with $d,k,1/\varepsilon$ and the separation $1/\gamma$.

\subsection{Learning SFD-FFD Mixture Models}

In this section, we will provide an algorithm for learning mixture models that contain both SFD and FFD components under some natural assumptions. To do that, we have to introduce the notion of FFD distributions, which will be the ``complement'' of the SFD components.

\begin{definition}
[Fast Fourier Decay]
Let $D$ be a probability distribution over $\R^d$. We say that $D$ satisfies the \emph{Fast Fourier Decay property (FFD)} with constants $c_1',c_2' > 0$ if the function \(R'(T)=\sup_{t:\|t\|_2 \ge T} |\phi_D(t)| \) satisfies that \[R'(T) \lesssim d^{-c_1'} T^{-c_2'}.\]

  \label{def:ffd}
\end{definition}

\subsubsection{Recovering the SFD part using Fourier}
\label{sec:sfd-proof}
In this section, we will show how to recover the means of the SFD components given samples from a mixture model that contains $k$ SFD components and $k'$ FFD components (whose Fourier decay is faster than that of the SFD part).

{
\begin{theorem}
    [Recovering the SFD means]
    \label{thm:recover-sfd}
    Let $\mathcal M$ be a mixture of \(k+k'\) distributions \(D_1,\dots, D_k,\) \(D_1', \dots, D_{k'}'\) over \(\R^d\), with means \(\mu_1,\dots, \mu_k,\mu_1',\dots, \mu_{k'}' \in \R^d\) and weights \(w_1,\dots, w_k, w_1',\dots, w_{k'}'\) that \(\sum_{j\in [k]} w_j + \sum_{j\in [k']} w_j' = 1\). Assume
    \begin{enumerate}
        \item \(D_1,\dots, D_k\) are $k$ translations of a distribution $D$
        over \(\R^d\) satisfying SFD, that is, there exist constants \(c_1, c_2 \ge 0\) such that  \(\inf_{t:\|t\|_2 \le T}|{\phi_{D}(t)}| \gtrsim d^{-c_1}T^{-c_2}\), and,
        \item $D_1',...,D'_{k'}$ satisfy FFD, that is, there exist constants \(c_1', c_2' \ge 0\) such that  \(\sup_{t:\|t\|_2 \ge T}|{\phi_{D_j'}(t)}| \lesssim d^{-c_1'}T^{-c_2'}\) for all \(j\in [k']\), \textbf{with $c_2' > c_2$.} 
    \end{enumerate}
Let \(\gamma = \min_{j\ne j' \in [k]} \|\mu_j - \mu_{j'} \|_2\) be the minimum separation among the SFD components, \(w_{\min} = \min_{j\in [k]} w_j\) be the minimum weight in the SFD part, and \(B = \max_{j\in [k]} \|\mu_j\|_2\) be the maximum norm of the SFD means. 
There is an algorithm that given \(\varepsilon ,\delta \in (0, 1)\) and \(n\) i.i.d.\ samples from \(\mathcal{M}\), outputs a list \(\{(\wh w_j,\wh\mu_j)\}_{j\in [k]}\) such that there is a permutation \(\pi\) on \([k]\) with \[
{\max_{j \in [k]}~
\| \mu_{j} - \wh \mu_{\pi(j)} \|_2 \leq \eps\,, \qquad\max_{j \in [k]} |w_j - \wh{w}_{\pi(j)}| \leq \eps}
\]
with probability at least $1-\delta$. 
The sample complexity is \[n =   {\poly_{c_1,c_2,c_1',c_2'}(d, 1/\gamma,B,1/w_{\min}, 1/\varepsilon )  \log(1/\delta)}  \]
{and the running time is $\poly(n).$}
\end{theorem}

}
\begin{proof}
The SFD part has $k$ components with weights $w_i$ and means $\mu_i$. Similarly, the FFD part has $k'$ components with weights $w_i'$ and means $\mu_i'$.
Let us compute the characteristic function of \(Y \sim \mathcal{M}\): \begin{align*}
    \E_{Y\sim \mathcal{M} } [e^{i \langle t, Y \rangle}] &=  \sum_{j\in [k]} w_j \phi_{D(\mu_j)}(t) + \sum_{j\in [k']} w_j' \phi_{D'_j(\mu_j')}(t) \\
    &=\sum_{j\in [k]} w_j e^{i \langle t, \mu_j \rangle} \phi_{D}(t) + \sum_{j\in [k']} w_j' e^{i \langle t, \mu_j' \rangle}\phi_{D'_j}(t). 
\end{align*}
Here recall that the SFD part consists of translations of $D$ while the FFD part consists of $D_1',...,D_{k'}'.$
The idea is to estimate 
\[\phi_{D}(t)^{-1} \E_{Y\sim \mathcal{M}} [e^{i \langle t, Y \rangle}] = \sum_{j\in [k]} w_je^{i \langle t, \mu_j \rangle} + \sum_{j\in [k']} w_j' e^{i \langle t, \mu_j' \rangle} \frac{\phi_{D'_j} (t)}{ \phi_{D}(t)}. \] 
Now, if all of the \(D'_j\), \(j\in [k']\), have fast enough Fourier decay compared to \(D\), namely \(\frac{d^{c_1'} T^{c_2'}}{d^{c_1} T^{c_2}}\) grows fast enough, then the second summation above will vanish for large \(t\). 
However, when \(t = 0\), we have \(\phi_{D}(0) = \phi_{D'_j}(0) = 1\). 
The trick here is to shift the ball \(B_T^d(0)\) where we will query the signal. 
Note that for any \(v\in \R^d\), we have \begin{align*}
   \phi_{D}(t+v)^{-1} \E_{Y\sim \mathcal{M}} [e^{i \langle t+v, Y \rangle}] = \sum_{j\in [k]} w_je^{i \langle v, \mu_j \rangle}e^{i \langle t, \mu_j \rangle} + \sum_{j\in [k']} w_j' e^{i \langle t+v, \mu_j' \rangle} \frac{\phi_{D'_j} (t+v)}{ \phi_{D}(t+v)}.
\end{align*}
Let \(T > 0\) be the large enough duration which will be determined later, and set \(v\) to be an arbitrary vector with \(\|v\|_2 = 2T\). 
Therefore, for \(t\in B^d_T(0)\), we have \(T \le \|t+v\|_2 \le 3T\), which implies \(|\phi_{D}(t+v)| \gtrsim d^{-{c_1}} (3T)^{-c_2} \gtrsim d^{-c_1} T^{-c_2}\), and \(|\phi_{D'_j}(t+v)| \lesssim d^{-c_1'} T^{-c_2'}\), for all \(j\in [k']\). {Here we applied} the SFD property at time $3T$ and the FFD property at time $T$.

Following the notation in \cref{thm:d-sft}, let the true signal be \(x^\star(t) = \sum_{j\in [k]} w_j e^{i \langle v,\mu_j \rangle} e^{i \langle t,\mu_j \rangle} \). 
Given i.i.d.\ samples \(Y_1,Y_2,\dots, Y_n\) from \(\mathcal{M}\), let the signal we observe be \[x(t) = \phi_{D}(t+v)^{-1}\cdot  \frac1n \sum_{\ell=1}^n e^{i \langle t+v, Y_\ell \rangle}.\] Also, let \(g(t) = x(t) - x^\star(t)\) be the noise. 
Since \(|e^{i \langle t + v, Y \rangle}| = 1\) is bounded, by Hoeffding's inequality, \begin{align*}
    \Pr \left[ \left| \frac{1}{n} \sum_{\ell=1}^n e^{i \langle t + v, Y_\ell  \rangle} - \E[e^{i \langle t+v, Y \rangle}] \right| \ge s \right] \le e^{-\Omega(ns^2)}
\end{align*}
for any fixed \(t\in B^d_T(0)\).
Then, for any fixed \(t\in B^d_T(0)\), the noise \begin{align*}
    |g(t)| &= \left| \phi_{D}(t+v)^{-1}\cdot \frac{1}{n} \sum_{\ell=1}^n e^{i \langle t+v, Y_\ell \rangle} - \sum_{j\in [k]}w_j e^{i \langle v, \mu_j \rangle} e^{i \langle t, \mu_j \rangle} \right| \\
    &\le |\phi_{D}(t+v)|^{-1}\left| \frac{1}{n} \sum_{\ell=1}^n e^{i \langle t+v, Y_\ell \rangle} -\E_{Y\sim \mathcal{M}} [e^{i \langle t+v, Y \rangle}] \right| + \sum_{j\in [k']} w_j' \left| \frac{\phi_{D'_j}(t+v)}{\phi_{D}(t+v)} \right| \\
    &\le O \left( d^{c_1} T^{c_2}s + \frac{d^{c_1} T^{c_2}}{d^{c_1'} T^{c_2'}} \right)
\end{align*}
with probability at least \(1 -e^{-\Omega(ns^2)}\). 

Now,
suppose that the algorithm in \Cref{thm:d-sft} queries the signal \(x(t)\) at times \(t=t_1,t_2,\dots, t_N\). 
By the union bound, with probability at least \(1-N\cdot e^{-\Omega(ns^2)}\), \(|g(t_j)| \le O \left( d^{c_1}T^{c_2}s + {d^{c_1-c_1'} T^{c_2-c_2'}}  \right)\) for all \(j\in [N]\). 
Then, we can apply \cref{thm:d-sft}, setting \begin{align*}
    s &= \Theta \left( \sqrt{\frac{\log(N/\delta)}{n}} \right), \\
    \theta &= \frac{\varepsilon ^2}{100 \sum_{j\in [k]} |w_j|^2}, \\
    T &= C_T\max  \left\{
        \left( \frac {d^{c_1-c_1'}}{\varepsilon } \right)^{1/(c_2'-c_2)}, 
        \frac{d^3B}{\gamma w_{\min}},
        \frac{d^{5/2} \log(k/\theta)}{\gamma} \right\},
\end{align*}
for some absolute constant \(C_T>0\). 
In this case, the noise level in \cref{thm:d-sft} is \begin{align*}
    \mathcal{N}^2  &= \max_{j\in [N]} |g(t_j)|^2 + \theta \sum_{j\in [k]}|w_j|^2 \\
    &= O \left( d^{c_1}T^{c_2}s + \frac{d^{c_1-c_1'} }{T^{c_2'-c_2}}  \right)^2 + \frac{\varepsilon ^2}{100}\\
    &= O \left( \frac{d^{c_1}T^{c_2}\sqrt{\log(N/\delta)}}{\sqrt{n}} + \frac{d^{c_1-c_1'} }{T^{c_2'-c_2}}+ \varepsilon \right)^2
\end{align*}
and \cref{alg:d-sft} runs in 
\begin{align*}
    &O(k d \log(BT)\log(BT /\theta) \log(kd) \log(1/\delta) + k^3 d \log(1/\delta)^2)\\
    &= \wt O \left( kd \log \left(B/(\gamma w_{\min} \varepsilon ) \right)^2 \log(1/\delta) + k^3 d \log(1/\delta)^2 \right)
\end{align*} time with 
\begin{align*}
    N &= O(kd\log(BT)\log(k/\theta) \log(kd)\log(1/\delta))\\
    &=\wt O \left(kd \log(B/(\gamma w_{\min} \varepsilon ))\log(1/\varepsilon )\log(1/\delta)\right)
\end{align*} 
and outputs \(\{(\wt w_j, \wt\mu_j) \}_{j\in [k]}\) such that there is a permutation \(\pi\) on \([k]\) such that for all \(j\in [k]\) that \(|w_j| \ge \Omega(\mathcal{N})\), 
\begin{align*}
    \left| w_j e^{i \langle v, \mu_j \rangle} - \wt w_{\pi(j)} \right| &\le O(\mathcal{N}) \le \wt{O} \left( \frac{d^{c_1}T^{c_2}\sqrt{\log(k/\delta) + \log\log(B/(\gamma w_{\min} \varepsilon ))}}{\sqrt{n}} + \frac{d^{c_1-c_1'}}{ T^{c_2'-c_2}} + \varepsilon  \right)
\end{align*}
and \begin{align*}
    \left\| \mu_j - \wt \mu_{\pi(j)} \right\|_2 &\le 
    O \left( \frac{d^3 B \mathcal N}{ \gamma T |w_j|} \right).
\end{align*}
Since \[T \ge C_T\max \left\{\left( \frac {d^{c_1-c_1'}}{\varepsilon } \right)^{1/(c_2'-c_2)}, \frac{d^3B}{\gamma w_{\min}},\right\},\] we have \begin{align*}
    \frac{d^{c_1-c_1'}}{T^{c_2'-c_2}} \lesssim \varepsilon ,\qquad \frac{d^{3} B \mathcal{N}}{\gamma T |w_j|} \lesssim \mathcal{N},
\end{align*}
and thus \begin{align*}
    \left| w_j e^{i \langle v, \mu_j \rangle} - \wt w_{\pi(j)} \right| &\le \wt{O} \left( \frac{d^{c_1}T^{c_2}\sqrt{\log(k/\delta) + \log\log(B/(\gamma w_{\min} \varepsilon ))}}{\sqrt{n}} + \varepsilon  \right), \\
    \left\| \mu_j - \wt \mu_{\pi(j)} \right\|_2 &\lesssim \mathcal{N} \le \wt{O} \left( \frac{d^{c_1}T^{c_2}\sqrt{\log(k/\delta) + \log\log(B/(\gamma w_{\min} \varepsilon ))}}{\sqrt{n}} + \varepsilon  \right). 
\end{align*}
Then, by choosing \begin{align*}
    n &= \wt O \left(  \frac{d^{2c_1} T^{2c_2} (\log(k/\delta)+\log\log(B/(\gamma w_{\min} \varepsilon )))}{ \varepsilon ^2}\right)\\
    &= {O} \left( {\poly_{c_1,c_2,c_1',c_2'}(d, 1/\gamma,B,1/w_{\min}, 1/\varepsilon )  \log(1/\delta)} \right) ,
\end{align*}
where the degree of the polynomial depends on the constants $c_1,c_2,c_1',c_2'$,
we will have \[\mathcal{N} \lesssim \varepsilon , \qquad \left| w_j e^{i \langle v, \mu_j \rangle} - \wt w_{\pi(j)} \right| \lesssim \varepsilon, \qquad \left\| \mu_j - \wt \mu_{\pi(j)} \right\|_2  \lesssim \varepsilon .  \] 
Assume \(\varepsilon \lesssim w_{\min}\) so that for all \(j\in [k]\), \(|w_j| \gtrsim \mathcal{N}\), otherwise we can output \(\wh w_{\pi(j)} = 0\) if \(|w_j| \lesssim \varepsilon \). 
Therefore, we will get \(w_{\min} = \min_{j\in [k]}|w_j| \ge \Omega(\mathcal{N})\), and the error \(\max_{j\in [k]}\| \mu_j - \wt \mu_{\pi(j)} \|_2 \le \varepsilon \) and \(\max_{j\in [k]}| w_j e^{i \langle v, \mu_j \rangle} - \wt w_{\pi(j)} | \le \varepsilon\) with probability \(1-\delta\), in $\poly(n)$ time. 
Lastly, our algorithm will output \(\{(\wh w_j, \wh\mu_j) \}_{j\in [k]}\) as the estimate, where \(\wh w_j = |\wt w_j|\) and \(\wh \mu_j = \wt \mu_j\), so that \[\left| w_{j} - \wh w_{\pi(j)} \right| = \left| | w_j e^{i \langle v, \mu_j \rangle}| - |\wt w_{\pi(j)}| \right| \le \left| w_j e^{i \langle v, \mu_j \rangle} - \wt w_{\pi(j)} \right| \le \varepsilon  \] and \(\| \mu_j - \wh \mu_{\pi(j)} \|_2 \le \varepsilon \), for all \(j\in [k]\). 

\end{proof}

\subsubsection{Recovering the FFD part using SoS}
\label{sec:ffd-proof}
\paragraph{Background on SoS tools.} Before presenting our result for this section, we provide some required background. First, we will say that a distribution $D$ satisfies the resilience property ({adapted from \citet{steinhardt_resilience_2017}}) with parameters $n$ and $\Delta$ if given any set $T$ of $n$ i.i.d.\ samples, it holds that with high probability for any subset set $S \subseteq T$ of size $\alpha n$, the empirical mean over $T$ is $\Delta(\alpha)$-close to the true mean of $D$. Hence, resilience is a measure of stability for the mean of $D$ and is implied e.g., by distributions with good concentration properties.
\begin{definition}
[Resilience]
\label{def:resilience}
Let \(D\) be a distribution over \(\R^d\) with mean \(\mu\). We say \(D\) satisfies \((n, \Delta)\)-resilience for \(n:\R \times \R \to \R\) and \(\Delta: \R\to \R\), if for any \(\delta \in (0, 1)\) and {sufficiently small} \(\alpha \in (0,1)\) the following holds: for \(n=n(\delta, \alpha)\) i.i.d.\ samples \(x_1,\dots, x_{n}\) from \(D\), with probability at least \(1-\delta\), \[\max_{\substack {S\subseteq [n] \\ |S| = \alpha n}} \left\| \frac{1}{\alpha n} \sum_{i\in S} x_i - \mu \right\|_2 \le \Delta(\alpha).  \]
\end{definition}
{%
To illustrate the above definition, if the tail of the distribution \(D\) is sub-Weibull, i.e., the tail is of order \(e^{-t^{\beta}}\) for some \(\beta>0\), then \(D\) satisfies resilience property with \(\Delta(\alpha) = O( \ln(1/\alpha)^{1/\beta})\). We will prove the following lemma in \Cref{app:resilience}. 
\begin{lemma}
[Tail Decay $\Rightarrow$ Resilience]
\label{lem:resilience-subW}
Let \(D\) be a distribution over \(\R^d\) with mean \(\mu\). 
Suppose for some constants \(C_0, \sigma, \beta > 0\), \[\Pr_{X\sim D}[|\langle X-\mu, v \rangle| \ge t] \le C_0 \exp \left( - \left(t/\sigma \right)^{\beta} \right)\] for all \(v\in S^{d-1}\) and \(t > 0\), then \(D\) satisfies \(\left(\frac1{\alpha}(d+\log(1/\delta))^{O(\max\{1/\beta, 1\})},O\left(\sigma(\ln\frac{1}{\alpha})^{1/\beta}\right)\right)\)-resilience. 
\end{lemma}
Since sub-Gaussian and sub-exponential distributions are special cases of sub-Weibull distributions, we get the following corollary immediately. 

\begin{corollary}
Let \(D\) be a distribution over \(\R^d\) with mean \(\mu\). 
\begin{enumerate}
    \item If \(D\) is sub-Gaussian, that is, there is some constant \(\sigma>0\) that \(\Pr_{X\sim D}[| \langle X-\mu, v \rangle| \ge t] \lesssim \exp \left( -(t/\sigma)^2 \right)\) for all \(v\in S^{d-1}\) and \(t>0\) (e.g., Gaussian distribution with constant bounded covariance), then \(D\) satisfies \(\left(\frac{1}{\alpha} \poly(d,\log(1/\delta)), O\left(\sigma\sqrt{\ln(1/\alpha)}\right)\right)\)-resilience. 
    \item If \(D\) is sub-exponential, that is, there is some constant \(\sigma>0\) that \(\Pr_{X\sim D}[| \langle X-\mu, v \rangle| \ge t] \lesssim \exp \left( -t/\sigma \right)\) for all \(v\in S^{d-1}\) and \(t>0\) (e.g., Laplace distribution with constant bounded covariance), then \(D\) satisfies \(\left(\frac{1}{\alpha} \poly(d,\log(1/\delta)), O\left(\sigma{\ln(1/\alpha)}\right)\right)\)-resilience. 
\end{enumerate}
\end{corollary}
}

The second definition that we will need is that of certifiably-bounded distributions \cite{hopkins2018mixture,kothari_better_2017,kothari2018robust}. 
\begin{definition}
[Certifiably Bounded]
\label{def:cb}
Let \(D\) be a distribution over \(\R^d\) with mean \(\mu\). We say \(D\) is \((2t, B)\)-certifiably-bounded for \(t\in \N\) and \(B>0\), if there is a degree-\(2t\) sum-of-squares proof of the following polynomial inequality on \(v\): \[\E_{x\sim D} [\langle x-\mu,v \rangle^{2t}] \le B^{2t}\|v\|_2^{2t}.\]
\end{definition}
To see why this definition is relevant, recall that a distribution $D$ is $s$-sub-Gaussian if all
its linear projections have tail probabilities decaying at least as fast as Gaussian tails. In terms of moments this means that for any $t \geq 1$ and for all $v:$
\[
\E_{x \sim D}[\<x-\mu, v\>^{t}] \leq (C s \sqrt{t} \|v\|_2)^t
\]
for some universal constant $C$. The above definition can be seen as an algorithmic friendly notion of  sub-Gaussian distributions since it guarantees that up to power $2t$, there is a short certificate in the form of a sum of squares proof that the moment-boundedness holds in all directions $v$. Note that \Cref{def:cb} allows for more general tail behaviors that sub-Gaussian since it allows for a general function $B$ in the bound.

For a distribution $D$ that is certifiably bounded distribution up to power $O(t)$ and has sub-exponential tails, there is a SoS algorithm that runs in roughly $d^{O(t)}$ time and performs robust mean estimation, i.e., uses an $\alpha$-corrupted sample from $D$ and computes a mean that is $B \alpha^{1-1/2t}$-close to the mean of $D$, given that the corruption rate $\alpha \leq 1/4$. More formally,

\begin{theorem}
[Robust Mean Estimation, {\cite[Theorem 5.4]{kothari_better_2017}}]    
\label{thm:sos-robust-mean-est}
Let \(x_1,\dots, x_n \in \R^d\) be such that there exists a subset \(I \subseteq [n]\) of size \((1- \alpha ) n\) that \(\{x_i\}_{i\in I}\) are i.i.d.\ samples from a \((2t,B)\)-certifiably bounded and sub-exponential distribution with mean \(\mu \in \R^d\). Then, if \(\alpha \le 1/4\) and \(n \gtrsim (2d \log (dt/\delta))^t+d\log(1/\delta)/ \alpha ^2\), there is an algorithm that runs in \(n^{O(t)}\) time and outputs an estimate \(\wh\mu\) such that with probability at least \(1-\delta\), \(\|\wh\mu -\mu\|_2 \le O(B \alpha ^{1-1/2t})\). 
\end{theorem}

Observe that this guarantee on the error interpolates between the $\sqrt{\alpha}$ error for $t=1$ (bounded covariance distributions) and $\alpha$ error for $t = \infty$ (e.g., Gaussian distributions). The above tool can be also extended to the case where the corruption rate $\alpha$ is above $> 1/2$. In this regime, we are working in the \emph{list-decodable} setting, where the goal is to recover a list of means that contains a good estimation of the true one. The next theorem essentially implies an efficient procedure that gets as input a (potentially heavily) corrupted sample from a certifiably bounded and sub-exponential distribution and outputs a list of sets (each of which contains some of the given points) with the guarantee that one of their empirical means will be close to the true one. More formally,
\begin{theorem}
[{List-decodable Mean Estimation, \cite[Theorem 5.5, Proposition 5.9]{kothari_better_2017}}]
\label{thm:list-decode}
Let \(x_1,\dots, x_n \in \R^d\) be such that there exists a subset \(I \subseteq [n]\) of size \(\alpha n\) that \(\{x_i\}_{i\in I}\) are i.i.d.\ samples from a \((2t,B)\)-certifiably bounded and sub-exponential distribution with mean \(\mu \in \R^d\). Then, if \(n \gtrsim (2d \log(dt/\delta))^t/\alpha\), there is an algorithm that runs in \(n^{O(t)}\) time and outputs a list of sets \(S_1,\dots, S_m \subseteq [n]\), such that with probability \(1-\delta\), \(m \le \frac{4}{\alpha}\) and the following holds (let \(\wt\mu_j = \frac1{|S_j|}\sum_{i\in S_j} x_i \)): \begin{enumerate}
    \item \(|S_j| \ge \alpha n/4\) for all \(j\in [m]\). 
    \item \(S_j \cap S_{j'} = \varnothing\) for \(j\ne j'\in [m]\). 
    \item \(S_j\) satisfies some resilience property for all \(j\in [m]\), that is, any subset \(S'_j \subseteq S_j\) with \(|S'_j| \ge \beta n\) satisfies \[ \left\| \frac1{|S'_j| }\sum_{i\in S'_j} x_i - \wt \mu_j \right\|_2  \le O(B/\alpha^{1/t}+ B/\beta^{1/2t}).\] 
    \item There exists a \(j\in [m]\) such that \(\|\wt\mu_j - \mu\|_2 \le O(B/\alpha^{1/t})\). 
\end{enumerate}
\label{thm:sos-cluster}
\end{theorem}

Let us shortly explain how these algorithms work. The main technical contribution of \citet{kothari_better_2017,kothari2018robust} is an SoS toolbox for upper bounding the injective tensor norm $\sup_{\|v\| \leq 1} \frac{1}{n} \sum_{i} \<v,x_i\>^{2t}$ of the $2t$-th moments of samples $x_1,...,x_n$. Observe that this quantity is directly related to the moment bounds of \Cref{def:cb}. In particular, they show that the Sum-of-Squares framework gives a
polynomial time procedure for a dimension-free upper bound on the injective norms of i.i.d. arbitrary distributions that are certifiably bounded and sub-exponential distributions (e.g., for Poincar\'e distributions). Both the robust mean estimation result and the list-decodable algorithm are derived under this SoS framework.

In more detail, the starting point of the above procedures is a convex relaxation of the clustering objective that gets $n$ points from the mixture and asks, roughly speaking, for either a collection of means that makes the injective norms of order $2t$ small or gives a certificate that this is not possible. To do this efficiently, one has to relax the injective tensor norm objective to the problem of finding means $w_1,...,w_n$ such that
\[
\frac{1}{n} \sum_{i \in [n]} \wt \E_{\xi(v)}[\<v, x_i - w_i\>^{2t}] 
\]
is small for all \emph{pseudo-distributions} $\xi(v)$ over the unit sphere.
While this can be implemented efficiently via convex programming (see \cite[Section C]{kothari_better_2017}), one has to take into account the outliers but also re-run the clustering procedure multiple times in order to avoid dependencies on the norm of the means. This directly implies the robust mean estimation algorithm \cite[Theorem 5.4, Algorithm 2]{kothari_better_2017}. To do this, one needs to keep a weight $c_i$ to each of the points $x_i$ in order to estimate more accurate means. The weight $c_i$ essentially amounts for the failure of the convex relaxation to certify an upper bound on the ``injective norm'' and hence we have to downweight this point (e.g., it could be an outlier). In particular, one can show that the outlier removal algorithm of \cite{kothari_better_2017} downweights the bad points much
more than the good points, when the ``injective norm'' is large. Moreover, they show that if the value of ``injective norm'' is small, then the returned points $w_1,...,w_n$ form a clustering such that one of the clusters is centered close to the
true mean $\mu$, which implies the robust mean estimation algorithm. A more complicated procedure is required for the list-decodable case \cite[Section 5.4]{kothari_better_2017} 

\paragraph{Using the SoS tools.}
The above algorithms will be a crucial tool for our algorithm for recovering the FFD components. Before stating our result, let us describe it. Our algorithm assumes a target distribution $\mathcal{M}$ that can be written as
\[
\mathcal M = \sum_{i \in [k]} w_i D_i(\mu_i) + \sum_{i \in [k']} w_i' D_i'(\mu_i')\,.
\]  
The algorithm's inputs are 
\begin{enumerate}
    \item i.i.d. samples drawn from $\mathcal M$ and
    \item a list of predictions for the means $\{\mu_i\}_{i \in [k]}$ which are $\epsilon$-accurate (this list should be understood as the output of the SFD algorithm of \Cref{thm:recover-sfd}).
\end{enumerate}
The goal of the algorithm is to efficiently use this information to estimate the remaining means $\{\mu_i'\}_{i \in [k']}$ of the components $D_1',...,D_{k'}'.$ The idea is to use the list-decodable algorithm of \Cref{thm:list-decode} together with the robust mean estimation algorithm of \Cref{thm:sos-robust-mean-est} in the following manner. First, we will think of the samples from the components $D_1,..,D_k$ as ``corrupted observations'' and since we do not know how $k$ relates to $k'$, we have to use the list-decodable routine to get a list of estimations for the means of the distributions $D_1',...D_{k'}'$\footnote{{In particular, we will use the list-decodable algorithm of \Cref{thm:list-decode} once for any $D_j'$ for $j \in [k']$, given that $n$ is sufficiently large.}}. To do that, we have to assume that each distribution $D_i'$ is certifiably bounded and sub-exponential. Moreover, we have to use the resilience property on $D_1,...,D_k$ in order to ``remove'' these known components using the given ``predictions'' of the input. In total, we get the following general guarantee, which works as long as there is some non-trivial separation between the components we want to estimate.

\begin{theorem}
\label{thm:recover-ffd}
Let $\mathcal M$ be a mixture of \(k+k'\) distributions \(D_1,\dots, D_k,\)\(D_1', \dots, D_{k'}'\) over \(\R^d\), with means \(\mu_1,\dots, \mu_k,\mu_1',\dots, \mu_{k'}' \in \R^d\) and weights \(w_1,\dots, w_k, w_1',\dots, w_{k'}'\) that \(\sum_{j\in [k]} w_j + \sum_{j\in [k']} w_j' = 1\). Assume
    \begin{enumerate}
        \item $D_1,\dots, D_k$ satisfy \((n_{\text{r}},\Delta)\)-resilience, and,
        \item $D_1',...,D'_{k'}$ are \((2t, B)\)-certifiably-bounded and sub-exponential. 
    \end{enumerate}
Let \(\gamma_{\text{F}} = \min_{j\ne j' \in [k']} \| \mu_j' - \mu_{j'}'\|_2\) be the minimum separation between the $\{D_i'\}$ components, \(\gamma_{\text{SF}} = \min_{j\in [k], j'\in [k']} \|\mu_j - \mu_{j'}'\|_2\) be the minimum separation between some element in $\{D_i\}$ and some element in $\{D_i'\}$, \(w_{\min} = \min\{\min_{j\in [k]}w_j,\min_{j\in [k'] }w_j'\}\), and \(c_0, C_0\) be some absolute constants. If for some \(C_{\text{sep}} \ge C_0\), \(\gamma_{\text{F}}\ge C_{\text{sep}}B/w_{\min}^{1/t}\) and \(\gamma_{\text{SF}}\gtrsim C_{\text{sep}}B/w_{\min}^{1/t}+\Delta(c_0C_{\text{sep}}^{-2t}w_{\min})\), then there is an algorithm that given \(\delta\in (0, 1)\), \(n\) i.i.d.\ samples from \(\mathcal{M}\), and a list \(\{\wh\mu_j\}_{j\in [k]}\) such that there is a permutation \(\pi\) on \([k]\) with \[\max_{j\in [k]} \|\mu_j  -\wh\mu_{\pi(j)}\|_2 \le \varepsilon\] for some \(\varepsilon \le \gamma_{\text{SF}}/4 \), outputs a list \(\{ \wh\mu_j' \}_{j\in [k']}\) such that there is a permutation \(\pi'\) on \([k']\) with \[\max_{j\in [k']} \|\mu_j' - \wh\mu_{\pi'(j)}'\|_2 \le O(B C_{\text{sep}}^{1-2t}),\] with probability at least \(1-\delta\), as long as \[n \gtrsim\frac{\log((k+k')/\delta)+ (2d \log(dtk'/\delta))^t+n_r(\frac{\delta}{3k}, c_0 C_{\text{sep}}^{-2t}w_{\min})+d\log(k'/\delta)C_{\text{sep}}^{4t}}{w_{\min}},\] and the running time is \(n^{O(t)}\). 
\end{theorem}

The above guarantee is quite general and can be immediately used to estimate the parameters of FFD components given that we have some estimates for the SFD part (using the sparse Fourier transform) and some non-trivial separation assumptions. Let us comment on the separation. As it is expected we need to impose some non-trivial separation between the components $D_1,',...,D_{k'}'$. This separation reads as 
\[
\gamma_F \geq C_{sep} B/w_{\min}^{1/t}
\]
and, intuitively, this corresponds to a separation of order $\poly(k)$.
Moreover, we have to impose some separability between the components we have estimated (i.e., $D_1,...,D_k)$ and the target components. 
This separation reads as 
\[
\gamma_{SF} \geq C_{sep} B/w_{\min}^{1/t} + \Delta(c_0 C_{sep}^{-2t} w_{\min})
\]
and is needed in order to use resilience (and use the given input predictions); note that we make no assumption on the tail of $D_1,...,D_k$ and hence the separation $\gamma_{SF}$ needs to grow as the tail becomes heavier.

\begin{proof}
The algorithm for recovering the FFD means will be as follows. 
\begin{enumerate}
    \item Run the list-decodable algorithm in \cref{thm:sos-cluster}, which will output sets \(S_1,\dots, S_m \subseteq [n]\).
    \item Remove all the sets \(S_j\) with \(\|\wt\mu_j - \wh\mu_{j'}\| \le \gamma_{\text{SF}}\) for some \(j'\in [k]\) (recall that \(\wt\mu_j\) is the empirical mean among \(\{x_i\}_{i\in S_j}\) and {$\wh \mu_{j'}$ is the given prediction for the mean of some $D_{i}$, $i \in [k]$).}
    \item Merge all \(S_j\) whose empirical means are within \(\gamma_{\text{F}}/2\), and run the robust mean estimation in \cref{thm:sos-robust-mean-est} on each consolidated set to get the estimates \(\{\wh\mu_j'\}_{j\in [k']}\) .
\end{enumerate}

By standard Chernoff bounds and a union bound, one can show that with probability at least \(1-\delta/3\), at least \(0.9 w_{j}n\) points among \(x_1,\dots, x_n\) are sampled from \(D_j\), for each \(j\in [k]\), and at least \(0.9 w_{j}'n\) points are sampled from \(D_j'\), for each \(j\in [k']\), as long as \(n \gtrsim \log((k+k')/\delta)/w_{\min}\). Thus, we can apply \cref{thm:sos-cluster} on each \(D_j'\) with \(\alpha = 0.9 w_j'\), as long as \(n \gtrsim (2d \log(dtk'/\delta))^t/w_{\min}\). As a result, for each \(j \in [k']\), there exists a \(j'\in [m]\) such that \(\|\wt\mu_{j'} - \mu_{j}'\|_2 \le O(B/w_j'^{1/t}) \le O(B/w_{\min}^{1/t})\) . 

Meanwhile, since \(n \gtrsim n_r(\frac{\delta}{3k}, c_0 C_{\text{sep}}^{-2t} w_{\min})/w_{\min}\), with probability at least \(1-\delta/3\), any \(c_0 C_{\text{sep}}^{-2t} w_{\min}\) fraction of the points sampled from \(D_j\) has its empirical mean within distance \(\Delta(c_0 C_{\text{sep}}^{-2t} w_{\min})\) of \(\mu_j\), for each \(j\in [k]\). 

Given \cref{thm:sos-cluster}, we will repeat the proof in \cite[Section 5.5]{kothari_better_2017}, with an extra case for the components {$D_1,...,D_k$ for which we are given accurate predictions}. 

First, we can show that after Step 2 in the above process (i.e., after removing the sets in Step 2), all the {survival sets} \(S_j\) have their empirical mean \(\wt\mu_j\) to be close to \(\mu_{j'}'\) for some \(j'\in [k']\). For a given \(S_j\), since \(|S_j| \ge 0.9w_{\min} n/4\), by the pigeonhole principle, \(S_j\) must either (1) have at least \(w_{j'}w_{\min}  n/5\) points sampled from some \(D_{j'}\), or (2) have at least \(w_{j'}' w_{\min} n/5\) points sampled from some \(D_{j'}'\). 
By \cref{thm:sos-cluster}, the mean of these points is within distance \(O(B/w_{\min}^{1/t})\) of \(\wt\mu_j\).
For the former case, the mean of these points is within distance \(\Delta(w_{\min}/5)\) of \(\mu_{j'}\), and we have \(\|\wt\mu_j - \wh\mu_{\pi(j')}\|_2 \le O(B/w_{\min}^{1/t}) + \Delta(w_{\min}/5) + \varepsilon \le \gamma_{\text{SF}}/2\), which means we must have removed \(S_j\). 
For the latter case, \citet[Section 5.5]{kothari_better_2017} has proved that it will yield \(\|\wt\mu_j - \mu_{j'}'\|_2 \le O(B/w_{\min}^{1/t})\). 

Then, we can show that for each \(S_j\), most of the points in it come from a single component \(D_{j'}'\). Suppose for the sake of contradiction that (1) more than \(\frac14 \alpha w_{\min} w_{j''}n\) points are sampled from \(D_{j''}\), or (2) more than \(\frac14 \alpha  w_{\min} w_{j''}'n\) points are sampled from \(D_{j''}'\), with \(\alpha = c_0C_{\text{sep}}^{-2t}\). For the former case, by \cref{thm:sos-cluster}, the mean of these points is within distance \(O(B/ w_{\min}^{1/t}+B/(\alpha w_{\min}^2)^{1/2t})=O(C_{\text{sep}}B/ w_{\min}^{1/t})\) of \(\wt\mu_j\), and within distance \(\Delta(\alpha  w_{\min}/5)\) of \(\mu_{j''}\). Therefore, we have \(\|\wt\mu_j - \wh\mu_{\pi(j'')}\|_2 \le O(C_{\text{sep}}B/ w_{\min}^{1/t}) + \Delta(\alpha w_{\min} /5)+ \varepsilon  \le \gamma_{\text{SF}}/2\), which is a contradiction as we must have removed \(S_j\) in this case. For the latter case, \citet[Section 5.5]{kothari_better_2017} gives a contradiction. Thus, for each \(S_j\), at most \(\sum_{j''\in [k]} \frac{1}{4}\alpha w_{\min} w_{j''}n + \sum_{j''\in [k'] }\frac{1}{4}\alpha  w_{\min} w_{j''}n = \frac14 \alpha  w_{\min} n \le \alpha |S_j|\) points come from any components other than \(D_{j'}'\). 

Since all the \(S_j\) have means \(\wt \mu_j\) that satisfy \(\|\wt\mu_j - \mu_{j'}'\|_2 \le O(B/ w_{\min}^{1/t})\) for some \(j' \in [k']\), after merging all \(S_j\) whose means are within \(\gamma_{\text{F}}/2 \ge \frac{C_{\text{sep}}}2B/ w_{\min}^{1/t}>\frac{C_0}2B/ w_{\min}^{1/t}\), we will get \(k'\) new sets \(S_1',\dots, S_{k'}'\), such that there is a permutation \(\pi'\) on \([k']\) that all but an \(\alpha\) fraction of the points in \(S_j'\) are sampled from \(D_{\pi'(j)}'\), for all \(j\in [k']\). By \cref{thm:sos-robust-mean-est}, for each \(j\in [k']\), we can robustly estimate the mean of \(D_{\pi'(j)}'\) and get \(\wh\mu_{\pi'(j)}'\) that satisfies \(\|\mu_j' - \wh\mu_{\pi'(j)}'\|_2 \le O(B C_{\text{sep}}^{1-2t}), \) with probability at least \(1-\delta/3k'\), as long as \(n \gtrsim (2d \log (dtk'/\delta))^t/w_{\min}+d\log(k'/\delta)C_{\text{sep}}^{4t}/w_{\min}\). {Here, we used the fact that the outliers' fraction $\alpha$ is of order $C_{\mathrm{sep}}^{-2t}$.}

In summary, the algorithm uses \[n \gtrsim\frac{\log((k+k')/\delta)+ (2d \log(dtk'/\delta))^t+n_r(\frac{\delta}{3k}, c_0 C_{\text{sep}}^{-2t}w_{\min})+d\log(k'/\delta)C_{\text{sep}}^{4t}}{w_{\min}}\] samples, runs in time \(n^{O(t)}\), and outputs \(\{\wh\mu_j\}_{j\in [k']}\) such that there is a permutation \(\pi'\) on \([k']\) that \[\max_{j\in [k']} \|\mu_j' - \wh\mu_{\pi'(j)}'\|_2 \le O(B C_{\text{sep}}^{1-2t}), \] with probability at least \(1-\delta\). 

\end{proof}

\subsubsection{Putting all together}
Combing \cref{thm:recover-sfd} and \cref{thm:recover-ffd}, we immediately get the following result for learning mixtures of SFD and FFD distributions. To get the result, first we use \Cref{thm:recover-sfd} to get a list of predictions for the means of the SFD part and then use this list together with samples from the mixture, to recover the FFD components.

\begin{theorem}
\label{thm:recover-sfd-ffd}
Let $\mathcal M$ be a mixture of \(k+k'\) distributions \(D_1,\dots, D_k,\)\(D_1', \dots, D_{k'}'\) over \(\R^d\), with means \(\mu_1,\dots, \mu_k,\mu_1',\dots, \mu_{k'}' \in \R^d\) and weights \(w_1,\dots, w_k, w_1',\dots, w_{k'}'\) that \(\sum_{j\in [k]} w_j + \sum_{j\in [k']} w_j' = 1\). Assume
    \begin{enumerate}
        \item \(D_1,\dots, D_k\) are $k$ translations of a distribution $D$
        over \(\R^d\) satisfying SFD, that is, there exist constants \(c_1, c_2 \ge 0\) such that  \(\inf_{t:\|t\|_2 \le T}|{\phi_{D}(t)}| \gtrsim d^{-c_1}T^{-c_2}\), 
        \item $D_1',...,D'_{k'}$ satisfy FFD, that is, there exist constants \(c_1', c_2' \ge 0\) such that  \(\sup_{t:\|t\|_2 \ge T}|{\phi_{D_j'}(t)}| \lesssim d^{-c_1'}T^{-c_2'}\) for all \(j\in [k']\) with $c_2' > c_2$,
        \item $D$ satisfies \((n_{\text{r}},\Delta)\)-resilience, and 
        \item $D_1',...,D'_{k'}$ are \((2t, B_t)\)-certifiably-bounded and sub-exponential. 
    \end{enumerate}
Let \(\gamma_{\text{S}} = \min_{j\ne j' \in [k]} \|\mu_j - \mu_{j'} \|_2\), \(\gamma_{\text{F}} = \min_{j\ne j' \in [k']} \| \mu_j' - \mu_{j'}'\|_2\), \(\gamma_{\text{SF}} = \min_{j\in [k], j'\in [k']} \|\mu_j - \mu_{j'}'\|_2\), \(w_{\min} = \min\{\min_{j\in [k]}w_j,\min_{j\in [k'] }w_j'\}\), \(B = \max_{j\in [k]} \|\mu_j\|_2\), and \(c_0, C_0\) be some absolute constant. If for some \(C_{\text{sep}} \ge C_0\), \(\gamma_{\text{F}}\ge C_{\text{sep}}B_t/w_{\min}^{1/t}\) and \(\gamma_{\text{SF}}\gtrsim C_{\text{sep}}B_t/w_{\min}^{1/t}+\Delta(c_0C_{\text{sep}}^{-2t}w_{\min})\), then there is an algorithm that given \(\varepsilon ,\delta \in (0, 1)\) and \(n\) i.i.d.\ samples from \(\mathcal{M}\), outputs two lists \(\{\wh\mu_j\}_{j\in [k]}\) and \(\{\wh\mu_j'\}_{j\in [k']}\) such that there is a permutation \(\pi\) on \([k]\) and a permutation \(\pi'\) on \([k']\) with \[
\max_{j \in [k]}~
\| \mu_{j} - \wh \mu_{\pi(j)} \|_2 \leq \eps\,, \qquad\max_{j \in [k']} \| \mu_{j}' - \wh \mu_{\pi'(j)}' \|_2 \le O(B_tC_{\text{sep}}^{1-2t})
\]
with probability at least $1-\delta$, as long as \begin{align*}
    n \gtrsim {}&\poly_{c_1,c_2,c_1',c_2'}(d, 1/\gamma_{\text{S}},B,1/w_{\min}, 1/\varepsilon )  \log(1/\delta) \\&+ \frac{ (2d \log(dtk'/\delta))^t+n_r(\frac{\delta}{3k}, c_0 C_{\text{sep}}^{-2t}w_{\min})+d\log(k'/\delta)C_{\text{sep}}^{4t}}{w_{\min}}
\end{align*}
and the running time is \(n^{O(t)}\). 
\end{theorem}

{As an immediate corollary one can show that there is an algorithm that learns mixtures of $k$ Laplace components (SFD part) and $k'$ FFD distributions which are (i) $2t$-certifiably-bounded, (ii) sub-exponential, and (iii) whose characteristic function decays faster than the Laplace. 
The separation between the Laplace components is arbitrary $\gamma > 0$, the separation between the FFD components is $\poly(k)$, and the separation between Laplace and FFD is {$\poly(k)$}.
The estimates for the Laplace means can be done in time $\poly(d,k,1/\epsilon,1/\gamma)$, while the remaining means can be estimated in time (roughly) $d^{O(t)}$.}

{In particular, if the FFD part consists of spherical Gaussian distributions, then one can achieve vanishing error on the estimates of the FFD means, independent of the separation. For simplicity, we will assume that the SFD part consists of Laplace distributions and the FFD part consists of Gaussian distributions, both with identity covariance.
\begin{corollary}
\label{cor:learning-mix-laplace-gaussian}
Let \(\mathcal{M}\) be a mixture of \(k\) Laplace distributions \(\lap(\mu_j, I)\) and \(k'\) Gaussian distributions \(\cN(\mu_j', I)\), with means \(\mu_1,\dots, \mu_k,\mu_1',\dots, \mu_k' \in \R^d\) and weights \(w_1,\dots, w_k,w_1', \dots, w_k'\) that \(\sum_{j\in [k]} w_j + \sum_{j\in [k']} w_j'=1\). 
Let \(\gamma_{\text{S}} = \min_{j\ne j' \in [k]} \|\mu_j - \mu_{j'} \|_2\), \(\gamma_{\text{F}} = \min_{j\ne j' \in [k']} \| \mu_j' - \mu_{j'}'\|_2\), \(\gamma_{\text{SF}} = \min_{j\in [k], j'\in [k']} \|\mu_j - \mu_{j'}'\|_2\), \(w_{\min} = \min\{\min_{j\in [k]}w_j,\min_{j\in [k'] }w_j'\}\), and \(B = \max_{j\in [k]} \|\mu_j\|_2\).
Then for any \(\beta > 0\), there is a separation \(\gamma_0 = O(k'^\beta)\), such that if \(\gamma_{\text{F}} \ge \gamma_0\), \(\gamma_{\text{SF}} \ge \gamma_0\), and \(w_{\min} \ge 1/\poly(k')\), then there is an algorithm that given \(\varepsilon ,\delta \in (0, 1)\) and \(n\) i.i.d.\ samples from \(\mathcal{M}\), outputs two lists \(\{\wh\mu_j\}_{j\in [k]}\) and \(\{\wh\mu_j'\}_{j\in [k']}\) such that there is a permutation \(\pi\) on \([k]\) and a permutation \(\pi'\) on \([k']\) with \[
\max_{j \in [k]}~
\| \mu_{j} - \wh \mu_{\pi(j)} \|_2 \leq \eps\,, \qquad\max_{j \in [k']} \| \mu_{j}' - \wh \mu_{\pi'(j)}' \|_2 \le \varepsilon 
\]
with probability at least $1-\delta$, as long as \begin{align*}
    n \gtrsim {}&\poly(d^{1/\beta}, 1/\gamma_{\text{S}},B,k', 1/\varepsilon )  \log(k'/\delta) 
\end{align*}
and the running time is \(n^{O(1/\beta)}\). 
\end{corollary}
This result follows from the facts that \begin{enumerate}
    \item \(\lap(\mu_j, I)\) satisfies SFD with parameter \(c_1 = 0\) and \(c_2 = 2\), i.e., \(\inf_{t:\|t\|_2 \le T} |\phi_{\lap(\mu_j, I)}(t)| \gtrsim T^{-2}\), 
    \item \(\cN(\mu_j', I)\) satisfies FFD with parameter \(c_1'=0\) and any \(c_2' \ge 0\), i.e., \(\sup_{t:\|t\|_2 \ge T} |\phi_{\cN(\mu_j', I)}(t)| \lesssim T^{-c_2'}\) for any \(c_2' \ge 0\),
    \item \(\lap(\mu_j, I)\) is sub-exponential, and thus satisfies \((\frac{1}{\alpha}\poly(d, \log(1/\delta), O(\log(1/\alpha)))\)-resilience, and
    \item \(\cN(\mu_j', I)\) is \((2t, O(\sqrt{t}))\)-certifiably bounded for any \(t\in \Z_{>0}\) (see, e.g., \cite{hopkins2018mixture,kothari2018robust}) and sub-exponential, 
\end{enumerate}
so that one can apply \cref{thm:recover-sfd-ffd} (taking \(t = O(1/\beta)\)) to estimate the SFD means \(\mu_j\) up to \(\varepsilon \) and the FFD means \(\mu_j'\) up to \(1/\poly(k')\). This warm start enables us to apply the local convergence algorithm by \citet{regev_learning_2017} to improve the estimations of the FFD means to \(\varepsilon \) accuracy. We will discuss in \Cref{app:local-convergence} how to adapt their algorithm for our settings with the presence of Laplace components. 
}
\subsection{Moment-Matching for Mixtures Models under SFD}
\label{sec:moments}

In this section, we show that moment-based methods are not useful for {parameter estimation for mixture models} under the SFD condition. To illustrate our moment-matching result, we study mixtures of Laplace distributions. This lower bound is information-theoretic and builds on the 
pigeonhole argument of \citet{regev_learning_2017}. If we apply their argument directly, then we can show the existence of two mixtures of Laplaces with moments that are close in the symmetric injective tensor norm, defined as \[\| T \|_* = \max_{y\in \R^d, \|y\|_2=1} |\langle T, y^{\otimes \ell} \rangle|,\] for order-\(\ell\) tensor \(T \in \R^{d^\ell}\). 

However, we can actually show a stronger result by adapting their proof, that the moments could be close in the Frobenius norm, defined as the entrywise \(\ell_2\) norm of the tensor, \[\fnorm{T} = \left( \sum_{i_1,i_2,\dots, i_\ell}  T_{i_1,i_2,\dots, i_\ell}^2 \right)^{1/2},\] for order-\(\ell\) tensor \(T \in \R^{d^{\ell}}\). 
\begin{theorem}
[Moment Matching]
\label{thm:laplace-mixture-moment-matching}
For \(d = \Theta(\log k)\) and \(R = \Theta(\log k)\), there exist two uniform mixtures of Laplaces \(Y\) and \(\tilde Y\) such that \(\| \E Y^{\otimes r} - \E \tilde Y^{\otimes r} \|_{\mathrm{F}} \le k^{-\Omega(\log\log k)} \) for all \(r=1,2,\dots, R\), while their parameter distance is at least \(\Omega(\sqrt{\log k})\). 
\end{theorem}

We proceed with the proof.  Let us compute the moments of a single Laplace first. 

\begin{lemma}
Suppose \(X \sim \lap(\mu, I_d)\), then \[\E X^{\otimes r} = \sum_{\substack{0\le s \le r \\ 2 \mid s}} \frac{r!}{(r-s)!}\left( \frac 1 {\sqrt{2}} \right)^s \sym\left( \mu^{\otimes (r-s)}\otimes I_d^{\otimes \frac s 2} \right),\] where \(\sym T\) is the symmetrization of tensor \(T\), i.e., \(\sym T = \frac1{r!} \sum_{\sigma\in S_r} T^\sigma\) and \((T^\sigma)_{i_1,i_2,\dots, i_r} =  T_{i_{\sigma_1},i_{\sigma_2},\dots, i_{\sigma_r}}\). 
\end{lemma}

\begin{proof}
The idea is to expand the characteristic function as Taylor series and compare the coefficients. First, \begin{align*}
    \E \exp \left( i \langle t, X \rangle \right) = \sum_{r\ge 0} \frac{\E(i \langle t, X \rangle)^r}{r!} = \sum_{r\ge 0} \frac{i^r}{r!} \langle t^{\otimes r}, \E X^{\otimes r} \rangle. 
\end{align*}
Meanwhile, for \(X\sim \lap(\mu, I_d)\), \begin{align*}
    \E \exp \left( i \langle t, X \rangle \right) &= \frac{\exp (i \langle t, \mu \rangle)}{1 + \frac12 \|t\|_2^2} = \left( \sum_{k\ge 0} \frac{(i \langle t,\mu \rangle)^k}{k!} \right) \left( \sum_{k\ge0} \left( -\frac12 \langle t, t \rangle \right)^k \right) \\
    &= \left( \sum_{k\ge 0} \frac{i^k}{k!} \langle t^{\otimes k}, \mu^{\otimes k} \rangle  \right) \left( \sum_{k\ge0} \left( -\frac12 \right)^k\langle t^{\otimes k}, t^{\otimes k} \rangle  \right) \\
    &= \sum_{k,\ell \ge 0} \frac{i^k}{k!} \left( -\frac12 \right)^\ell \langle t^{\otimes (k+\ell)}, \mu^{\otimes k}\otimes t^{\otimes \ell} \rangle \\
    &= \sum_{k,\ell \ge 0} \frac{i^k}{k!} \left( -\frac12 \right)^\ell \langle t^{\otimes (k+2\ell)}, \mu^{\otimes k}\otimes I_d^{\otimes \ell} \rangle \\
    &= \sum_{r\ge 0} \sum_{\substack{0\le s \le r \\ 2 \mid s}} \frac{i^{r-s}}{(r-s)!} \left( -\frac12\right)^{s/2} \langle t^{\otimes r}, \mu^{\otimes(r-s)}\otimes I_d^{\otimes \frac s2} \rangle. 
\end{align*}
Thus, we have \begin{align*}
    \langle t^{\otimes r}, \E X^{\otimes r} \rangle = \left\langle t^{\otimes r}, \sum_{\substack{0\le s \le r \\ 2 \mid s}} \frac{r!}{(r-s)!} \left(\frac1{\sqrt2}\right)^{s}  \mu^{\otimes(r-s)}\otimes I_d^{\otimes \frac s2} \right\rangle. 
\end{align*}
Since \(\E X^{\otimes r}\) is symmetric, \[\E X^{\otimes r} = \sum_{\substack{0\le s \le r \\ 2 \mid s}} \frac{r!}{(r-s)!}\left( \frac 1 {\sqrt{2}} \right)^s \sym\left( \mu^{\otimes (r-s)}\otimes I_d^{\otimes \frac s 2} \right).\]
\end{proof}

We will also need the following facts for the proof. 

\begin{fact}
\label{fact:sym-norm-ineq}
For order-\(k\) tensor \(T\), \(\left\| \sym T \right\|_{\mathrm{F}} \le \|T\|_{\mathrm{F}}\). 
\end{fact}
\begin{proof}
First, note that for \(\sigma \in S_k\) \begin{align*}
    \fnorm{T^\sigma}^2 &= \sum_{i_1,\dots, i_k} (T^\sigma)_{i_1,\dots, i_k}^2 = \sum_{i_1,\dots, i_k} T_{i_{\sigma_1},\dots, i_{\sigma_k}}^2 = \sum_{i_1,\dots, i_k} T_{i_{1},\dots, i_{k}}^2 = \fnorm{T}. 
\end{align*}
Then by the triangle inequality, \begin{align*}
\left\| \sym T \right\|_{\mathrm{F}} &= \fnorm{\frac{1}{k!} \sum_{\sigma\in S_k} T^{\sigma}} \le \frac{1}{k!} \sum_{\sigma\in S_k} \fnorm{T^\sigma} = \fnorm{T}. 
\end{align*}
\end{proof}

\begin{fact}
\label{fact:tensor-id-norm}
For order-\(k\) tensor \(T \in \R^{d^k}\), \(\| T \otimes I_d^{\otimes \ell} \|_F = d^{\ell/2}\|T\|_F\).
\end{fact}
\begin{proof}
By definition, \begin{align*}
    \| T \otimes I_d^{\otimes \ell} \|_F^2 &= \sum_{\substack{i_1,\dots, i_k \\ j_1,j_2,\dots, j_{2\ell-1},j_{2\ell}}}(T_{i_1,\dots, i_k} \mathbf{1}[j_1=j_2]\cdots \mathbf{1}[j_{2\ell-1}=j_{2\ell}])^2 \\
    &=\sum_{{i_1,\dots, i_k}}T_{i_1,\dots, i_k}^2 \sum_{j_1}  \sum_{j_3} \cdots \sum_{j_{2\ell-1}} 1 \\
    &= \|T\|_F^2 \cdot d^{\ell}.
\end{align*}
Therefore, \(\| T \otimes I_d^{\otimes \ell} \|_F = d^{\ell/2}\|T\|_F\). 
\end{proof}

We now use the following lemma, which roughly speaking guarantees that if $\mathcal F$ is a large enough collection of sets $\{\mu_1,...,\mu_k\}$, then there exist two sets in \(\mathcal{F}\) such that their tensor powers match. 

\begin{lemma}[{\cite[Lemma~3.6]{regev_learning_2017}}]
\label{lem:dirac-moment-matching-rv17}
Consider a collection \(\mathcal{F}\) of sets of vectors \(\{\mu_j\}_{j\in [k]}\), where \(\mu_j\in \R^d\) satisfies \(\|\mu_j\|_2 \le \sqrt{d}\) for all \(j\in [k]\). Then for any \(R \ge d\), if \(|\mathcal{F}| > \frac 1\eta \exp (\frac52(2eR/d)^d R \log(5d))\), it holds that for at least \((1-\eta)\) fraction of the sets \(\{\mu_1,\dots, \mu_k\}\in\mathcal{F}\), there is another  \(\{\tilde\mu_1,\dots, \tilde\mu_k\}\in\mathcal{F}\) satisfying that for \(r=1,2,\dots, R\), \[\left\| \frac1k \sum_{j=1}^k \mu_j^{\otimes r} - \frac1k \sum_{j=1}^k \tilde{\mu}_j^{\otimes r} \right\|_{\mathrm{F}} \le (d+1)^{-2R}. \]
\end{lemma}
\begin{remark}
The original proof in \citet{regev_learning_2017} showed the tensor powers match in the symmetric injective tensor norm. But the same proof works for the Frobenius norm as well. 
\end{remark}

We will apply the above lemma which holds for arbitrary collections of vectors to the special case where these vectors are the means of a mixture of Laplaces. 
\begin{lemma}
\label{lem:lap-moment-matching}
Under the same notation of \cref{lem:dirac-moment-matching-rv17}, let $\{\mu_1,...,\mu_k\}$ and $\{\wt \mu_1,..., \wt \mu_k\}$
be as in \Cref{lem:dirac-moment-matching-rv17}, i.e.,
for \(r=1,2,\dots, R\), \[\left\| \frac1k \sum_{j=1}^k \mu_j^{\otimes r} - \frac1k \sum_{j=1}^k \tilde{\mu}_j^{\otimes r} \right\|_{\mathrm{F}} \le (d+1)^{-2R}. \]
Let \(Y\) be the uniform mixture of \(k\) Laplaces \(\lap(\mu_j, I_d)\), \(j\in [k]\), and \(\tilde Y\) be the uniform mixture of \(k\) Laplaces \(\lap(\tilde \mu_j, I_d)\), \(j\in [k]\), then for \(r = 1,2,\dots, R\), \(\|\E Y^{\otimes r} - \E \tilde Y^{\otimes r}\|_{\mathrm{F}} \le \sqrt{R} \left( \frac{R}{\sqrt 2 e d^{7/4}} \right)^R\). 
\end{lemma}
\begin{proof}
Compute {\allowdisplaybreaks\begin{align*}
\|\E Y^{\otimes r} - \E \tilde Y^{\otimes r}\|_{\mathrm{F}} &= \left\| \frac 1 k \sum_{j\in [k]} \sum_{\substack{0\le s \le r \\ 2 \mid s}} \frac{r!}{(r-s)!}\left( \frac 1 {\sqrt{2}} \right)^s \sym\left( \mu_j^{\otimes (r-s)}\otimes I_d^{\otimes \frac s 2} \right) \right.\\
&\left.\quad - \frac 1 k \sum_{j\in [k]} \sum_{\substack{0\le s \le r \\ 2 \mid s}} \frac{r!}{(r-s)!}\left( \frac 1 {\sqrt{2}} \right)^s \sym\left( \tilde \mu_j^{\otimes (r-s)}\otimes I_d^{\otimes \frac s 2} \right) \right\|_{\mathrm{F}} \\
&= \left\| \sum_{\substack{0\le s \le r \\ 2 \mid s}} \frac{r!}{(r-s)!}\left( \frac 1 {\sqrt{2}} \right)^s \sym\left(\frac 1 k \sum_{j\in [k]} \left(\mu_j^{\otimes (r-s)}-\tilde\mu_j^{\otimes (r-s)}\right)\otimes I_d^{\otimes \frac s 2} \right) \right\|_{\mathrm{F}} \\
&\le \sum_{\substack{0\le s \le r \\ 2 \mid s}} \frac{r!}{(r-s)!}\left( \frac 1 {\sqrt{2}} \right)^s d^{s/4}\left\|\frac 1 k \sum_{j\in [k]} \left(\mu_j^{\otimes (r-s)}-\tilde\mu_j^{\otimes (r-s)}\right) \right\|_{\mathrm{F}} \\
&\le (d+1)^{-2R} \sum_{0\le s \le r} \frac{r!}{s!} \left( \frac{d^{1/4}}{\sqrt 2} \right)^{r-s} \\
&\lesssim (d+1)^{-2R} r! \left( \frac{d^{1/4}}{\sqrt{2}} \right)^r \\
&\lesssim \sqrt{R} \left( \frac{R}{\sqrt 2 e d^{7/4}} \right)^R. 
\end{align*}    }
\end{proof}

We also need the following lemma to lower bound the parameter distance of the mixtures. 

\begin{lemma}[{\cite[Claim~3.4]{regev_learning_2017}}]
\label{lem:separation}
Let \(x_1,\dots,x_N\) be chosen independently and uniformly from the ball of radius \(r\) in \(\R^d\). Then for any \(0 < \gamma < 1\), with probability at least \(1-N^2\gamma^d\), we have that for all \(i\ne j\), \(\|x_i-x_j \|_2 \ge \gamma r\).
\end{lemma}

\begin{proof}[Proof of \cref{thm:laplace-mixture-moment-matching}]
As in \cite{regev_learning_2017}, we first choose \(N\) points \(x_1,x_2,\dots, x_N\) independently and uniformly at random from the ball of radius \(\sqrt{d}\) in \(\R^d\). Then let the collection \(\mathcal{F}\) be all the sets of \(k\) distinct points, so \(|\mathcal{F}| = \binom{N}{k}\). 
There exists constants \(c_1,c_2<c_3,\gamma<1\), such that when \(N = c_1 k\), \(d = c_2 \log k\), and \(R = c_3 \log k\), it holds that \(\binom{N}{k} \ge (\frac N k)^k \ge \frac 12 \exp (\frac52(2eR/d)^d R \log(5d))\) and \(N^2 \gamma^d < 1\). Thus, by \cref{lem:lap-moment-matching}, there exists two uniform mixtures of Laplaces \(Y\) and \(\tilde Y\) such that for \(r = 1,2,\dots, R\), \begin{align*}
\|\E Y^{\otimes r} - \E \tilde Y^{\otimes r}\|_{\mathrm{F}} \le \sqrt{R} \left( \frac{R}{\sqrt 2 e d^{7/4}} \right)^R \le k^{-\Omega(\log\log k)}. 
\end{align*}
Meanwhile, since \(Y\) and \(\tilde Y\) are different, there exists a component \(\mu_j\) in \(Y\) that is not in \(\tilde Y\). By \cref{lem:separation}, for all \(\tilde j \in [k]\), \(\|\mu_j - \tilde \mu_{\tilde j}\|_2 \ge \gamma r =\Omega(\sqrt d) = \Omega(\sqrt{\log k}) \).
\end{proof}

\section{Application II: Estimation with Noise-Oblivious Adversaries}
\label{sec:noise-obl}

In this section, we provide our consistent estimator for high-dimensional mean estimation for general distributions \(D\) in the noise-oblivious model.
Recall that, under the setting of \Cref{def:subset-of-signals}, \(D(\mu)\) denotes the translation of distribution \(D\) that has mean \(\mu\), and the input of the algorithm can be viewed as \(n\) independent random variables, with a \((1-\alpha)\) fraction being sampled from \(D(\mu)\), and the rest \(\alpha\) fraction being sampled from \(D(z_k)\), where \(z_k\) is chosen by the adversary, for \(k=1,2,\dots, \alpha n\).

\begin{theorem}
\label{thm:noise-obi-mean-estimation-full}
Consider the \(d\)-dimensional mean estimation problem in the setting of \Cref{def:subset-of-signals} with distribution $D(\mu)$ with  true mean $\mu \in \R^d$ such that \(\|\mu\|_2 \le B\) for some \(B > 0\).
Define $R(T) := \sup_{t\in B_T^d(0)}|\phi_{D}(t)|^{-1}$ for any $T>0$. 
If {the corruption rate} \(\alpha \le \alpha_0\) for some absolute constant \(\alpha_0>0\), 
then there is an algorithm that gets as input accuracy $\eps, \delta \in (0,1)$ and computes an estimate $\wh \mu \in \R^d$ such that $\| \mu - \wh \mu\|_2 < \eps$ with probability $1-\delta$. The algorithm uses \(O\left(R(Cd^3 B/\varepsilon)^2 (\log d+ \log\log( B/\varepsilon))\log(1/\delta)\right)\) i.i.d.\ samples and runs in \[ 
\wt O\left(\left( R(Cd^3 B/\varepsilon)^2+\log(B/\varepsilon)\right)d\log(B/\varepsilon)\log(1/\delta) + d \log(1/\delta)^2\right)\]
time.
\end{theorem}

\begin{proof}
For a sample \(Y_j\) generated by one of the distributions, say \(D(z)\), we have for \(t\in \R^d\)
\begin{equation}
\label{eq:cf-one-sample}
\E[e^{i \langle t, Y_j \rangle}] = \phi_{D(z)} (t) = e^{i \langle t, z \rangle}\phi_{D}(t). 
\end{equation}
Given a set of samples $\{Y_j\}_{j \in [n]}$ of size $n$ generated according to \Cref{def:subset-of-signals}, averaging \cref{eq:cf-one-sample} over \(j=1,2,\dots, n\), we have 
\[
\frac 1 n\sum_{j=1}^n \E[e^{i\langle t, Y_j \rangle}] = (1-\alpha) e^{i\langle t, \mu \rangle} \phi_{D}(t) + \frac 1 n \sum_{k=1}^{\alpha n} e^{i\langle t, z_k \rangle} \phi_{D}(t).
\]
Again, the idea is to estimate 
\[
\phi_{D}(t)^{-1}\cdot \frac 1 n\sum_{j=1}^n \E[e^{i\langle t, Y_j \rangle}] = (1-\alpha) e^{i\langle t, \mu \rangle} + \frac 1 n \sum_{k=1}^{\alpha n} e^{i\langle t, z_k \rangle },
\]
and apply the sparse Fourier transform on the estimation to recover the true mean \(\mu\). Here, we can view the noise as being not only from the estimation error, but also from the contamination, so that the true signal is \(1\)-sparse, i.e., \((1-\alpha) e^{i\langle t, \mu \rangle}\). 

Following the notation in \Cref{thm:d-sft}, let the true signal be \(x^\star(t) = (1-\alpha)e^{i \langle t, \mu \rangle}\). The observed signal is \[x(t) = \phi_{D}(t)^{-1}\cdot \frac 1 n\sum_{j=1}^n e^{i\langle t, Y_j \rangle}, \] which is from the empirical average of the characteristic function. Also, let \(g(t) = x(t) - x^\star(t)\) be the noise. Then \begin{align*}
    g(t) &= x(t) - x^\star(t) \\
    &= \phi_{D}(t)^{-1}\cdot \frac 1 n\sum_{j=1}^n e^{i\langle t, Y_j \rangle} - (1-\alpha) e^{i\langle t, \mu \rangle} \\
&= \underbrace{\phi_{D}(t)^{-1}\left(\frac 1 n\sum_{j=1}^n e^{i\langle t, Y_j \rangle}- \frac 1 n\sum_{j=1}^n \E[e^{i\langle t, Y_j \rangle}]  \right) }_{ g_1(t)} + \underbrace{\frac 1 n \sum_{k=1}^{ \alpha n} e^{i  \langle t, z_k \rangle}}_{g_2(t)}.
\end{align*}

For \(g_1(t)\), we can use concentration inequalities to bound the difference between the empirical average and the expectation. 
Since \(|e^{i  \langle t, Y_j \rangle }| = 1\) is bounded, by Hoeffding's inequality, \begin{align*}
    \Pr \left[ \left| \frac 1 n\sum_{j=1}^n e^{i\langle t, Y_j \rangle}- \frac 1 n\sum_{j=1}^n \E[e^{i\langle t, Y_j \rangle}] \right| \ge s \right] \le e^{-\Omega(n s^2)}
\end{align*}
{for any fixed time $t\in \R^d$.}
Suppose the algorithm in \Cref{thm:d-sft} queries the signal \(x(t)\) on \(t=\{t_1,t_2,\dots, t_N\}\). 
For such $t$, with probability at least \(1-e^{-\Omega(ns^2)}\),
\begin{align*}
|g_1(t)|&=|\phi_{D}(t)|^{-1}\left|\left(\frac 1 n\sum_{j=1}^n e^{i\langle t, Y_j \rangle}- \frac 1 n\sum_{j=1}^n \E[e^{i\langle t, Y_j \rangle}]  \right) \right| \le  s\cdot R(T). 
\end{align*}
By the union bound, with probability at least \(1-N\cdot e^{-\Omega(ns^2)}\), \(|g_1(t_j)| \le s\cdot R(T)\) for all \(j\in [N]\).
Meanwhile, for any \(t\in \R^d\), \begin{align*}
    |g_2(t)| &\le \frac{1}{n} \sum_{k=1}^{\alpha n} \left| e^{i \langle t, z_k \rangle} \right| = \frac{\alpha n}{n} = \alpha. 
\end{align*}

{Now we are ready to apply \Cref{thm:d-sft}.}
Set \(s = \Theta \left( \sqrt{\frac{\log (N/\delta_1)}{n}} \right)\). Then the probability of success is \(1 - N \cdot e^{-\Omega(ns^2)} = 1-\delta_1\) for some failure probability $\delta_1$. 

We can apply \Cref{thm:d-sft}, by setting \(k=1\), \(\gamma = O(1)\), $\delta = \delta_2$ for some failure probability $\delta_2$, \(\theta\) to be some small enough constant, and \(n=cR(T)^2  {\log(N/\delta_1)}\) for large enough constant \(c>0\).
Thus, the number of samples needed by the algorithm in \Cref{thm:d-sft} is \(N = O(d\log(B T)\log (d)\log(1/\delta_2))\), and the noise level
\begin{align*}
\mathcal{N}^2 &= \max_{j\in [N]} |g(t_j)|^2 + \theta (1-\alpha)^2 \\
&=\max_{j\in [N]} 2|g_1(t_j)|^2 +2 |g_2(t_j)|^2+\theta (1-\alpha)^2 \\
&\le 2s^2 R(T)^2 +2 \alpha^2+\theta (1-\alpha)^2 \\
&\le 2/c + 2\alpha^2 + \theta(1-\alpha)^2. 
\end{align*}
For small enough constant \(\alpha \), we have \(|w_1| = 1- \alpha= \Omega(\mathcal N)\), and thus the algorithm in \Cref{thm:d-sft} outputs \(\hat\mu\) such that \(\|\mu - \hat\mu\|_2 \le O \left( \frac{d^3 B}{T} \right)\) with probability \(1-\delta_2\) in time \(O(d\log (BT)^2\log(d)\log(1/\delta_2)+d \log(1/\delta_2)^2)\).
Since each estimation of \(x(t_j)\) requires \(O(n)\) time to compute, \(j = 1, 2, \dots, N\), the overall running time is \begin{align*}
    &O(d\log (BT)^2\log(d)\log(1/\delta_2)+d \log(1/\delta_2)^2 + nN) \\
    &\le \wt O (d \log(BT)^2 \log(1/\delta_2) + d \log(1/\delta_2)^2 + R(T)^2 d \log(BT)\log(1/\delta_1)\log(1/\delta_2) ).
\end{align*} 
Take \(\delta_1\) and \(\delta_2\) to be some small enough constant, then the algorithm uses \(n = O(R(T)^2 (\log d+ \log\log(BT)))\) samples and \(\wt O(d\log(BT)^2 + R(T)^2d\log(BT))\) time, and succeeds with constant probability. 

To make the error \(\|\mu-\wh\mu\|_2 \le \varepsilon \), we will need \(T = Cd^3 B/\varepsilon \) for some constant \(C > 0\), and thus the sample complexity is \[n = O\left(R(Cd^3 B/\varepsilon)^2 \left(\log d+ \log\log( B/\varepsilon)\right)\right)\] and the time complexity is \[\wt O\left(\left( R(Cd^3 B/\varepsilon)^2+\log(B/\varepsilon)\right)d\log(B/\varepsilon)\right).  \]

To boost the success probability from constant to \(1 - \delta\), one can apply \cref{lem:boosting}, so that \(\|\mu-\wh\mu\|_2 \le \varepsilon \) with probability \(1 - \delta\), using \[O\left(R(Cd^3 B/\varepsilon)^2 (\log d+ \log\log( B/\varepsilon))\log(1/\delta)\right)\] samples and \[\wt O\left(\left( R(Cd^3 B/\varepsilon)^2+\log(B/\varepsilon)\right)d\log(B/\varepsilon)\log(1/\delta) + d \log(1/\delta)^2\right)\] time. 
\end{proof}

Moreover, if we posits that \(D\) satisfies some general assumptions (e.g., bounded covariance), then the dependency on \(B\) can be removed by first roughly estimating the mean (e.g., up to \(O(\sqrt{\alpha})\)) and then running our Fourier-based algorithm on the samples subtracted by the estimate. 
\begin{corollary}
\label{cor:noise-obl-mean-estimation-after-median}
Under the same notation in \Cref{thm:noise-obi-mean-estimation-full}, if \(\alpha \le \alpha_0\) for some absolute constant \(\alpha_0>0\), and \(D\) has covariance matrix \(\Sigma \preceq \sigma^2 I\) for some constant \(\sigma\), then the algorithm uses \[\wt O\left(\left(R(Cd^3 /\varepsilon)^2 (\log d+ \log\log( 1/\varepsilon)) + d\right)\log(1/\delta)\right)\] i.i.d.\ samples and runs in \[\wt O\left(\left(\left( R(Cd^3 /\varepsilon)^2+\log(1/\varepsilon)\right)d\log(1/\varepsilon) + d^2\right)\log(1/\delta) + d \log(1/\delta)^2\right)\] time. 
\end{corollary}
\begin{proof}
\citet[Theorem 1.3]{cheng-robust-2019} gave a robust mean estimation algorithm for distributions with bounded covariance, which outputs \(\wt\mu\) that \(\|\mu-\wt\mu\|_2 \le O(\sigma \sqrt{\alpha})\) with constant probability using \(\wt O(d/\alpha)\) samples and \(\wt O(d^2/\poly(\alpha))\) time, if \(\alpha \le 1/4\). 
Note that if \(\alpha < 1/4\), we can set \(\alpha = 1/4\) by viewing some of the inliers being picked by the adversary. 
Therefore, there is an algorithm that outputs \(\wt\mu\) that \(\|\mu-\wt\mu\|_2 \le O(1)\) with constant probability using \(\wt O(d)\) samples and \(\wt O(d^2)\) time. 
Subtracting \(\wt\mu\) from all the sample, we will have the true mean be bounded by \(O(1)\) and run (one round of) the algorithm in \Cref{thm:noise-obi-mean-estimation-full} with \(B = O(1)\). Similarly, we can repeat the whole process \(O(\log(1/\delta))\) times to boost the success probability from constant to \(1 - \delta\). 
\end{proof}

\begin{corollary}
\label{cor:noise-obl-mean-estimation-Gaussian}
Under the same notation in \Cref{thm:noise-obi-mean-estimation-full}, if \(\alpha \le \alpha_0\) for some absolute constant \(\alpha_0>0\), and \(D\) is the standard Gaussian distribution, then the algorithm uses \(2^{O(d/ \varepsilon ^2)}\log(1/\delta)\) samples and runs in \(2^{O(d/ \varepsilon^2 )}\log(1/\delta)\) time. 
\end{corollary}
\begin{proof}
It suffices to estimate each coordinate of \(\mu\) up to \(\varepsilon / \sqrt{d}\) to get \(\varepsilon \) error in \(\ell_2\) distance. 
For standard Gaussian distribution, the marginal distribution on each coordinate is a one-dimensional standard Gaussian, with characteristic function \(\phi_{\cN(0, 1)}(t) = e^{-t^2/2}\). 
Thus, \(R(T) = \sup_{t\in B_T^1(0)} |\phi_{\cN(0, 1)}(t)|^{-1} = e^{T^2/2}\). 
To estimate one coordinate of \(\mu\) up to \(\varepsilon /\sqrt{d}\) with probability \(1-\delta/d\), by \Cref{cor:noise-obl-mean-estimation-after-median}, the sample complexity is \begin{align*}
\wt O\left(R(C\sqrt{d}/\varepsilon)^2  \log\log( 1/\varepsilon)\log(d/\delta)\right) = 2^{O(d/ \varepsilon ^2)} \log(d/\delta),
\end{align*}
and the time complexity is \begin{align*}
\wt O\left(\left( R(C\sqrt{d} /\varepsilon)^2+\log(1/\varepsilon)\right)\log(1/\varepsilon) \log(d/\delta) \right) = 2^{O(d/ \varepsilon ^2)} \log (d/\delta). 
\end{align*}
Note that the \(\log(1/\delta)^2\) term in the time complexity in \cref{cor:noise-obl-mean-estimation-after-median} is not needed, as when \(d=1\), one can simply take the median during boosting. 

By a union bound, we will have the estimate \(\wh\mu\) satisfies \(\|\mu-\wh\mu\|_2 \le \sqrt{d}\cdot \varepsilon /\sqrt{d} = \varepsilon \) with probability \(1-\delta\), using \(2^{O(d/ \varepsilon ^2)}\log(d/\delta) = 2^{O(d/ \varepsilon ^2)}\log(1/\delta)\) samples and \(d 2^{O(d/ \varepsilon ^2)}\log(d/\delta) = 2^{O(d/ \varepsilon ^2)}\log(1/\delta)\) time. 
\end{proof}

\begin{corollary}
\label{cor:noise-obl-mean-estimation-Laplace}
Under the same notation in \Cref{thm:noise-obi-mean-estimation-full}, if \(\alpha \le \alpha_0\) for some absolute constant \(\alpha_0>0\), and \(D\) is the Laplace distribution with variance \(1\), then the algorithm uses \(\widetilde O(d^2\log(1/\delta) / \varepsilon ^4)\) samples and runs in \(\widetilde O(d^3\log(1/\delta) / \varepsilon ^4)\) time. 
\end{corollary}
\begin{proof}
Since for multivariate Laplace distribution, the marginal distribution on each coordinate is a one-dimensional Laplace distribution, the analysis is analogous to that of \cref{cor:noise-obl-mean-estimation-Gaussian}. 
However, the characteristic function \(\phi_{\lap(0, 1)}(t) =\frac{1}{1 + t^2/2}\). Thus, \(R(T) = \sup_{t\in B^1_T(0)} |\phi_{\lap(0, 1)}(t)|^{-1} = O(T^2)\). For estimating one coordinate, the sample complexity is \begin{align*}
\wt O\left(R(C\sqrt{d}/\varepsilon)^2  \log\log( 1/\varepsilon)\log(d/\delta)\right) = \wt O(d^2 \log(1/\delta)/ \varepsilon ^4),
\end{align*}
and the time complexity is \begin{align*}
\wt O\left(\left( R(C\sqrt{d} /\varepsilon)^2+\log(1/\varepsilon)\right)\log(1/\varepsilon) \log(d/\delta) \right) = \wt O(d^2 \log(1/\delta)/ \varepsilon ^4). 
\end{align*}
Again, in total the algorithm uses \(\wt O(d^2 \log(1/\delta)/ \varepsilon ^4)\) samples and \(\wt O(d^3 \log(1/\delta)/ \varepsilon ^4)\) time. 
\end{proof}

\renewcommand*{\bibfont}{\small}

\printbibliography

\appendix
\section{Boosting for Mixture Models: Proof of \cref{lem:boosting}}
\label[appendix]{app:boosting}
\begin{proof}[Proof of \Cref{lem:boosting}]
We run $R =O(\log(1/\delta))$ independent copies of $A$ with input accuracy $\epsilon' = \min\{\varepsilon /3,\gamma/16\},$ and greedily perform a clustering algorithm on the set of all \(kR\) pairs of weights and points, denoted by \(M\). Let \(M_w\) and \(M_\mu\) denote the sets of the first (weight) and the second (points) element of the pairs in \(M\), respectively. The boosting algorithm is shown in \cref{alg:boosting}. 

\begin{algorithm}[!ht]
\centering
\begin{algorithmic}[1]
\Require Algorithm \(A(\varepsilon ', \varepsilon _w)\), target accuracy \(\varepsilon \) and confidence \(\delta\). 
\Ensure Estimation $\{(\wh w_j, \wh \mu_j)\}_{j \in [k]}. $
\State \(\varepsilon ' \gets \min\{\varepsilon /3, \gamma/16\}\). 
\State \(M \gets \varnothing\). 
\For{\(\ell \gets 1,\dots, R\)} 
\State \(\{(\wh w_j^{(\ell)}, \wh \mu_j^{(\ell)}) \} \gets A(\varepsilon ', \varepsilon _w)\).
\If{\(\min_{i\ne j} \|\wh\mu_i^{(\ell)} - \wh\mu_j^{(\ell)}\|_2 > \gamma/2\)}
\State \(M \gets M \cup \{(\wh w_j^{(\ell)}, \wh \mu_j^{(\ell)}) \}_{j\in [k]}\).  \label{alg:boost:line:Madd}
\EndIf
\EndFor
\For{\(j \gets 1,\dots, k\)} \label{alg:boost:line:for}
\State Choose \(\wh\mu_j \in M_\mu\) such that \(|M_\mu \cap B^d_{2 \varepsilon' }(\wh\mu_j)| \ge \frac{3}{5}R\). \Comment{\(M_\mu := \{\wh \mu:(\wh w, \wh \mu)\in M\}\)} \label{alg:boost:line:choose-mu}
\State \(\wh w_j \gets \mathrm{median}\{\wh w: (\wh w, \wh \mu) \in M, \|\wh\mu_j - \wh\mu\|_2 \le 4 \varepsilon'  \}\). \label{alg:boost:line:choose-w}
\State Remove from \(M\) the subset \(\{(\wh w, \wh\mu): \|\wh\mu_j - \wh \mu\|_2 \le 6 \varepsilon ' \} \). \label{alg:boost:line:remove}
\EndFor \label{alg:boost:line:endfor}
\State \Return \(\{(\wh w_j, \wh\mu_j)\}_{j\in [k]}\). 
\end{algorithmic}
    \caption{Boosting}
    \label{alg:boosting}
\end{algorithm}

We will say that round \(\ell\) is ``good'' if there is a permutation \(\pi\) such that \(\max_{j\in [k]} \|\mu_j - \wh\mu_{\pi(j)}^{(\ell)} \|_2 \le \varepsilon ' \) and \(\max_{j\in [k]} |w_j - \wh w_{\pi(j)}| \le \varepsilon _w\), and we will call it ``bad'' otherwise. Then \(\Pr[\ell \text{ is good}] \ge 2/3\) {by the success probability of $A$}. 
Let \(S = \sum_{\ell=1}^R \mathbf{1}[\ell \text{ is good}]\). 
Since we are running independent copies of \(A\), by Hoeffding's inequality, \(\Pr[S \le \frac{3}{5}R] \le \exp (-2 (\frac{2}{3}R - \frac{3}{5}R)^2/R) = \exp(- \frac{2}{225}R)\). 
Thus, choosing \(R = \frac{225}{2}\log(1/\delta)\), we have \(\Pr[S \ge  \frac{3}{5}R] \ge 1-\delta\). Suppose this happens. We will use the following two lemmas.

\begin{lemma}
\label{lem:boost-close}
Suppose \(S \ge \frac{3}{5}R\). If \(\wh\mu \in M_\mu\) satisfies \(\|\wh\mu - \mu_j\|_2 \le \varepsilon '\) for some \(j\in [k]\), then \(|M_\mu\cap B_{2 \varepsilon' }^d(\wh\mu)| \ge \frac{3}{5}R\). 
\end{lemma}
\begin{proof}
For each good \(\ell\), we have \(\max_{j\in [k]} \|\mu_j - \wh\mu_{\pi(j)}^{(\ell)} \|_2 \le \varepsilon ' \) for some permutation \(\pi\), and thus for \(i\ne j\in [k]\), \begin{align*}
\left\| \wh\mu_{\pi(i)}^{(\ell)} - \wh\mu_{\pi(j)}^{(\ell)} \right\|_2 &\ge \left\| \mu_i - \mu_j \right\|_2 - \left\| \mu_i- \wh\mu_{\pi(i)}^{(\ell)} \right\|_2 - \left\| \mu_j- \wh\mu_{\pi(j)}^{(\ell)} \right\|_2 \\
&\ge \gamma - 2 \varepsilon ' > \gamma/2. 
\end{align*}
Therefore, \(\{(\wh w_j^{(\ell)}, \wh\mu_j^{(\ell)})\}_{j\in [k]}\) will be added to the set \(M\) in line~\ref*{alg:boost:line:Madd}. 
Since \(S \ge \frac{3}{5}R\), for each true \(\mu_j\), \(j\in [k]\), \(|M_\mu \cap B^d _{\varepsilon' } (\mu_j)| \ge \frac{3}{5}R\). And thus for each \(j\in [k]\), if \(\|\wh\mu - \mu_j\|_2 \le \varepsilon '\), then \[\left| M_\mu\cap B_{2 \varepsilon' }^d(\wh\mu) \right| \ge \left|M_\mu \cap B^d _{\varepsilon' } (\mu_j)\right| \ge \frac{3}{5}R. \]
\end{proof}
\begin{lemma}
\label{lem:boost-far}
Suppose \(S \ge \frac{3}{5}R\). If \(\wh\mu\in M_\mu\) satisfies \(\|\wh\mu - \mu_j\|_2 > 3 \varepsilon '\) for all \(j\in [k]\), then \(|M_\mu \cap B^d_{2 \varepsilon '}(\wh\mu)| \le \frac{2}{5}R\)
\end{lemma}
\begin{proof}
We will prove this by contradiction. Suppose there exists such a \(\wh\mu\in M_\mu\) that \(\|\wh\mu - \mu_j\|_2 > 3 \varepsilon '\) for all \(j\in [k]\) and \(|M_\mu\cap B^d_{2 \varepsilon '}(\wh\mu)| > \frac{2}{5}R\). Then all the \(\wh\mu'\in M_\mu \cap B^d_{2 \varepsilon '}(\wh\mu)\) must from some bad round, since for any \(j\in [k]\), \begin{align*}
    \left\| \wh\mu' - \mu_j \right\|_2 \ge \left\| \wh\mu - \mu_j \right\|_2 - \left\| \wh\mu' - \wh\mu \right\|_2 > 3 \varepsilon ' - 2 \varepsilon ' = \varepsilon '. 
\end{align*}
Since there are at most \(\frac{2}{5}R\) bad rounds \(\ell\) that have the result \(\{(\wh w_j^{(\ell)},\wh\mu_j^{(\ell)})\}_{j\in [k]}\) added into the set \(M\), by the pigeonhole principle, there exist distinct \(\wh\mu', \wh\mu'' \in M_\mu\cap B^d_{2 \varepsilon '}(\wh\mu)\) from the same round \(\ell\). However, in this case \begin{align*}
    \|\wh\mu' - \wh\mu''\|_2 \le \|\wh\mu' - \wh\mu\|_2 + \|\wh\mu - \wh\mu''\|_2 \le 4 \varepsilon ' \le \gamma/2,
\end{align*}
which means in round \(\ell\), the result \(\{(\wh w_j^{(\ell)},\wh\mu_j^{(\ell)})\}_{j\in [k]}\) will not be added into \(M\). This is a contradiction. 
\end{proof}
We are ready to show the correctness of \cref{alg:boosting}, particularly the \textbf{for} loop in lines~\ref*{alg:boost:line:for}~to~\ref*{alg:boost:line:endfor}. 
We will show by induction that, there is a permutation \(\pi\) such that in the \(j\)-th iteration, \begin{enumerate}
    \item the \(\wh\mu_j\) chosen in line~\ref*{alg:boost:line:choose-mu} will satisfy \(\|\wh\mu_j - \mu_{\pi(j)}\|_2 \le 3 \varepsilon ' \le \varepsilon \); \label{enum:IH-mu}
    \item the \(\wh w_j\) chosen in line~\ref*{alg:boost:line:choose-w} will satisfy \(|\wh w_j - w_{\pi(j)}| \le \varepsilon _w\); \label{enum:IH-w}
    \item after line~\ref*{alg:boost:line:remove}, \(M_\mu\cap B^d_{3 \varepsilon '}(\mu_{\pi(j)}) = \varnothing\); \label{enum:IH-clean}
    \item after line~\ref*{alg:boost:line:remove}, for \(j' \in [ k] \backslash \{\pi(j'')\}_{j''\in [j]}\), \(M_\mu \cap B^d_{\varepsilon '}(\mu_{j'})\) will not be removed. \label{enum:IH-safe}
\end{enumerate} 

When \(j = 1\), the above four statements are proved as follows. 
\begin{enumerate}
\item From the proof of \cref{lem:boost-close}, we know for each true \(\mu_{j'}\), \(j'\in [k]\), \(|M_\mu \cap B^d_{\varepsilon '}(\mu_{j'})| \ge \frac{3}{5}R\), and thus for some \(\wh\mu\in M_\mu\), \(|M_\mu \cap B^d_{2 \varepsilon '}(\wh\mu)| \ge \frac{3}{5}R\). 
Hence, we can indeed pick in line~\ref*{alg:boost:line:choose-mu} a \(\wh\mu_j \in M_\mu\) that \(|M_\mu \cap B^d_{2 \varepsilon '}(\wh\mu_j)| \ge \frac{3}{5}R > \frac{2}{5}R\), and by \cref{lem:boost-far}, there is a \(j'\in [k]\) that \(\|\wh\mu_j - \mu_{j'}\|_2 \le 3 \varepsilon ' \le \varepsilon \). 
Such \(j'\) will be unique, as otherwise \(\min_{i\ne j}\|\mu_i- \mu_j\|_2 \le 6 \varepsilon ' < \gamma\). 
Let \(\pi(j) = j'\). 
\item Since \(B^d_{\varepsilon '}(\mu_{\pi(j)}) \subseteq B^d_{4 \varepsilon '}(\wh\mu_j)\), and at least \(\frac{3}{5}R\) points in \(B^d_{\varepsilon '}(\mu_{\pi(j)})\) are from some good rounds, the set \(W:=\{\wh w: (\wh w, \wh\mu)\in M, \|\wh \mu_j - \wh\mu\|_2 \le 4 \varepsilon '\}\) contains at least \(\frac{3}{5}R\) weights \(\wh w\) that \(|\wh w - w_{\pi(j)}| \le \varepsilon _w\). Meanwhile, \(|W| \le R\). Otherwise, since there are at most \(R\) rounds that have added the result into \(M\), there will be distinct \((\wh w', \wh\mu'), (\wh w'', \wh\mu'') \) that come from the same round \(\ell\), such that \(\|\wh\mu' - \wh\mu''\|_2 \le 8 \varepsilon ' \le \gamma/2 \), which means \(\{(\wh w_j^{(\ell)}, \wh\mu_j^{(\ell)}) \}_{j\in [k]}\) will not be added into \(M\), which is a contradiction. Therefore, \(\wh w_j = \mathrm{median}(W)\) will satisfy that \(|\wh w_j - w_{\pi(j)}| \le \varepsilon _w\). 
\item In line~\ref*{alg:boost:line:remove}, we remove \(\{(\wh w, \wh\mu): \|\wh\mu_j - \wh \mu\|_2 \le 6 \varepsilon ' \} \) from \(M\), while every \(x\in B^d_{3 \varepsilon '}(\mu_{\pi(j)})\) will satisfy \(\|\wh \mu_j - x\|_2 \le \|\wh \mu_j - \mu_{\pi(j)}\|_2 + \|\mu_{\pi(j)} - x\|_2 \le 6 \varepsilon '\). 
\item Meanwhile, we will not remove any points in \(B^d_{\varepsilon '}(\mu_{j''})\) for \(j'' \ne \pi(j)\), because otherwise if \(x \in B^d_{\varepsilon '}(\mu_{j''})\) is removed, then \begin{align*}
    \left\| \mu_{\pi(j)} - \mu_{j''} \right\|_2 \le \left\| \mu_{\pi(j)} - \wh\mu_{j} \right\|_2 + \left\| \wh\mu_{j} - x \right\|_2 + \left\| x - \mu_{j''} \right\|_2 \le 3 \varepsilon ' + 6 \varepsilon ' + \varepsilon' < \gamma,
\end{align*}
which is a contradiction. 
\end{enumerate}

When \(j \ge 2\), assume the four statements hold for all the previous iterations, the statements are proved as follows. 
\begin{enumerate}
\item From part~\labelcref{enum:IH-safe} of the induction hypothesis and the proof of \cref{lem:boost-close}, for \(j' \in [k] \backslash \{\pi(j'') \}_{j'' \in [j-1]}\), \(|M_\mu \cap B^d_{\varepsilon '}(\mu_{j'})| \ge \frac35R\), and thus there is some \(\wh\mu\in M_\mu\) that \(|M_\mu \cap B^d_{2 \varepsilon '}(\wh \mu)| \ge \frac35R\). Hence, we can indeed pick in line~\ref*{alg:boost:line:choose-mu} a \(\wh\mu_j \in M_\mu\) that \(|M_\mu \cap B^d_{2 \varepsilon '}(\wh\mu_j)| \ge \frac{3}{5}R > \frac{2}{5}R\), and by \cref{lem:boost-far}, there is a \(j'\in [k]\) that \(\|\wh\mu_j - \mu_{j'}\|_2 \le 3 \varepsilon ' \le \varepsilon \). 
Similarly, such \(j'\) will be unique. 
And from part~\labelcref{enum:IH-clean} of the induction hypothesis, \(j' \ne \pi(j'')\) for \(j'' \in [j-1]\). 
Therefore, it is valid to let \(\pi(j) = j'\). 
\item Since at least \(\frac{3}{5}R\) points in \(B^d_{\varepsilon '}(\mu_{\pi(j)})\) are from good rounds, similarly as in the case for \(j=1\), \(|\wh w_j - w_{\pi(j)}| \le \varepsilon _w\). 
\item[3, 4.] Also similarly, in line~\ref*{alg:boost:line:remove}, we will remove any points \(\wh\mu\in M_\mu\) that \(\|\wh\mu - \mu_{\pi(j)}\|_2 \le 3 \varepsilon '\), without removing any points in \(B^d_{\varepsilon '}(\mu_{j''})\) for \(j'' \ne \pi(j)\). 
\end{enumerate}

Therefore, we have showed that with probability \(1-\delta\), \cref{alg:boosting} outputs \(\{(\wh w_j, \wh\mu_j) \}_{j\in [k]}\) such that \(\max_{j\in [k]} \| \wh\mu_j - \mu_{\pi(j)}\|_2 \le \varepsilon \) and \(\max_{j\in [k]} \| \wh w_j - w_{\pi(j)}\|_2 \le \varepsilon_w\). The algorithm uses \(n(\varepsilon ')R = O(n(\varepsilon ')\log(1/\delta))\) samples. For the running time, since it takes \(O(|M|^2d)\) time in line~\ref*{alg:boost:line:choose-mu} to find such \(\wh\mu_j\), \(O(|M|)\) times in line~\ref*{alg:boost:line:choose-w} to find the median, and \(O(|M|)\) time in line~\ref*{alg:boost:line:remove} to remove the subset, the algorithm runs in \(T(\varepsilon ')R + O(k(kR)^2d) = O(T(\varepsilon ')\log(1/\delta) + k^3d \log(1/\delta)^2)\). 
\end{proof}

\section{Resilience from Sub-Weibull tails: Proof of \cref{lem:resilience-subW}}
\label[appendix]{app:resilience}
Without loss of generality, assume \(\mu = 0\) and \(\sigma = 1\), and for simplicity assume \(C_0 = 1\). That is, for all \(v\in S^{d-1}\) and \(t>0\), \[\Pr_{X\sim D} [|\langle X,v \rangle | \ge t] \le \exp (-t^{\beta}).\] 
Given \(\delta,\alpha \in (0, 1)\) and \(n\) i.i.d.\ samples \(x_1,\dots, x_n\). 
Let \[M(v) = \max_{\substack {S\subseteq [n] \\ |S| = \alpha n}} \left\langle \frac{1}{\alpha n} \sum_{i\in S} x_i, v \right\rangle = \max_{\substack {S\subseteq [n] \\ |S| = \alpha n}} \frac 1{\alpha n} \sum_{i\in S} \langle x_i, v \rangle\] and \(M = \sup_{v\in S^{d-1}} M(v)\), then 
\[M = \sup_{v\in S^{d-1}} \max_{\substack {S\subseteq [n] \\ |S| = \alpha n}}\left\langle \frac{1}{\alpha n} \sum_{i\in S} x_i, v \right\rangle =  \max_{\substack {S\subseteq [n] \\ |S| = \alpha n}} \sup_{v\in S^{d-1}}\left\langle \frac{1}{\alpha n} \sum_{i\in S} x_i, v \right\rangle = \max_{\substack {S\subseteq [n] \\ |S| = \alpha n}} \left\| \frac{1}{\alpha n} \sum_{i\in S} x_i \right\|, \] 
for which we want to find an upper bound \(\Delta(\alpha)\). 
Let \(N\subseteq S^{d-1}\) be an \(\varepsilon \)-net of the unit sphere \(S^{d-1}\), then \(|N| \le (1+2/ \varepsilon )^{d}\). 
Take \(\varepsilon =1/2\), we have \(|N| \le 5^d\) and for any \(u \in S^{d-1}\), there exists \(v\in N\) that \(\|u-v\|_2 \le 1/2\). 
Then, for any \(S\subseteq [n]\) with \(|S| = \alpha n\), \begin{align*}
\left\langle u, \frac{1}{|S|} \sum_{i\in S} x_i \right\rangle& = \left\langle v, \frac{1}{|S|} \sum_{i\in S} x_i \right\rangle + \left\langle u-v, \frac{1}{|S|} \sum_{i\in S} x_i \right\rangle \\
&\le \left\langle v, \frac{1}{|S|} \sum_{i\in S} x_i \right\rangle +  \| u-v\|_2  \left\|\frac{1}{|S|}\sum_{i\in S} x_i  \right\|_2 \\
&\le M(v) + \frac{1}{2} M.
\end{align*}
Take the supreme over \(u \in S^{d-1}\) and the maximum over \(S\) on the LHS, we have \(M \le M(v) + \frac{1}{2}M\), which implies \(M \le 2M(v)\). Therefore, we only need to upper bound \(M(v)\) for finitely many \(v\in N\). 

Fix \(v \in S^{d-1}\), let \(Y =\langle X, v \rangle\) for \(X \sim D\), and \(y_i = \langle x_i, v \rangle\) for \(i\in [n]\), which can be viewed as \(n\) i.i.d.\ samples from \(Y\). 
Moreover, let \(y_{(i)}\) be the \(i\)-th smallest element among \(y_1,\dots, y_n\). 
Thus, \[M(v) = \max_{S\subseteq[n], |S| = \alpha n} \frac1{\alpha n} \sum_{i\in S} y_i = \frac{1}{\alpha n} \sum_{i=n-\alpha n+1}^n y_{(i)}\] (assuming \(\alpha n\) is an integer for simplicity). Also, note that we now have \(\Pr_{Y}[Y \ge t] \le \exp(-t^{\beta})\) for all \(t>0\). 

Let \(t_0 = (\ln\frac1\alpha )^{1/\beta}\), \(L_j = [2^j t_0, 2^{j+1} t_0)\), \(N_j = \left| \{i:y_i \in L_j \} \right|\), for \(j \in \Z_{\ge 0}\), and \(C>0\) be some large enough absolute constant. 
Suppose \(N_j \le \frac{C\alpha n}{3^j}\) for all \(j\ge 0\), then \begin{align*}
    M(v) &= \frac{1}{\alpha n} \left( \alpha n \cdot t_0 + \sum_{j\ge 0} N_j \cdot 2^{j+1} t_0 \right) \\
    &\le t_0 + \frac{t_0}{\alpha n} \sum_{j \ge 0 } \frac{C\alpha n}{3^j} \cdot 2^{j+1} \\
    &\le t_0 + 6C t_0.
\end{align*}
Therefore, by the union bound, \begin{align*}
    \Pr[M(v) > (6C+1) t_0] \le \Pr\left[\exists j\ge 0, N_j > \frac{C\alpha n}{3^j}\right] \le \sum_{j \ge 0} \Pr \left[ N_j \ge \frac{C\alpha n}{ 3^j} \right].
\end{align*}
Since \[\Pr[y_i \in L_j] \le \Pr[y_i \ge 2^j t_0] \le \exp (-(2^j t_0)^{\beta}) = \alpha^{2^{j\beta}},\] by the multiplicative Chernoff bound, \[\Pr\left[N_j \ge \frac{C \alpha n}{3^j}\right] \le \begin{cases}
    \left( \frac{en \alpha^{2^{j\beta}}}{C\alpha n /3^{j}} \right)^{C\alpha n/3^j}, & C\alpha n /3^j > 1, \\
    en\alpha^{2^{j\beta}}, & C\alpha n/3^j \le 1. 
\end{cases}\]
The second case is because \(N_j\) is an integer, and thus \(\Pr[N_j \ge t] = \Pr[N_j\ge 1]\) if \(0 < t \le 1\). 

When \(C\alpha n/3^j > 1\), i.e., \(j < \ln(C\alpha n)/\ln 3=:j^\star\), we have for sufficiently small \(\alpha>0\), \begin{align*}
\left(\frac{en \alpha^{2^{j\beta}}}{C\alpha n /3^{j}} \right)^{C\alpha n/3^j}\le \left(\frac{e \alpha^{2^{j\beta/2}-1}}{C} \right)^{C\alpha n/3^j}\le  \left( \frac eC \right)^{C\alpha n/3^j}\alpha^{C\alpha n\cdot \frac{2^{j\beta/2}-1}{3^j}}. 
\end{align*}
Let \(h(x) = \frac{2^{x\beta/2}-1}{3^x}\) for \(x\ge 0\). 
If \(\beta \ge 2 \log_2 3\), then \(h(x)\) is increasing, and thus \begin{align*}
    \sum_{j=0}^{j^\star} \left( \frac eC \right)^{C\alpha n/3^j}\alpha^{C\alpha n\cdot \frac{2^{j\beta/2}-1}{3^j}} \le \left( \frac eC \right)^{C\alpha n} + \sum_{j=1}^{j^\star} \alpha ^{C\alpha n \cdot \frac{2^{\beta/2}-1}{3}} \le \left( \frac eC \right)^{C\alpha n} + \frac{\ln(C\alpha n)}{\ln 3} \alpha^{C\alpha n \frac{2}{3}} \le \exp \left( -\Omega(\alpha n) \right).
\end{align*}
If \(\beta < 2\log_2 3\), then by calculating the derivative, \(h(x)\) is increasing on \(x\in [0, x^\star]\) and is decreasing on \(x\in [x^\star, +\infty)\), where \(x^\star =\frac{\ln\ln 3 - \ln(\ln 3-\frac\beta2 \ln2)}{\frac\beta2 \ln2}\). 
Thus, \begin{align*}
h(x) &\ge \min\{h(1), h(j^\star)\} = \min \left\{ \frac{2^{\beta/2}-1}{3} , \left( \frac{2^{\beta/2}}{3} \right)^{\ln(C\alpha n)/\ln 3} - \frac1 {3^{\ln(C\alpha n)/\ln 3}}\right\} = \Omega \left( (C\alpha n)^{\frac{\beta}{2\log_2 3} -1} \right)
\end{align*}
for \(x\in [1,j^\star]\), and \begin{align*}
\sum_{j=0}^{j^\star} \left( \frac eC \right)^{C\alpha n/3^j}\alpha^{C\alpha n\cdot \frac{2^{j\beta/2}-1}{3^j}} \le \left( \frac eC \right)^{C\alpha n} + \sum_{j=1}^{j^\star} \alpha^{\Omega \left( (C\alpha n)^{\frac{\beta}{2\log_2 3}} \right)} \le \exp \left( -\left( \alpha n \right)^{\Omega(\beta)} \right). 
\end{align*}
When \(j\ge j^\star\), since \[\alpha^{2^{j\beta}} = \alpha^{2^{j^\star\beta} \cdot 2^{(j-j^\star)\beta}} \le \alpha^{2^{j^\star \beta} (1+(j-j^\star)\beta\ln 2)},\]
we have 
\begin{align*}
\sum_{j\ge j^\star} en\alpha^{2^{j\beta}} \le en\sum_{j\ge j^\star} \alpha^{2^{j^\star \beta} (1+(j-j^\star)\beta\ln 2)} \le en \alpha^{2 ^{j^\star \beta}} \frac{1}{1 - \alpha^{2^{j^\star\beta}\cdot \beta\ln 2}} \le 2en \alpha^{(C\alpha n)^{\beta/\log_2 3}} \le \exp (- (\alpha n)^{\Omega(\beta)}). 
\end{align*}
Combining the two cases, we have \begin{align*}
    \Pr[M(v) > (6C+1)t_0] \le \sum_{j\ge 0} \Pr \left[ N_j \ge \frac{C\alpha n}{3^j} \right] &\le \begin{cases}
    \exp \left( -(\alpha n)^{\Omega(\beta)} \right), & \beta < 2\log_2 3, \\
    \exp \left( -\Omega(\alpha n)  \right), & \beta \ge 2\log_2 3 
\end{cases} \\
&\le \exp \left( - (\alpha n)^{\Omega(\min\{\beta, 1\})} \right). 
\end{align*}

By a union bound over the \(\frac12\)-net \(N\), we get \begin{align*}
    \Pr[M > (12C+2)t_0] \le \Pr[\exists v\in N, M(v) > (6C+1)t_0] \le 5^d \exp \left( - (\alpha n)^{\Omega(\min\{\beta, 1\})} \right).
\end{align*}
Hence, to make \(\Pr[M > (12C + 2)t_0] \le \delta\), we only need \(n \ge \frac{1}{\alpha} \left( d+\log(1/\delta) \right)^{O(\max\{1/\beta, 1\})}\). Here \(\Delta(\alpha) = (12C+2)t_0 = O(\ln(1/\alpha)^{1/\beta})\). 

\section{Local Convergence for Gaussians}
\label[appendix]{app:local-convergence}
In this section, we will briefly describe the iterative algorithm by \citet{regev_learning_2017} for local convergence when the warm start estimations are accurate up to \(1/\poly(k')\) error, and then explain how to generalize the algorithm so that it still works in the presence of Laplace components as in \cref{cor:learning-mix-laplace-gaussian}. 

Suppose for now we only have Gaussian components \(\cN(\mu_j', I)\) with weights \(w_j'\), \(j\in [k']\), and we have rough estimates \(\wt \mu_j'\) such that \(\|\mu_j' - \wt\mu_j'\|_2 \le 1/\poly(k') \). 
In their algorithm, they consider the input mixture distribution restricted to some regions \(S_j\) so that it has large mass from the \(j\)-th component, and relatively small mass from all the others. 
For each \(j\in [k']\), \[S_j = \left\{x\in \R^d : \forall \ell\in [k'] \backslash \{j\}, \left| \langle x-\wt\mu_j', e_{j\ell}' \rangle \right| \lesssim \sqrt{\log k'}, \text{ and } \|x-\wt\mu_j'\|_2 \lesssim \sqrt{d}+\sqrt{\log k'} \right\},\] where \(e_{j\ell}'\) is the unit vector along \(\wt\mu_j' - \wt\mu_\ell'\). 
Then they set up a non-linear equation system where the true means are the solution, and solve it by the Newton method. Specifically, the equation system is \[F_j(\bar\mu_1',\dots, \bar\mu_{k'}'):=\sum_{j=1}^{k'} w_i' \int_{y\in S_j} (y-\wt\mu_j') \cdot \frac{1}{(2\pi)^{d/2}}\exp \left( - \frac{\|y-\bar\mu_j'\|_2^2}{2} \right)\mathrm{d} y = u_j, \] where \(u_j\) is the sample mean of the input distribution restricted on \(S_j\), after subtracting \(\wt\mu_j\), which is indeed equal to LHS when \(\bar\mu_j'\) is the true mean for \(j\in [k']\). 
Let \(F\) denote \((F_1,\dots, F_{k'})\) and \(u\) denote \((u_1,\dots, u_{k'})\), the Newton method will have the iterative update as \begin{align*}
    \mu'^{(0)} &= \wt \mu', \\
    \mu'^{(t+1)} &= \mu'^{(t)} + (\nabla F(\mu'^{(t)}))^{-1} (u - F(\mu'^{(t)})).
\end{align*}
Note that the rough estimate \(\wt\mu_j'\) will be used both to define \(S_j\) and as the initialization. 
Also, note that in each iteration, one can estimate the integrals in \(F\) and \(\nabla F\) by generating samples from \(\cN(\mu'^{(t)}, I)\) and estimate \(u\) by the input samples. 
Thus, the accuracy of the final estimation is guaranteed by the robust version of the Newton method, as long as \(\delta \| (\nabla F)^{-1}\|\|\nabla ^2 F\| \le 1/2\), where \(\delta=1/\poly(k)\) is the accuracy of the initial estimation, and \(\|\cdot\|\) is the operator norm. 
To upper bound \(\|(\nabla F)^{-1}\|\), they show that \(\nabla F\) has some diagonal dominance property. 
This is from the standard fact of Gaussian tails, as from the definition of \(S_j\), the mass of \(\cN(\mu_j', I)\) outside \(S_j\) will be \(1/\poly(k)\) small; meanwhile the total mass of the other components inside \(S_j\) will be \(1/\poly(k)\) if the minimum separation \(\gamma_{\text{F}}=\Omega(\sqrt{\log k})\) (in their settings) or even \(\exp(-\poly(k))\) if the minimum separation is \(\gamma_{\text{F}}=\poly(k)\) (in our settings). 

Now if we have additionally Laplace components \(\lap(\mu_j, I)\) with weights \(w_j\), \(j\in [k]\), we will need to modify the definition of \(S_j\), otherwise the Laplace components could have large mass on \(S_j\), e.g., some \(\mu_j\) could even lie in \(S_{j'}\). 
As a result, we only need to add linear constraints to exclude the Laplace regions, similarly as how the linear constraints in the original definition of \(S_j\) exclude the other Gaussian components. 
Specifically, we will define 
\begin{align*}
S_j = \Bigl\{ x\in \R^d : &\forall \ell\in [k], \left| \langle x-\wh\mu_j, e_{j\ell} \rangle \right| \lesssim \sqrt{\log k'}, \\
&\forall \ell\in [k'] \backslash \{j\}, \left| \langle x-\wt\mu_j', e_{j\ell}' \rangle \right| \lesssim \sqrt{\log k'} , \text{ and } \|x-\wt\mu_j'\|_2 \lesssim \sqrt{d}+\sqrt{\log k'} \Bigr\},
\end{align*}
where \(e_{j\ell}\) is the unit vector along \(\wt\mu_j' - \wh\mu_\ell\). 
Then the guarantee for the Newton method is still valid from the following facts. 
\begin{enumerate}
    \item \(\cN(\mu_j', I)\) still has small mass outside \(S_j\), since \(k = \poly(k')\), which follows from the assumption \(w_{\min} \ge 1/\poly(k')\) in \cref{cor:learning-mix-laplace-gaussian}. 
    \item \(\lap(\mu_{j'}, I)\) has only \(\exp(-\poly(k'))\) mass inside \(S_j\), since Laplace distributions have exponential tail and we assume the separation between the Gaussian and Laplace components \(\gamma_{\text{SF}}=\poly(k')\). 
\end{enumerate}

\end{document}